\pdfoutput=1
\documentclass{jfm}

            \newif\ifdraft \newif\ifcolorfigs
            \drafttrue \colorfigstrue
            \newif\ifJFM \newif\ifblog  \newif\ifthesis
            \blogfalse \thesisfalse \JFMtrue

 \draftfalse     % For web and JFM version, no comments
\colorfigstrue

\ifJFM
  \usepackage{upmath}
\fi

\usepackage{ifpdf}
\ifpdf
  \usepackage[pdftex]{color,graphicx}
  \usepackage{epstopdf}
\else
  \usepackage[dvips]{color,graphicx}
\fi

\usepackage{float}
\usepackage{natbib}
\usepackage{amsmath,amsfonts,amssymb,amsbsy,amscd,amsgen}
\usepackage{url}
\usepackage{subfigure}
\usepackage{ifthen}

\usepackage[active]{srcltx}

\ifcolorfigs
  \graphicspath{{./figs/}}  %% directories with color graphics files
  \usepackage[pdftex,colorlinks]{hyperref} %% hyperlinks, LAST package called
\else
  \graphicspath{{./figs-bw/}}    %% BW graphics files paper copy, no hyperlinks
\fi

\ifdraft    % display comments in text
   
   \newcommand{\PC}[1]{$\footnotemark\footnotetext{PC: #1}$}
   
   \newcommand{\DV}[1]{$\footnotemark\footnotetext{DV: #1}$}
   
   \newcommand{\JFG}[1]{$\footnotemark\footnotetext{JG: #1}$}
   
   \newcommand{\JH}[1]{$\footnotemark\footnotetext{JH: #1}$}

   \newcommand{\ES}[1]{$\footnotemark\footnotetext{ES: #1}$}
   
   \newcommand{\file}[1]{$\footnotemark\footnotetext{{\bf file} #1}$}
   \newcommand{\mycomment}[2]{\noindent \textbf{\underline{#1}}: \emph{#2}}
\else   % drop comments
   
   \newcommand{\PC}[1]{}
   \newcommand{\JFG}[1]{}
   \newcommand{\DV}[1]{}
   \newcommand{\JH}[1]{}

   \newcommand{\ES}[1]{}
   
   \newcommand{\file}[1]{}
   \newcommand{\mycomment}[2]{}
   \newcommand{\edit}[1]{#1}               % for the journal

\fi %%%%% COMMENTS END %%%%%%%%%%%%%%%

\ifcolorfigs            %% switch for color/BW figure captions
  \newcommand{\colorcomm}[2]{#1}
  \newcommand{\weblink}[1]{{\tt \href{http://#1}{#1}}}
  \newcommand{\HREF}[2]{{\href{#1}{#2}}}

\else
  \newcommand{\colorcomm}[2]{#2}
  \newcommand{\weblink}[1]{{\tt #1}}
  \newcommand{\HREF}[2]{{#2}}

\fi

\ifJFM
 \newcommand{\citecomm}[2]{#2}  % J Fluid Mech style citation - no names
\else
 \newcommand{\citecomm}[2]{#1}  % physics style citation - with names
\fi

\newcommand{\rf}     [1] {~\cite{#1}}

\newcommand{\refeq}  [1] {(\ref{#1})}

\newcommand{\reffig} [1] {figure~\ref{#1}}

\newcommand{\refFig} [1] {Figure~\ref{#1}}

\newcommand{\reftab} [1] {table~\ref{#1}}

\newcommand{\refsect}[1] {\S\,\ref{#1}}
\newcommand{\refsects}[2] {\S\,\ref{#1} and \S\,\ref{#2}}

\newcommand{\beq}{\begin{equation}}
\newcommand{\continue}{\nonumber \\ }
\newcommand{\nnu}{\nonumber}
\newcommand{\eeq}{\end{equation}}
\newcommand{\ee}[1] {\label{#1} \end{equation}}

\newcommand{\bea}{\begin{eqnarray}}
\newcommand{\eea}{\end{eqnarray}}
\newcommand{\barr}{\begin{array}}
\newcommand{\earr}{\end{array}}

\newsavebox{\bartName}

{\hspace*{\fill}\nolinebreak[1]\usebox{\bartName}\vspace*{1ex}\end{minipage}}

\newcommand{\ie}{{i.e.}}        % APS
\newcommand{\eg}{{e.g.}}        % APS

\ifJFM                          % J Fluid Mech macros

\renewcommand\eg{e.g.\ }
\else
\fi

\defcitealias{GHCW07}{GHC}

\newcommand{\NS}{Navier-Stokes}
\newcommand{\NSe}{Navier-Stokes equations}

\newcommand{\Reynolds}{\textit{Re}}  % Reynolds number
\newcommand{\pC}{plane Couette}

\newcommand{\pCf}{plane Couette flow}
\newcommand{\PCf}{Plane Couette flow}

\newcommand{\eqv}{equilibrium}
\newcommand{\Eqv}{Equilibrium}
\newcommand{\eqva}{equilibria}
\newcommand{\Eqva}{Equilibria}

\newcommand{\reqv}{traveling wave}

\newcommand{\reqva}{traveling waves}
\newcommand{\Reqva}{Traveling waves}
\newcommand{\reqvD}{traveling-wave}

\newcommand{\po}{periodic orbit}

\newcommand{\hec}{heteroclinic connection}

\newcommand{\cohStr}{coherent state}
\newcommand{\recurrStr}{recurrent coherent state}

\newcommand{\stateDsp}{state-space}
\newcommand{\StateDsp}{State-space}

\newcommand{\statesp}{state space}
\newcommand{\bu}{\ensuremath{{\bf u}}}
\newcommand{\bv}{\ensuremath{{\bf v}}}
\newcommand{\bff}{\ensuremath{{\bf f}}}

\newcommand{\hbu}{\tilde{{\bf u}}}

\newcommand{\be}{{\bf e}}
\newcommand{\bx}{{\bf x}}
\newcommand{\ex}{{\hat{\bf x}}} % unit vectors

\newcommand{\ez}{{\hat{\bf z}}}
\newcommand{\bnabla}{\ensuremath{\bf \nabla}}
\newcommand{\GPCF}{\ensuremath{\Gamma}} % Hoyle notation, equivariant symmetry group

\newcommand{\tNS}{\ensuremath{{\text{NS}}}}

\newcommand{\tEQ}{\ensuremath{{\text{EQ}}}}
\newcommand{\tLM}{\ensuremath{{\text{EQ}_0}}}
\newcommand{\tLB}{\ensuremath{{\text{EQ}_1}}}
\newcommand{\tUB}{\ensuremath{{\text{EQ}_2}}}
\newcommand{\tNNB}{\ensuremath{{\text{EQ}_3}}}
\newcommand{\tNB}{\ensuremath{{\text{EQ}_4}}}

\newcommand{\tEQzero}{\ensuremath{{\text{EQ}_0}}}
\newcommand{\tEQone}{\ensuremath{{\text{EQ}_1}}}
\newcommand{\tEQtwo}{\ensuremath{{\text{EQ}_2}}}
\newcommand{\tEQthree}{\ensuremath{{\text{EQ}_3}}}
\newcommand{\tEQfour}{\ensuremath{{\text{EQ}_4}}}
\newcommand{\tEQfive}{\ensuremath{{\text{EQ}_5}}}
\newcommand{\tEQsix}{\ensuremath{{\text{EQ}_6}}}
\newcommand{\tEQsev}{\ensuremath{{\text{EQ}_7}}}
\newcommand{\tEQeight}{\ensuremath{{\text{EQ}_8}}}
\newcommand{\tEQnine}{\ensuremath{{\text{EQ}_9}}}

\newcommand{\tEQten}{\ensuremath{{\text{EQ}_{10}}}}
\newcommand{\tEQelev}{\ensuremath{{\text{EQ}_{11}}}}
\newcommand{\tEQtwel}{\ensuremath{{\text{EQ}_{12}}}}
\newcommand{\tEQthirt}{\ensuremath{{\text{EQ}_{13}}}}

\newcommand{\tTWone}{\ensuremath{{\text{TW}_1}}}  % spanwise
\newcommand{\tTWtwo}{\ensuremath{{\text{TW}_2}}} % Divakar D1, lower streamwise
\newcommand{\tTWDone}{\ensuremath{{\text{TW}_2}}} % Divakar D1, lower streamwise
\newcommand{\tTWthree}{\ensuremath{{\text{TW}_3}}}  % upper streamwise

\colorcomm{\definecolor{orangina}{rgb}{0.9,0.7,0}}{}
\newcommand{\sLM}{\ensuremath{\odot}}

\newcommand{\sLB}{\colorcomm{{\Large \ensuremath{\color{blue}\circ}}}
                            {{\Large \ensuremath{\circ}}}}
\newcommand{\sUB}{\colorcomm{{\Large \ensuremath{\color{blue}\bullet}}}
                            {{\Large \ensuremath{\bullet}}}}

\newcommand{\sNNB}{\colorcomm{{\scriptsize \ensuremath{\color{red}\square}}}
                            {{\scriptsize \ensuremath{\square}}}}
\newcommand{\sNB}{\colorcomm{{\scriptsize \ensuremath{\color{red}\blacksquare}}}
                            {{\scriptsize \ensuremath{\blacksquare}}}}

\newcommand{\sEQone}{\colorcomm{{\Large \ensuremath{\color{blue}\circ}}}
                            {{\Large \ensuremath{\circ}}}}
\newcommand{\sEQtwo}{\colorcomm{{\Large \ensuremath{\color{blue}\bullet}}}
                            {{\Large \ensuremath{\bullet}}}}

\newcommand{\sEQthree}{\colorcomm{{\scriptsize \ensuremath{\color{red}\square}}}
                            {{\scriptsize \ensuremath{\square}}}}
\newcommand{\sEQfour}{\colorcomm{{\scriptsize \ensuremath{\color{red}\blacksquare}}}
                            {{\scriptsize \ensuremath{\blacksquare}}}}

\newcommand{\sEQfive}{\colorcomm{\ensuremath{\color{green}\lozenge}}
                            {\ensuremath{\lozenge}}}
\newcommand{\sEQsix}{\colorcomm{\ensuremath{\color{green}\blacklozenge}}
                            {\ensuremath{\blacklozenge}}}

\newcommand{\sEQsev}{\colorcomm{\ensuremath{\triangleleft}}
                            {\ensuremath{\triangleleft}}}
\newcommand{\sEQeight}{\colorcomm{\ensuremath{\blacktriangleleft}}
                            {\ensuremath{\blacktriangleleft}}}

\newcommand{\sEQnine}{\colorcomm{\ensuremath{\color{orangina}\bigstar}}
                            {\ensuremath{\bigstar}}}

\newcommand{\sEQten}{\colorcomm{\ensuremath{\color{magenta}\triangledown}}
                            {\ensuremath{\triangledown}}}
\newcommand{\sEQelev}{\colorcomm{\ensuremath{\color{magenta}\blacktriangledown}}
                            {\ensuremath{\blacktriangledown}}}

\newcommand{\sEQtwel}{\colorcomm{\ensuremath{\color{magenta}\triangle}}
                            {\ensuremath{\triangle}}}
\newcommand{\sEQthirt}{\colorcomm{\ensuremath{\color{magenta}\blacktriangle}}
                            {\ensuremath{\blacktriangle}}}

\newcommand{\sTWone}{\colorcomm{{\large \ensuremath{\color{blue}\triangleright}}}
                            {{\large \ensuremath{\triangleright}}}}

\newcommand{\uEQ}{\ensuremath{\bu_{\text{\tiny EQ}}}}

\newcommand{\huUB}{\ensuremath{\hbu_{\text{\tiny EQ2}}}}

\newcommand{\uREQV}{\ensuremath{\bu_{\text{\tiny TW}}}}

\newcommand{\isotropyG}[1]{\ensuremath{H_{\text{\tiny #1}}}}

\newcommand{\bCell}{\ensuremath{\Omega}}
\newcommand{\bNarrow}{\ensuremath{\Omega_{\text{\tiny GHC}}}}
\newcommand{\bHKW}{\ensuremath{\Omega_{\text{\tiny{HKW}}}}}

\newcommand{\pd}[2]{\frac{\partial #1}{\partial #2}}
\newcommand{\Norm}[1]{\|{#1}\|}

\renewcommand\Re{{\rm Re\,}}

\newcommand{\shift}{\ensuremath{\ell}}
\newcommand{\trHalf}[1]{\tau_{#1}}    % 1/2 cell translation
\newcommand{\trDiscr}[2]{\tau_{#1}^{#2}}    % discrete cell translation 1/4, ...
\newcommand{\ssp}{a}            % state space point
\newcommand\vField{\ensuremath{{\bf F}}} % yet another Gibson statesp vel field
\newcommand\vCM{mean velocity}  % or `center-of-mass velocity?' `bulk momentum?'
\newcommand{\eigExp}[1][]{
\ifthenelse{\equal{#1}{}}{\ensuremath{\lambda}}{\ensuremath{\lambda^{(#1)}}}
                        }
\newcommand{\eigRe}[1][]{
\ifthenelse{\equal{#1}{}}{\ensuremath{\mu}}{\ensuremath{\mu^{(#1)}}}
                        }
\newcommand{\eigIm}[1][]{
  \ifthenelse{\equal{#1}{}}{\ensuremath{\omega}}{\ensuremath{\omega^{(#1)}}}
            }

\newcommand{\Wmnfld}[2]{%
\ifthenelse{\equal{#2}{}}{\ensuremath{W_{#1}}\!}
                         {\ensuremath{W^{#1}_{\text{\tiny #2}}}\!} %Negative space is screwing up spacing in text
                        }

\newcommand{\beUBg}[1]{\ensuremath{\be_{#1}}}

\ifblog
\newenvironment{offset}
               {\list{}{\listparindent 3em%
                        \advance\rightmargin -3em}%
                \item\relax}
               {\endlist}
\newtheorem{exmple}{\noindent\small\textsf{\textbf{Example}}}[chapter]

\newtheorem{rmark}{{\small\textsf{\textbf{Remark}}}}[chapter]

\newtheorem{exerc}{\textsf{\textbf{Exercise}}}[chapter]

\newcommand{\solution}[3]{
        {\noindent\small
         \textsf{\textbf{Solution \ref{#1}~-~#2}} %LABEL - TITLE
         \slshape\sffamily{#3}                    %TEXT
         }
         \vskip  1ex  %4mm
\fi % end of blog switch

\title[Equilibria and traveling waves of plane Couette flow]
{Equilibrium and traveling-wave solutions of plane Couette flow}
\author[J. F. Gibson, J. Halcrow, and P. Cvitanovi\'c]
{
J.\ns F.\ns G\ls I\ls B\ls S\ls O\ls N,
\ns
J.\ns  H\ls A\ls L\ls C\ls R\ls O\ls W,
\ns
\and
P.\ns C\ls V\ls I\ls T\ls A\ls N\ls O\ls V\ls I\ls \'C
}
\affiliation{
 School of Physics,
 Georgia Institute of Technology,
 Atlanta, GA  30332, USA
}

\pubyear{2009}
\begin{document}

\maketitle

We present ten new equilibrium solutions to plane Couette flow
in small periodic cells at low Reynolds number \textit{Re} and
two new traveling-wave solutions.
\edit{ The solutions are continued
under changes of \textit{Re} and spanwise period. We provide a
partial classification of the isotropy groups of plane Couette
flow and show which kinds of solutions are allowed by each
isotropy group.}
We find two complementary visualizations particularly revealing. Suitably
chosen sections of their $3D$-physical space velocity fields are helpful in
developing physical intuition about coherent structures observed in low
\textit{Re} turbulence. Projections of these solutions and their unstable
manifolds from their $\infty$-dimensional state space onto suitably chosen
2- or 3-dimensional subspaces reveal their interrelations and the role they
play in organizing turbulence in wall-bounded shear flows.

% These unstable
% flow-invariant solutions can only be computed numerically,
% but they are `exact' in the sense that they converge to
% solutions of the Navier-Stokes equations as the numerical resolution increases.

\section{Introduction}
\label{s:intro}

\begin{figure}
\begin{center}
  \includegraphics[width=0.80\textwidth]{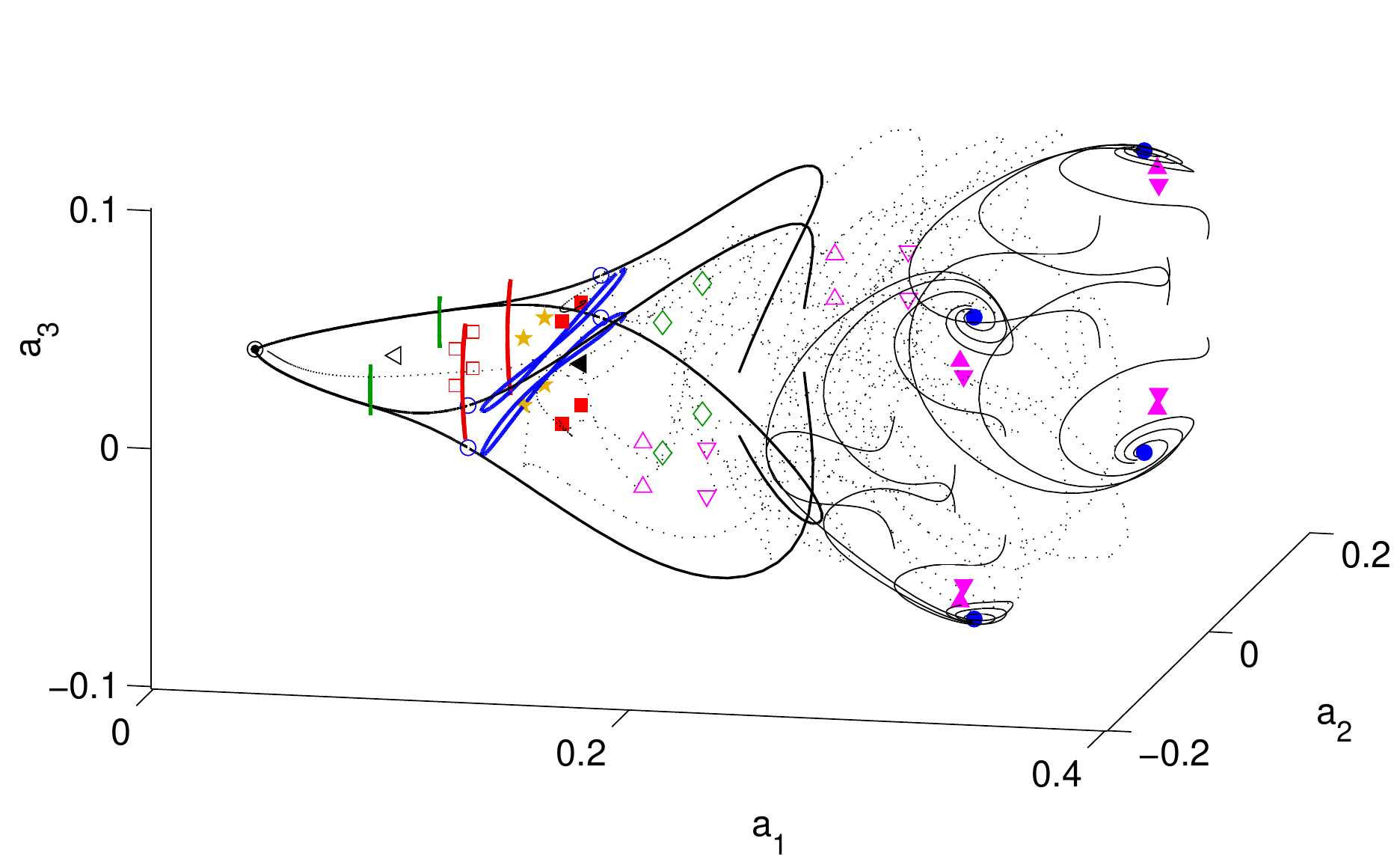}
\end{center}
 \caption{
A 3-dimensional projection of the $\infty$-dimensional \statesp\ of
plane Couette flow in the periodic cell \bNarrow\ at $\Reynolds = 400$,
showing all \eqva\ and \reqva\ discussed in \refsect{s:eqba}. \Eqva\ are
marked:
\sLM\ \tLM\ (laminar flow),
\sLB\ \tLB,
\sUB\ \tUB,
\sNNB\ \tNNB,
\sNB\ \tNB,
\sEQfive\ \tEQfive,
\sEQsev\ \tEQsev,
\sEQeight\ \tEQeight,
\sEQnine\ \tEQnine,
\sEQten\ \tEQten,
\sEQelev\ \tEQelev,
\sEQtwel\ \tEQtwel, and
\sEQthirt\ \tEQthirt.
\Reqva\ trace out closed orbits:
the spanwise-traveling \tTWone\ (\colorcomm{blue}{horizontal} loops),
streamwise \tTWDone\ (\colorcomm{green}{short vertical} lines), and
\tTWthree\ (\colorcomm{red}{longer vertical} lines). In this projection
the latter two streamwise {\reqva} appear as line segments.
The $\tLB \to \tLM$ \hec s and the $S$-invariant portion of the $\tLB$ and
$\tUB$ unstable manifolds are shown with \colorcomm{black}{solid} lines.
The cloud of dots are temporally equispaced points on a long transiently
turbulent trajectory, indicating the natural measure. The projection is
onto the translational basis \refeq{globalUBframe} constructed from \eqv\
\tUB.
}
 \label{f:WalR400all}
\end{figure}

In \citet{GHCW07} (henceforth referred to as \citetalias{GHCW07}) we
\edit{formed visualizations of the $\infty$-dimensional state-space
dynamics of moderate \Reynolds\ turbulent flows, using \edit{precisely
calculated}
\eqv\ solutions of the \NSe\ to define dynamically invariant, intrinsic, and
representation independent coordinate frames.} These
\edit{\stateDsp\ portraits} (\reffig{f:WalR400all}) offer a visualization
\edit{of the {\em dynamics} of transitionally turbulent flows},
complementary to $3D$ visualizations
\edit{of the spatial features of velocity fields} (\reffig{f:bigbox}).
Side-by-side animations of the two
visualizations illustrate their complementary strengths (see
\cite{GibsonMovies} online simulations). In these animations, $3D$
spatial visualization of instantaneous velocity fields helps elucidate
the physical processes underlying the formation of unstable coherent
structures, such as the Self-Sustained Process (SSP) theory of
\cite{Waleffe90,W95a,W97}. Running concurrently, the $\infty$-dimensional
\stateDsp\ representation enables us to track the unstable manifolds
of \eqva\ {and} the heteroclinic connections between
them \citep{GHCV08}, and {provides us with} new insight into the
nonlinear \statesp\ geometry and dynamics of moderate \Reynolds\
wall-bounded flows.

Here we continue our investigation of \eqv\ and \reqvD\ solutions of
\NSe, \edit{presenting ten new {\eqv} solutions and two new \reqva}
{of \pCf,}
and \edit{continuing} the solutions as functions of \Reynolds\ and
periodic cell size $[L_x,2,L_z]$.
\edit{
Nagata found the first pair of nontrivial \eqva\ \citep{N90} and the
first \reqv\ in \pCf\ \citep{N97}.
\citecomm{Clever and Busse\rf{CB92}}{\cite{CB92}}
found closely related \eqva\ in plane Couette flow with Rayleigh-B\'enard convection.
       }
\edit{
\cite{Cherh97} reported two-dimensional \eqva\ of plane Couette
but these were later shown to be artifacts of the truncation
\citep{Rincon07,Ehren08}.
       }
\cite{W98,W03} computed the Nagata \eqva\ guided by the SSP theory
\edit{
and showed that they were insensitive to the boundary conditions at
the wall.
       }
Other \reqva\ were computed by \cite{Visw07a} and \cite{JKSNS05}.
\citecomm{Schmiegel\rf{Schmi99}}{\cite{Schmi99}} computed and
investigated a large number of \eqva. His 1999 Ph.D.\ provides a
wealth of ideas and information on solutions to plane Couette flow,
and in many  regards the published literature  is still catching up
this work. \citetalias{GHCW07} added the dynamically important
${\bf u}_{\text{NB}}$ \eqv\ (labeled \tNB\ in this paper).
\edit{
Parallel theoretical advances have been made in channel and pipe flows,
with the discovery of \reqva\ in channel flows \citep{W01},
and \reqva\ \citep{FE03,WK04,Pringle07} and relative periodic
orbits \citep{duguet08} in pipes. Moreover, \reqva\ have been
observed experimentally in turbulent pipe flow \citep{science04}.
We refer the reader to \citetalias{GHCW07} for a more detailed review.
}

We review  \pCf\  in \refsect{s:review}. The main
advances reported in this paper are \edit{(\refsect{s:symm})
a classification of plane Couette symmetry groups that support
equilibria, (\refsect{s:eqba}) the determination of a number of new
\eqva\ and \reqva\ of plane Couette flow, and (\refsect{s:bifurRe},
\refsect{s:aspect}) continuation of these solutions in Reynolds number
and spanwise aspect ratio.}
Outstanding challenges are discussed in \refsect{s:conclusions}.
Detailed numerical results such as stability eigenvalues and symmetries
of corresponding eigenfunctions are given in \cite{HalcrowThesis}, while
the complete data sets for the invariant solutions can be downloaded from
\HREF{http://channelflow.org}{{\tt channelflow.org}}.

\section{\PCf\ -- a review}
\label{s:review}

%%%%%%%%%%%%%%%%%%%%%%%%%%%%%%%%%%%%%%%%%%%%%%%%%%%%%%%%

\begin{figure}
\begin{center}
  \includegraphics[width=0.75\textwidth]{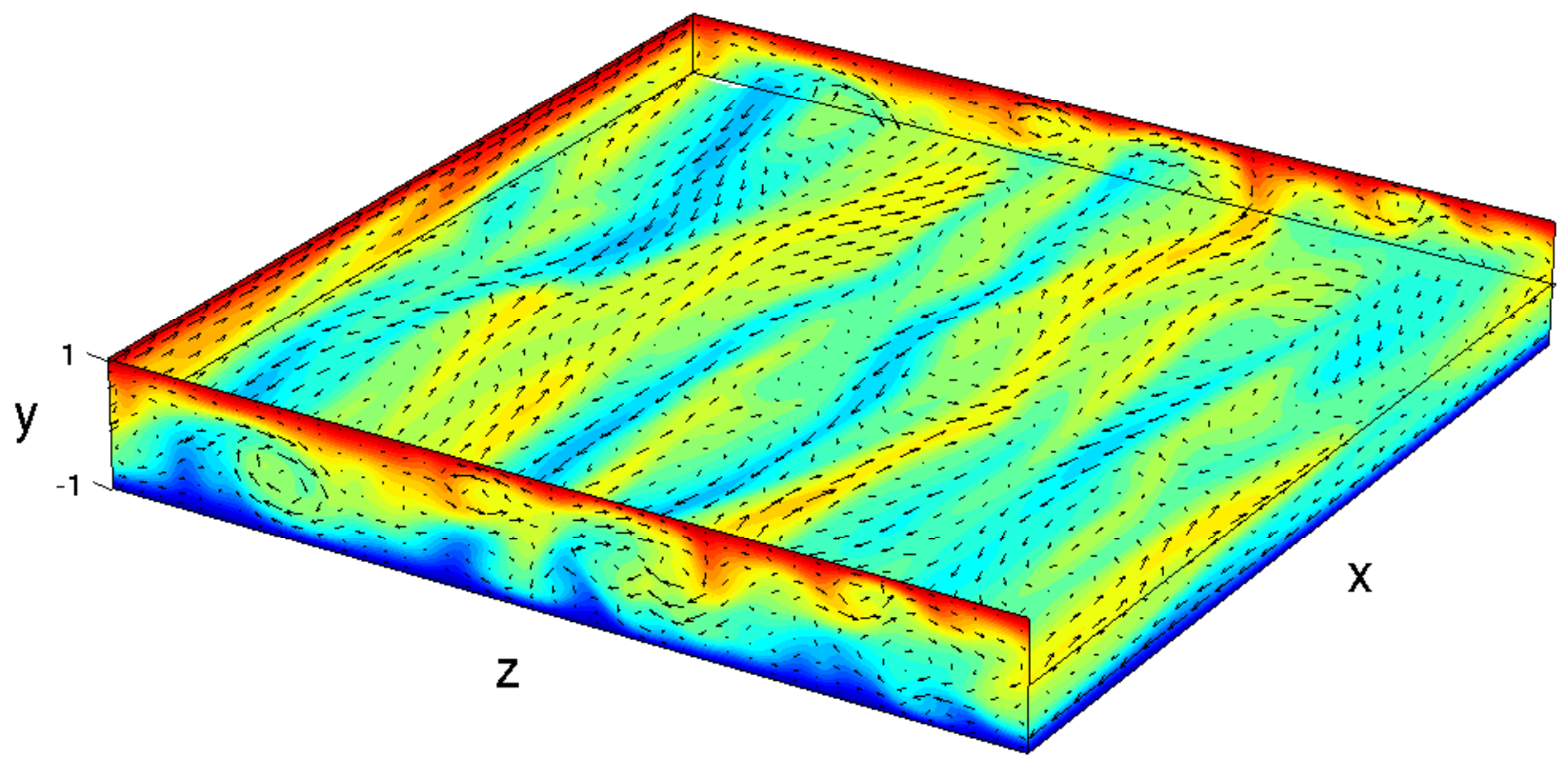}
\end{center}
 \caption{
A snapshot of a typical turbulent state in a large aspect-ratio cell
$[L_x, 2, L_z] = [15,2,15]$, $\Reynolds = 400$. The walls at
$y= \pm1$ move away/towards the viewer at equal and opposite
velocities $U=\pm1$. The \colorcomm{color}{grayscale} indicates
the streamwise ($u$, or $x$ direction) velocity of the fluid:
\colorcomm{red}{light grey} shows fluid moving at $u=+1$,
\colorcomm{blue}{bottom}, at $u=-1$.
The \colorcomm{colormap}{shading} as a function of $u$ is indicated
by the laminar {\eqv} in \reffig{f:eqbaboxes}.
Arrows indicate in-plane velocity in the respective planes:
$[v,w]$ in $(y,z)$ planes, etc. The top half of the fluid is
cut away to show the $[u,w]$ velocity in the $y=0$ midplane.
See
\cite{GibsonMovies} for movies of the time evolution
of such states.
}
 \label{f:bigbox}
\end{figure}
%%%%%%%%%%%%%%%%%%%%%%%%%%%%%%%%%%%%%%%%%%%%%%%%%%%%%%%%

\PCf\ is comprised of an incompressible viscous fluid confined between
two infinite parallel plates moving in opposite directions at constant
velocities, with no-slip boundary conditions imposed at the
walls. The plates move along the streamwise or $x$ direction, the
wall-normal direction is $y$, and the spanwise direction is $z$. The
fluid velocity field is $\bu(\bx)=[u,v,w](x,y,z)$. We define the Reynolds
number as $\Re = U h / \nu$, where $U$ is half the relative velocity of
the plates, $h$ is half the distance between the plates, and $\nu$ is the
kinematic viscosity. After non-dimensionalization, the plates are positioned
at $y = \pm 1$ and move with velocities $\bu = \pm 1\, \ex$, and the
\NSe\ are
\beq
    \frac{\partial { \bu}}{\partial t}
    + { \bu}\cdot \bnabla \bu
=
    - { \bnabla} p
    + \frac{1}{\Reynolds}
        { \bnabla}^2 { \bu } \,, \quad \nabla \cdot \bu = 0
\,.
\ee{NavStokesDev}
We seek spatially periodic {\eqv} and {\reqvD} solutions to
\refeq{NavStokesDev} for the domain $\bCell = [0, L_x] \times [-1, 1]
\times [0, L_z]$ (or $\bCell = [L_x, 2, L_z]$), with periodic
boundary conditions in $x$ and $z$. Equivalently, the spatial
periodicity of solutions can be specified in terms of
their fundamental wavenumbers $\alpha$ and $\gamma$. A given
solution is compatible with a given domain if
\edit{ $\alpha = m 2\upi/L_x$
and $\gamma = n 2\upi/L_z$} for integer $m,n$. In this study the
spatial mean of the pressure gradient is held fixed at zero.

Most of this study is conducted at $\Reynolds = 400$ in one of the two
small aspect-ratio cells
\begin{align}
\bNarrow~  &= [2\upi/1.14, 2, 2\upi/2.5]
           &\approx [5.51, 2, 2.51]
           &\approx [190, 68, 86] \;\;\; \text{wall units}
    \continue
\bHKW      &= [2 \upi/1.14, 2, 2 \upi/1.67]
           &\approx [5.51, 2, 3.76]
           &\approx [190, 68, 128] \; \text{wall units},
    \label{cellHKW}
\end{align}
where the wall units are in relation to a mean shear rate of $\langle
\partial u/ \partial y \rangle = 2.9$ in non-dimensionalized units
computed for a large aspect-ratio simulation at $\Reynolds = 400$.
Empirically, at this Reynolds number the \bHKW\ cell sustains turbulence
for \edit{very long times} \citep{HaKiWa95},
\edit{
whereas the \bNarrow\ cell exhibits only short-lived transient
turbulence \citepalias{GHCW07}. The $z$ length scale $L_z = 4 \upi/5$
of \bNarrow\ was chosen as a compromise between the $L_z = 6 \upi/5$ of
\bHKW\ and its first harmonic $L_z/2 = 3 \upi/5$ \citep{W02}.
}
Unless stated otherwise, all calculations
are carried out for $\Re = 400$ and the $\bNarrow$ cell. In the notation
of this paper, the solutions presented in\rf{N90} have wavenumbers
$(\alpha, \gamma) = (0.8, 1.5)$ and fit in the cell $[2 \upi/0.8, 2,
2\upi/1.5] \approx [7.85, 2, 4.18]$.%
\footnote
    {Note also that Reynolds number in \rf{N90} is based on the
    full wall separation and the relative wall velocity, making it a
    factor of four larger than the Reynolds number used in this paper.
    }
\edit{
\cite{W03} showed that these solutions first appear at critical Reynolds
number of 127.7 and $(\alpha, \gamma) = (0.577, 1.15)$.
}
\cite{Schmi99}'s study of plane Couette solutions and their
bifurcations was conducted in the cell of size
$\bCell = \; [4\upi, 2,2\upi] \approx [12.57, 2, 6.28]$.

Although the aspect ratios studied in this paper are small, the
$3D$ states explored by \eqva\ and their unstable manifolds explored
here are strikingly similar to typical states in larger aspect-ratio cells,
such as \reffig{f:bigbox}.
\cite{KKR71} observed that streamwise instabilities give rise to
pairwise counter-rotating rolls whose spanwise separation is
approximately 100 wall units. These rolls, in turn, generate streamwise
streaks of high and low speed fluid, by convecting fluid alternately away
from and towards the walls. The streaks have streamwise instabilities
whose length scale is roughly twice the roll separation.
These `coherent structures' are prominent in numerical and experimental
observations (see \reffig{f:bigbox} and \cite{GibsonMovies} animations),
and they motivate our investigation of how {\eqv} and {\reqvD}
solutions of Navier-Stokes change with $\Reynolds$ and cell size.

Fluid states are characterized by their energy $E = \frac{1}{2} \Norm{\bu}^2$
and energy dissipation rate $D = \Norm{\bnabla \times \bu}^2$, defined in terms
of the inner product and norm
\begin{align}
 (\bu, \bv)  &= \frac{1}{V}
                \int_\bCell \! d \bx \;
                       \bu \cdot \bv \,, \;\;
  \Norm{\bu}^2  = (\bu, \bu)
\,.
\label{innerproduct}
\end{align}
The rate of energy input is
$I = 1/(L_x L_z) \int \int dx dz \, \partial u / \partial y$,
where the integral is  taken over the upper and lower walls at $y = \pm 1$.
Normalization of these quantities is set so that $I=D=1$ for laminar flow and
$\dot{E} = I - D$. It is often convenient to consider fields as differences
from the laminar flow, since these differences constitute a vector space,
\edit{and thus can be added together, multiplied by scalars, etc. We
indicate such differences with tildes: $\hbu = \bu - y \hat{{\bf x}}$. Note
that the total velocity field  $\bu$ does {\em not} form a vector space:
the sum of any two total plane Couette velocity fields violates the $u =
\pm 1$ boundary conditions at the moving walls.}

\section{Symmetries and isotropy subgroups}
\label{s:symm}

On an infinite domain and in the absence of boundary conditions, the Navier-Stokes
equations are equivariant under any $3D$~translation, $3D$~rotation, and
$\bx \to -\bx$, $\bu \to -\bu$ inversion through the origin \citep{frisch}.
In plane Couette flow, the counter-moving walls restrict the rotation
symmetry to rotation by $\upi$ about the $z$-axis. We denote this rotation
by $\sigma_{x}$ and the inversion through the origin by $\sigma_{xz}$.
\edit{The suffixes indicate
which of the homogeneous directions $x,z$ change sign and simplify the
notation for the group algebra of rotation, inversion, and translations
presented in \refsects{s:flipnshift}{s:67-fold}.}
The $\sigma_{xz}$ and $\sigma_x$ symmetries generate a discrete dihedral group
$D_1 \times D_1 = \{e,\sigma_x,\sigma_{z},\sigma_{xz}\}$ of order 4, where
\begin{align}
\sigma_x    \, [u,v,w](x,y,z) &= [-u,-v,w](-x,-y,z) \nnu \\
\sigma_z    \, [u,v,w](x,y,z) &= [u, v,-w](x,y,-z)  \label{sigma}\\
\sigma_{xz} \, [u,v,w](x,y,z) &= [-u,-v,-w](-x,-y,-z) \nnu
\,.
\end{align}

The walls also restrict the translation symmetry to $2D$ in-plane
translations. With periodic boundary conditions, these translations
become the $SO(2)_x \times SO(2)_z$ continuous two-parameter
group of streamwise-spanwise translations
\begin{align}
\tau(\shift_x, \shift_z) [u, v, w](x,y,z) &= [u, v, w](x+\shift_x, y, z+ \shift_z)
\,.
\label{translation}
\end{align}
The equations of plane Couette flow are thus equivariant under the group
$\GPCF = O(2)_x \times O(2)_z = D_{1,x} \ltimes SO(2)_{x} \times D_{1,z}
\ltimes SO(2)_z$, where $\ltimes$ stands for a semi-direct product,
$x$  subscripts indicate streamwise translations
and sign changes in $x,y$, and $z$ subscripts indicate spanwise translations
and sign changes in $z$.

The solutions of an equivariant system can satisfy all of
the system's symmetries, a proper subgroup of them, or
have no symmetry at all. For a given solution $\bu$, the
subgroup that contains all symmetries that fix $\bu$ (that satisfy
$s \bu = \bu$) is called the isotropy (or stabilizer) subgroup of $\bu$.
\citep{hoyll06,MarRat99, golubitsky2002sp, GL-Gil07b}. For example, a typical
turbulent trajectory $\bu(\bx,t)$ has no symmetry beyond the identity,
so its isotropy group is $\{e\}$. At the other extreme is the laminar
{\eqv}, whose isotropy group is the full plane Couette symmetry
group $\GPCF$.

In between, the isotropy subgroup  of the Nagata \eqva\ and most
of the \eqva\ reported here is $S = \{e, s_1, s_2, s_3\}$, where
\begin{align}
s_1 \, [u, v, w](x,y,z) &= [u, v, -w](x+L_x/2, y, -z) \nnu \\
s_2 \, [u, v, w](x,y,z) &= [-u, -v, w](-x+L_x/2,-y,z+L_z/2) \label{shiftRot}\\
s_3 \, [u, v, w](x,y,z) &= [-u,-v,-w](-x, -y, -z+L_z/2) \nnu
\,.
\end{align}
These particular combinations of flips and shifts match the symmetries
of instabilities of streamwise-constant streaky flow \citep{W97,W03}
and are well suited to the wavy streamwise streaks observable in
\reffig{f:bigbox}, with suitable choice of $L_x$ and $L_z$.
But $S$ is one choice among a number of intermediate isotropy
groups of $\GPCF$, and other subgroups might also play an
important role in the turbulent dynamics. In this section we
provide a partial classification of the isotropy groups of $\GPCF$, sufficient
to classify all currently known invariant solutions and to guide
the search for new solutions with other symmetries. We focus on
isotropy groups involving at most half-cell shifts. The main result is
that among these, up to conjugacy in spatial translation, there
are only  five isotropy groups in which we should expect to find {\eqva}.

\subsection{Flips and half-shifts}
\label{s:flipnshift}

A few observations will be useful in what follows. First, we note the
key role played by the \edit{rotation and reflection} symmetries $\sigma_x$
and $\sigma_z$ \refeq{sigma} in the classification of solutions and
their isotropy groups. The equivariance of plane Couette flow under
continuous translations allows for {\reqvD} solutions, \ie,
solutions that are steady in a frame moving with a constant velocity
in $(x,z)$. In {\statesp}, {\reqva} either trace out
circles or wind around tori, and these sets are both
continuous-translation and time invariant. The sign changes under
$\sigma_x$, $\sigma_{z}$, and $\sigma_{xz}$, however, imply particular
centers of symmetry in $x$, $z$, and both $x$ and $z$, respectively,
and thus fix the translational phases of fields that are fixed by these
symmetries. Thus the presence of $\sigma_x$ or $\sigma_z$ in an
isotropy group prohibits {\reqva} in $x$ or $z$, and the
presence of $\sigma_{xz}$ prohibits any form of {\reqv}. Guided
by this observation, we will seek {\eqva} only for isotropy subgroups
that contain the $\sigma_{xz}$ inversion symmetry.

Second, the periodic boundary conditions impose discrete
translation symmetries of $\tau(L_x, 0)$ and $\tau(0, L_z)$ on
velocity fields. In addition to this full-period translation symmetry,
a solution can also be fixed under a rational translation, such as
$\tau(m L_x/n, 0)$ or a continuous translation $\tau(\ell_x, 0)$.
If a field is fixed under continuous translation, it is
constant along the given spatial variable. If it is fixed under rational
translation $\tau(m L_x/n, 0)$, it is fixed under $\tau(m L_x/n,
0)$ for $m \in [1, n-1]$ as well, provided that $m$ and $n$ are
relatively prime. For this reason the subgroups of the
continuous translation $SO(2)_x$ consist of the discrete cyclic groups
$C_{n,x}$ for $n=2,3,4,\ldots$ together with the trivial subgroup $\{e\}$
and the full group $SO(2)_x$ itself, and similarly for $z$. For rational
shifts $\ell_x/L_x = m/n$ we simplify the notation a bit by rewriting
\refeq{translation} as
\begin{align}
\trDiscr{x}{m/n} = \tau(m L_x/n, 0) \,,\;
\trDiscr{z}{m/n} = \tau(0,m L_z/n) \,.
\label{translDisc}
\end{align}
Since $m/n = 1/2$ will loom large in what follows, we omit exponents of $1/2$:
\beq
    \trHalf{x} = \trDiscr{x}{1/2}
    \,,\;
    \trHalf{z} = \trDiscr{z}{1/2}
    \,,\;
    \trHalf{xz} = \trHalf{x} \trHalf{z}
\,.
\label{tauHalf}
\eeq
If a field $\bu$ is fixed under a rational shift $\tau(L_x/n)$,
it is periodic on the smaller spatial domain $x \in [0,L_x/n]$.
For this reason we can exclude from our searches all \eqv\
whose isotropy subgroups contain
rational translations in favor of \eqva\ computed on smaller domains.
However, as we need to study bifurcations into
states with wavelengths longer than the initial state,
the linear stability computations
need to be carried out in the full $[L_x,2,L_z]$ cell.
For example, if \tEQ{}\ is an \eqv\ solution in the
$\bCell_{1/3}= [L_x/3,2,L_z]$ cell, we refer to the
same solution repeated thrice in $\bCell = [L_x,2,L_z]$
as ``spanwise-tripled'' or
$3 \times \tEQ{}$. Such solution is by construction fixed under the
$C_{3,x} = \{e,\trDiscr{x}{1/3},\trDiscr{x}{2/3}\}$ subgroup.

Third, some isotropy groups are conjugate to each other under
symmetries of the full group $\GPCF$. Subgroup $H'$ is conjugate to $H$
if there is an $s \in \GPCF$ for which $H' = s^{-1} H s$. In spatial terms,
two conjugate isotropy groups are equivalent to each other under a coordinate
transformation. A set of conjugate isotropy groups forms a conjugacy class.
It is necessary to consider only a single representative of each conjugacy
class; solutions belonging to conjugate isotropy groups can be generated by
applying the symmetry operation of the conjugacy.

In the present case conjugacies under spatial translation symmetries are
particularly important. Note that $O(2)$ is not an abelian group, since
reflections $\sigma$ and translations $\tau$ along the same axis do not
commute \citep{Harter93}. Instead we have $\sigma \tau  = \tau^{-1} \sigma$.
Rewriting this relation as $\sigma \tau^{2} = \tau^{-1} \sigma \tau$, we
note that
\bea
\sigma_x \tau_x(\ell_x, 0)
  &=& \tau^{-1}(\ell_x/2, 0) \, \sigma_x \, \tau(\ell_x/2, 0)
\,.
\label{origShift}
\eea
The right-hand side of \refeq{origShift} is a similarity
transformation that translates the origin of coordinate system. For
$\shift_x = L_x/2$ we have
\beq
\trDiscr{x}{-1/4} \, \sigma_{x} \, \trDiscr{x}{1/4}
        = \sigma_{x} \trHalf{x}
\label{origQuartShift}
\,,
\eeq
and similarly for the spanwise shifts / reflections. Thus for
each isotropy group containing the shift-reflect $\sigma_x \tau_x$
symmetry, there is a simpler conjugate isotropy group in which
$\sigma_x \tau_x$ is replaced by $\sigma_x$ (and similarly
for $\sigma_z \tau_z$ and $\sigma_z$). We choose as the representative
of each conjugacy class the simplest isotropy group, in which all such
reductions have been made. However, if an isotropy group contains both
$\sigma_x$ and $\sigma_x \tau_x$, it cannot be simplified this way,
since the conjugacy simply interchanges the elements.

Fourth, for $\shift_x = L_x$, we have
$\trHalf{x}^{-1} \, \sigma_{x} \, \trHalf{x} = \sigma_x \,,$
so that, in the special case of half-cell shifts,
$\sigma_x$ and $\tau_x$ commute. For the same reason,
$\sigma_z$ and $\tau_z$ commute, so the order-16 isotropy subgroup
\beq
G = D_{1,x} \times C_{2,x} \times D_{1,z} \times C_{2,z} \subset \GPCF
\ee{order16subgrp}
is abelian.

\subsection{The 67-fold path}
\label{s:67-fold}

We now undertake a partial classification
of the lattice of isotropy subgroups of plane Couette flow. We focus on
isotropy groups involving at most half-cell shifts, with $SO(2)_x \times SO(2)_z$
translations restricted to order 4 subgroup of spanwise-streamwise
translations \refeq{tauHalf} of half the cell length,
\beq
T = C_{2,x}\times C_{2,z}
  =  \{e,\trHalf{x},\trHalf{z},\trHalf{xz}\}
\,.
\label{tauD2}
\eeq
All such isotropy subgroups of $\GPCF$ are contained
in the subgroup $G$  \refeq{order16subgrp}. Within $G$, we look for the
simplest representative of each conjugacy class, as described above.

Let us first enumerate all subgroups $\isotropyG{} \subset G$.
The subgroups can be of order
$|\isotropyG{}| = \{1,2,4,8,16\}$.
A subgroup is generated by multiplication of a set of
generator elements, with the choice of
generator elements unique up to a permutation of subgroup
elements.
A subgroup of order $|\isotropyG{}| =  2$ has only one generator,
since every group element is its own inverse. There are 15
non-identity elements in $G$ to choose from, so there are 15 subgroups
of order 2.
Subgroups of order 4 are generated by multiplication of two
group elements. There are 15 choices for the first and 14
choices for the second. However, each order-4 subgroup
can be generated by $3 \cdot 2$ different choices of generators.
For example, any two of $\tau_x, \tau_z, \tau_{xz}$ in any order
generate the same group $T$. Thus there are $(15 \cdot 14)/(3 \cdot 2) = 35$
subgroups of order 4.

Subgroups of order 8 have three generators.  There are
15 choices for the first generator, 14 for the second, and 12 for the
third. There are 12 choices for the third
generator and not  13, since if it were chosen to be the product of the
first two generators, we would get a subgroup of order 4.
Each order-8 subgroup can be generated
by $7 \cdot 6 \cdot 4$ different choices of three generators, so there are
$(15 \cdot 14 \cdot 12)/(7 \cdot 6 \cdot 4) = 15$ subgroups of order 8.
In summary: there is the group $G$ itself, of order 16,
15 subgroups of order 8, 35 of order 4, 15 of
order 2, and 1 (the identity) of order 1,
or 67 subgroups in all \citep{HalcrowThesis}.
This is whole lot of isotropy subgroups to juggle; fortunately,
the observations of \refsect{s:flipnshift} show that there
are only 5 {\em distinct conjugacy classes} in which we can expect
to find \eqva.

The 15 order-2 groups fall into 8 distinct conjugacy
classes, under conjugacies between $\sigma_x \tau_x$ and $\sigma_x$
and $\sigma_z \tau_z$ and $\sigma_z$. These conjugacy classes are
represented by the 8 isotropy groups generated individually by the 8
generators
$\sigma_x,\, \sigma_z,\, \sigma_{xz},\, \sigma_x \tau_z,\,  \sigma_z \tau_x,\,
\tau_x,\, \tau_z,\,$ and $\tau_{xz}$. Of these, the latter three imply
periodicity on smaller domains. Of the remaining five,
$\sigma_x$ and $\sigma_x \tau_z$ allow {\reqva} in $z$,
$\sigma_z$ and $\sigma_z \tau_x$ allow {\reqva} in $x$.
Only a single conjugacy class, represented by the isotropy
group
\beq
  \{e,\sigma_{xz}\}
\,,
\ee{S3subgrp}
breaks both continuous translation symmetries. Thus, of all
order-2  isotropy groups, we expect only this group to have {\eqva}.
\tEQnine, \tEQten, and \tEQelev\ described below are examples of \eqva\
with isotropy group $\{e,\sigma_{xz}\}$.

Of the 35 subgroups of order 4, we need to identify those that
contain $\sigma_{xz}$ and thus support {\eqva}. We choose
as the simplest representative of each conjugacy class the isotropy
group in which $\sigma_{xz}$ appears in isolation.
Four isotropy subgroups of order 4 are generated by picking
$\sigma_{xz}$ as the first generator, and $\sigma_{z},\, \sigma_{z}
\trHalf{x},\, \sigma_{z} \trHalf{z},\,$ or $\sigma_{z} \trHalf{xz}$
as the second generator (\emph{R} for reflect-rotate):
\begin{align}
 R~~  &=  \{e, \sigma_x, \sigma_z, \sigma_{xz}\}
      ~~~~~~~\; = \{e,\sigma_{xz}\} \times \{e,\sigma_{z}\} \nnu\\
 R_x~ &=  \{e,\sigma_x \trHalf{x}, \sigma_z \trHalf{x}, \sigma_{xz}\}
      ~~ = \{e,\sigma_{xz}\} \times \{e,\sigma_{x}\trHalf{x}\}
        \label{subg4RR} \\
 R_z~ &=  \{e, \sigma_x \trHalf{z}, \sigma_z \trHalf{z}, \sigma_{xz}\}
      ~~\; = \{e,\sigma_{xz}\} \times \{e,\sigma_{z}\trHalf{z}\}
        \nnu\\
 R_{xz} &= \{e, \sigma_x \trHalf{xz}, \sigma_z \trHalf{xz}, \sigma_{xz}\}
        = \{e,\sigma_{xz}\} \times \{e,\sigma_{z}\trHalf{xz}\}
        \simeq S \,. \nnu
\end{align}
These are the only isotropy groups of order 4 containing $\sigma_{xz}$
and no isolated translation elements. Together with $\{e,\sigma_{xz}\}$,
these 5 isotropy subgroups represent the 5 conjugacy classes in
which expect to find {\eqva}.

The $R_{xz}$ isotropy subgroup is particularly important, as the \cite{N90}
{\eqva} belong to this conjugacy class \citep{W97,CB97,W03}, as do
most of the solutions reported here. The \emph{NBC} isotropy subgroup of
\cite{Schmi99} and $S$ of \cite{GHCW07} are conjugate to $R_{xz}$ under
quarter-cell coordinate transformations. In keeping with previous literature,
we often represent this conjugacy class with
$S = \{e, s_1, s_2, s_3\} = \{e, \sigma_z \tau_x, \sigma_x \tau_{xz},
\sigma_{xz} \tau_z\}$ rather than the simpler conjugate group $R_{xz}$.
Schmiegel's \emph{I} isotropy group is conjugate to our $R_{z}$; \cite{Schmi99}
contains many examples of $R_z$-isotropic \eqva. $R$-isotropic {\eqva} were found
by \rf{TuckBar03} for plane Couette flow in which the translation symmetries
were broken by a streamwise ribbon. We have not searched for $R_x$-isotropic
solutions, and are not aware of any published in the literature.

The remaining subgroups of orders 4 and 8 all involve $\{e,\tau_i\}$ factors
and thus involve states that are periodic on half-domains.
For example, the isotropy subgroup of \tEQsev\ and \tEQeight\ studied below is
$S  \times \{e, \tau_{xz}\} \simeq R  \times \{e, \tau_{xz}\}$,
and thus these are doubled states of solutions on half-domains. For
the detailed count of all 67 subgroups, see \cite{HalcrowThesis}.

\subsection{{\StateDsp} visualization}
\label{s:visualStatSp}

\citetalias{GHCW07} presents a method for visualizing low-dimensional
projections of trajectories in the infinite-dimensional {\statesp} of
the Navier-Stokes equations. Briefly, we construct an orthonormal basis
$\{\be_1, \be_2, \cdots, \be_n\}$ that spans a set of physically
important fluid states $\hbu_A$, $\hbu_B$, $\dots$, such as {\eqv} states
and their eigenvectors, and we project the evolving fluid state $\hbu(t) =
\bu(t) - y \hat{\bx}$ onto this basis using the $L^2$ inner product \refeq{innerproduct}. This produces a low-dimensional projection
\beq
\ssp(t) =(\ssp_1, \ssp_2, \cdots, \ssp_n, \cdots)(t)
    \,,\qquad
\ssp_n(t) = (\hbu(t), \be_n)
\,,
\ee{intrSspTraj}
which can be viewed in $2d$ planes $\{\be_m, \be_n\}$ or in
$3d$ perspective views $\{\be_{\ell},\be_m, \be_n\}$. The
{\stateDsp} portraits are {\em dynamically intrinsic}, since the
projections are defined in terms of intrinsic solutions of the
equations of motion, and {\em representation independent}, since the
inner product \refeq{innerproduct} projection is independent of the
numerical or experimental representation of the fluid state data. Such
bases are effective because moderate-\Reynolds\ turbulence explores a
small repertoire of unstable coherent structures (rolls, streaks,
their mergers), so that the trajectory $a(t)$ does not stray far from
the subspace spanned by the key structures.

There is no {\em a priori} prescription for picking a `good' set of
basis fluid states, and construction of $\{\be_n\}$ set requires
some experimentation. Let the $S$-invariant subspace be the
flow-invariant subspace of states $\bu$ that are fixed under $S$;
this consists of all states whose isotropy group is $S$ or contains
$S$ as a subgroup. The plane Couette system at hand has a total of
29 known \eqva\ within the $S$-invariant subspace:
four translated copies each of \tLB~-~\tEQsix, two translated
copies of \tEQsev\ (which have an additional $\tau_{xz}$ symmetry),
plus the laminar \eqv\ \tLM\ at the origin. As shown in
\citetalias{GHCW07}, the dynamics of different regions of {\statesp}
can be elucidated by projections onto basis sets constructed from
combinations of {\eqva} and their eigenvectors.

In this paper we present global views
of all invariant solutions in terms of the orthonormal `translational
basis' constructed in \citetalias{GHCW07} from the four translated copies
of \tUB:
\begin{align}
 & \qquad\qquad\qquad\qquad\qquad\qquad
              ~~~~~~ \trHalf{x} ~~\, \trHalf{z}    ~\, \trHalf{xz}
    \nnu\\
\beUBg{1} &= c_1 (1 + \trHalf{x} + \trHalf{z} + \trHalf{xz})
      \, \huUB      ~~~~  +  ~~  +     ~~   +
    \nnu\\
\beUBg{2} &= c_2 (1 + \trHalf{x} - \trHalf{z} - \trHalf{xz})
      \, \huUB      ~~~~  +  ~~  -     ~~   -
    \label{globalUBframe}\\
\beUBg{3} &= c_3 (1 - \trHalf{x} + \trHalf{z} - \trHalf{xz})
      \, \huUB      ~~~~  -  ~~  +     ~~   -
     \nnu
     \\
\beUBg{4} &= c_4 (1 - \trHalf{x} - \trHalf{z} + \trHalf{xz})
      \, \huUB      ~~~~  -  ~~  -     ~~   +
    \nnu
\,,
\end{align}
where $c_n$ is a normalization constant determined by
$\Norm{\beUBg{n}} = 1$.
The last 3 columns indicate the symmetry of the basis vector
under half-cell translations; e.g.\ $\pm1$ in the $\trHalf{x}$ column
implies $\trHalf{x} \beUBg{j} = \pm \beUBg{j}$.

\begin{figure}
 \centering
\includegraphics[width=0.44\textwidth]{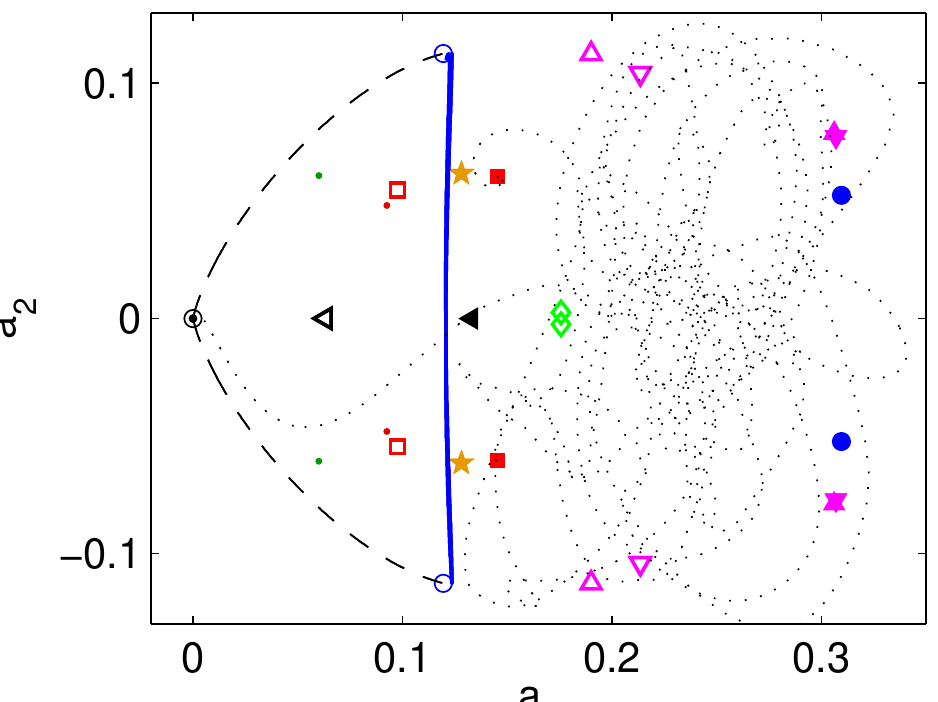}~%
\includegraphics[width=0.44\textwidth]{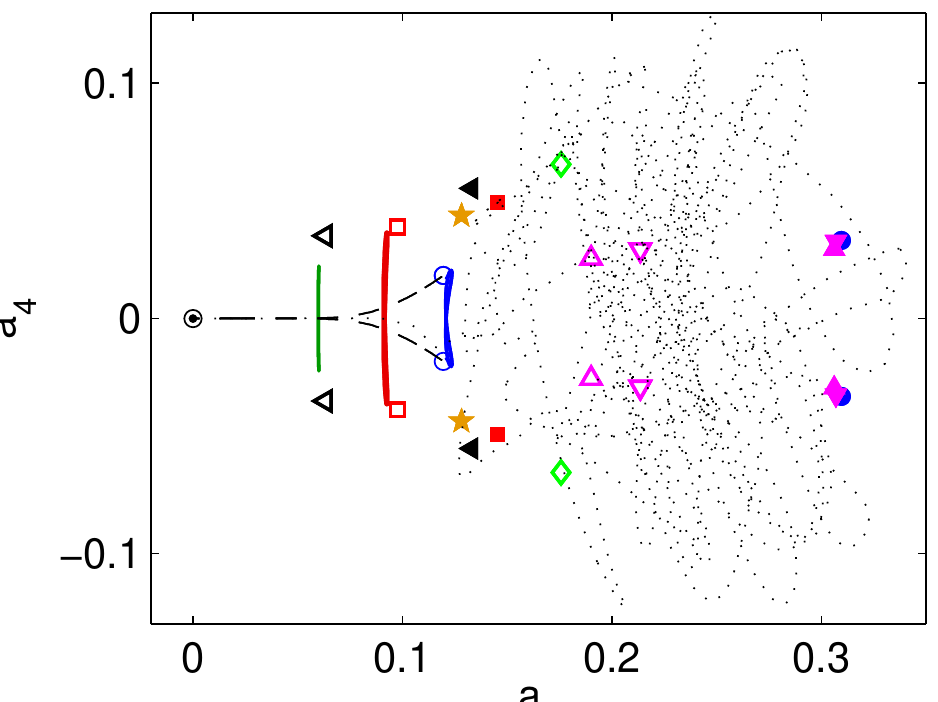}%
 \\
\includegraphics[width=0.44\textwidth]{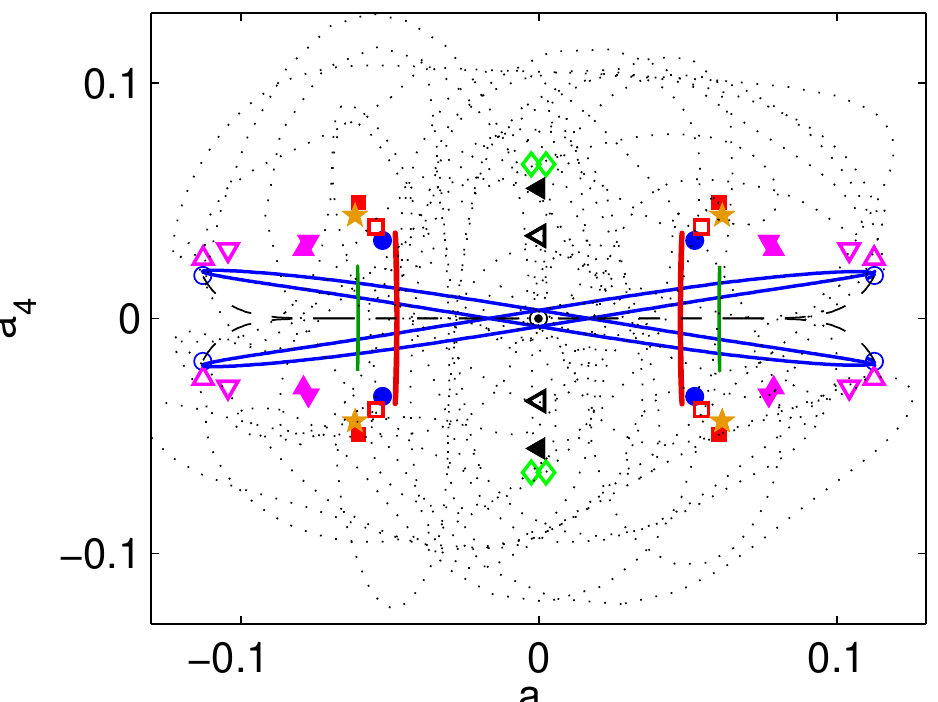}~%
\includegraphics[width=0.44\textwidth]{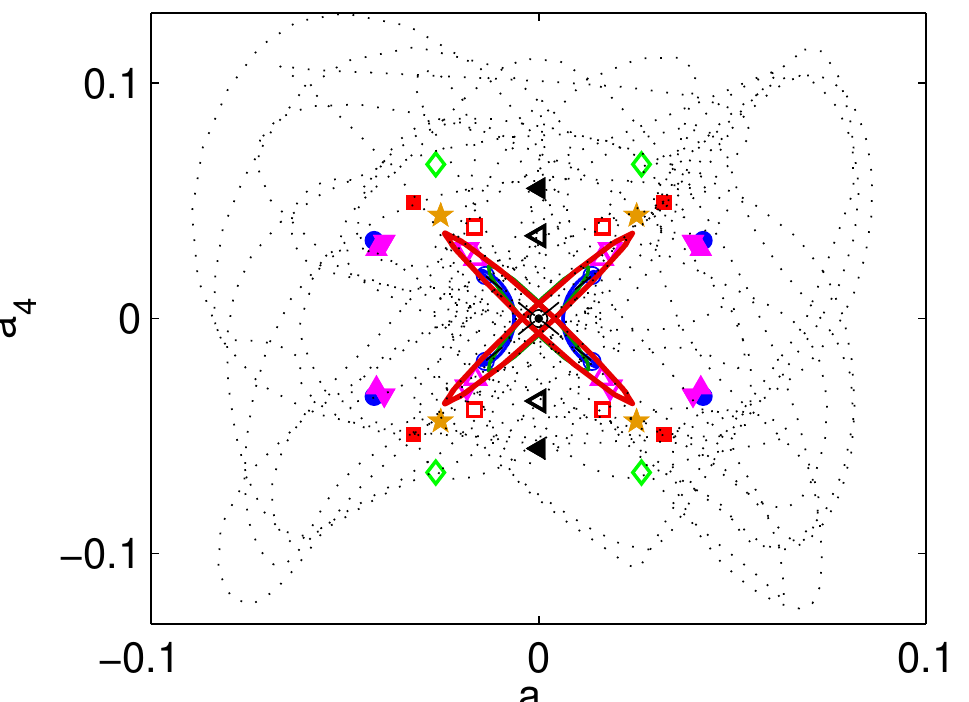}
\caption[State space portraits of known \eqva.]{
Four projections of \eqva, \reqva\  and their half-cell shifts
onto translational basis \refeq{globalUBframe} constructed from
\eqv\  \tNB. \Eqva\ are marked \sLM\ \tLM, \sLB\ \tLB, \sUB\ \tUB,
\sNNB\ \tNNB, \sNB\ \tNB, \sEQfive\ \tEQfive, \sEQsev\ \tEQsev,
\sEQnine\ \tEQnine, \sEQten\ \tEQten, \sEQelev\ \tEQelev,
\edit{\sEQtwel\ \tEQtwel, and \sEQthirt\ \tEQthirt}.
\Reqva\ trace out closed loops. In some projections the loops appear as
line segments or points. \tTWone \colorcomm{(blue)}{} is a spanwise-traveling,
symmetry-breaking bifurcation off \tLB, so it passes close to different
translational phases of \sLB\ \tLB. Similarly, \tTWthree\ \colorcomm{(red)}{}
bifurcates off \sNNB\ \tNNB\ and so passes near its translations. \tTWDone\
\colorcomm{(green)}{} was not discovered through bifurcation (see
\refsect{s:eqba}); it appears as the shorter, isolated line segment in
$(a_1,a_4)$ and $(a_2, a_4)$. The $\tLB \to \tLM$ relaminarizing \hec s
are marked by dashed lines.  A long-lived transiently turbulent trajectory
is plotted with a dotted line. The \tNB-translational basis was chosen
here since it displays the shape of \reqva\ more clearly than the
projection on the \tUB-translational basis of \reffig{f:WalR400all}.
    }
\label{f:reqvaPortrait}
\end{figure}

\section{\Eqva\ and \reqva\ of \pCf}
\label{s:eqba}

We seek {\eqv} solutions to \refeq{NavStokesDev} of the form
$\bu(\bx,t) = \uEQ(\bx)$ and {\reqvD} or relative {\eqv}
solutions of the form $\bu(\bx,t) = \uREQV(\bx- {\bf c} t)$ with ${\bf
c}=(c_x,0,c_z)$. Let $\vField_{\tNS}(\bu)$ represent the \NSe\
\refeq{NavStokesDev} for the given geometry, boundary conditions, and
Reynolds number, and $\bff^t_{\tNS}$ its time-$t$ forward map
\begin{align}
\pd{\bu}{t} = \vField_{\tNS}(\bu)
    \,, \qquad
    \bff_{\tNS}^t(\bu) = \bu + \int_0^t \! d \tau\,  \vField_{\tNS}(\bu(\tau))
\,.
\label{symbolicNS}
\end{align}
Then for any fixed $T > 0$, \eqva\ satisfy $\bff^T(\bu) - \bu = 0$ and
\reqva\ satisfy $\bff^T(\bu) - \tau \, \bu = 0$, where $\tau = \tau(c_x T,
c_z T)$. When $\bu$ is approximated with a finite spectral expansion and
$\bff^t$ with a numerical simulation algorithm, these equations become a
set of nonlinear equations in the expansion coefficients for $\bu$ and,
in the case of \reqva, the wave velocities $(c_x,0,c_z)$.

\cite{Visw07b} presents an algorithm for computing solutions to these
equations based on Newton search, Krylov subspace methods,
and an adaptive `hookstep' trust-region limitation to the Newton steps.
This algorithm can provide highly accurate solutions from even poor
initial guesses. The high accuracy stems from the use of Krylov subspace
methods, which can be efficient with $10^5$ or more spectral expansion
coefficients. The robustness with respect to initial guess stems from the
hookstep algorithm. The hookstep limitation restricts steps to a radius $r$ of
estimated validity for the local linear approximation to the Newton equations.
As $r$ increases from zero, \edit{the direction of the hookstep varies smoothly
from the gradient of the residual within the Krylov subspace} to the Newton step, so
that the hookstep algorithm
behaves as a gradient descent when far away from a solution and as
the Newton method when near, thus greatly increasing the algorithm's region
of convergence around solutions, compared to the Newton method \citep{DS}.

The choice of initial guesses for the search algorithm is one of the main
differences between this study and previous calculations of {\eqva} and
\reqva\ of shear flows. Prior studies have used homotopy, that is, starting
from a solution to a closely related problem and following it through small
steps in parameter space to the problem of interest. \Eqva\ for \pCf\ have
been continued from Taylor-Couette flow \citep{N90}, Rayleigh-B\'enard flow
\citep{CB97}, and from plane Couette with imposed body forces \citep{W98}.
\Eqva\ and \reqva\ have also been found using
``edge-tracking'' algorithms, that is, by  adjusting the magnitude of a
perturbation of the laminar flow until it neither decays to laminar nor
grows to turbulence, but instead converges toward a nearby weakly unstable
solution (\cite{IT01,SYE05,Visw07a,SGLDE08}). In this study, we take
as initial guesses samples of velocity fields generated by long-time simulations
of turbulent dynamics. The intent is to find the
{\em dynamically most important}
solutions, by sampling the turbulent flow's natural measure.

We discretize $\bu$ with a spectral expansion of the form
\beq
\bu(\bx) = \sum_{j=-J}^J \sum_{k=-K}^K \sum_{\ell=0}^L
   \bu_{jkl} \,T_{\ell}(y) \,
   e^{2\upi i (jx/L_x + kz/L_z)}\, ,
\ee{CFDexpansion}
where the $T_{\ell}$ are Chebyshev polynomials. Time integration of $\bff^t$ is
performed with a primitive-variables Chebyshev-tau algorithm with tau
correction, influence-matrix enforcement of boundary conditions, and
third-order semi-implicit backwards-differentiation time stepping,
and dealiasing
in $x$ and $z$ (\cite{Kleiser80,Canuto88,Peyret02}). We eliminate
from the search space the linearly dependent spectral coefficients of $\bu$
that arise from incompressibility, boundary conditions,
and complex conjugacies that arise from the real-valuedness
   of velocity fields.
\edit{Our codes for Navier-Stokes integration, Newton-hookstep search,
parametric continuation, and eigenvalue calculation are available for download from
\weblink{channelflow.org} website \citep{channelflow}, along with a
database of all solutions described in this paper.} For further details
on the numerical methods see \citetalias{GHCW07} and \cite{HalcrowThesis}.

Solutions presented in this paper use spatial discretization
\refeq{CFDexpansion} with $(J,K,L) = (15,15,32)$ (or $32\!\times\!33\!\times\!32$
gridpoints) and roughly 60k expansion coefficients, and integration is
performed with a time step of $\Delta t = 0.03125$.
The estimated accuracy of each solution is listed in \reftab{t:eqbtable5}.
As is clear from \cite{Schmi99} Ph.D.\ thesis, ours is almost certainly an incomplete
inventory; while for any finite \Reynolds, finite aspect-ratio cell the number of
distinct \eqv\ and \reqv\ solutions is finite, we know of no way of
determining or bounding this number.
It is difficult to compare our solutions directly to those of
Schmiegel since those solutions were computed in a
$[4 \upi, 2, 2 \upi]$ cell (roughly twice our cell size in both span and
streamwise directions) and with lower spatial resolution (2212 independent
expansion functions versus our 60k for a cell of one-fourth the volume).

\subsection{\Eqv\ solutions}
\label{s:eqbSols}

\begin{figure}
\centering
{\includegraphics[width=0.44\textwidth]{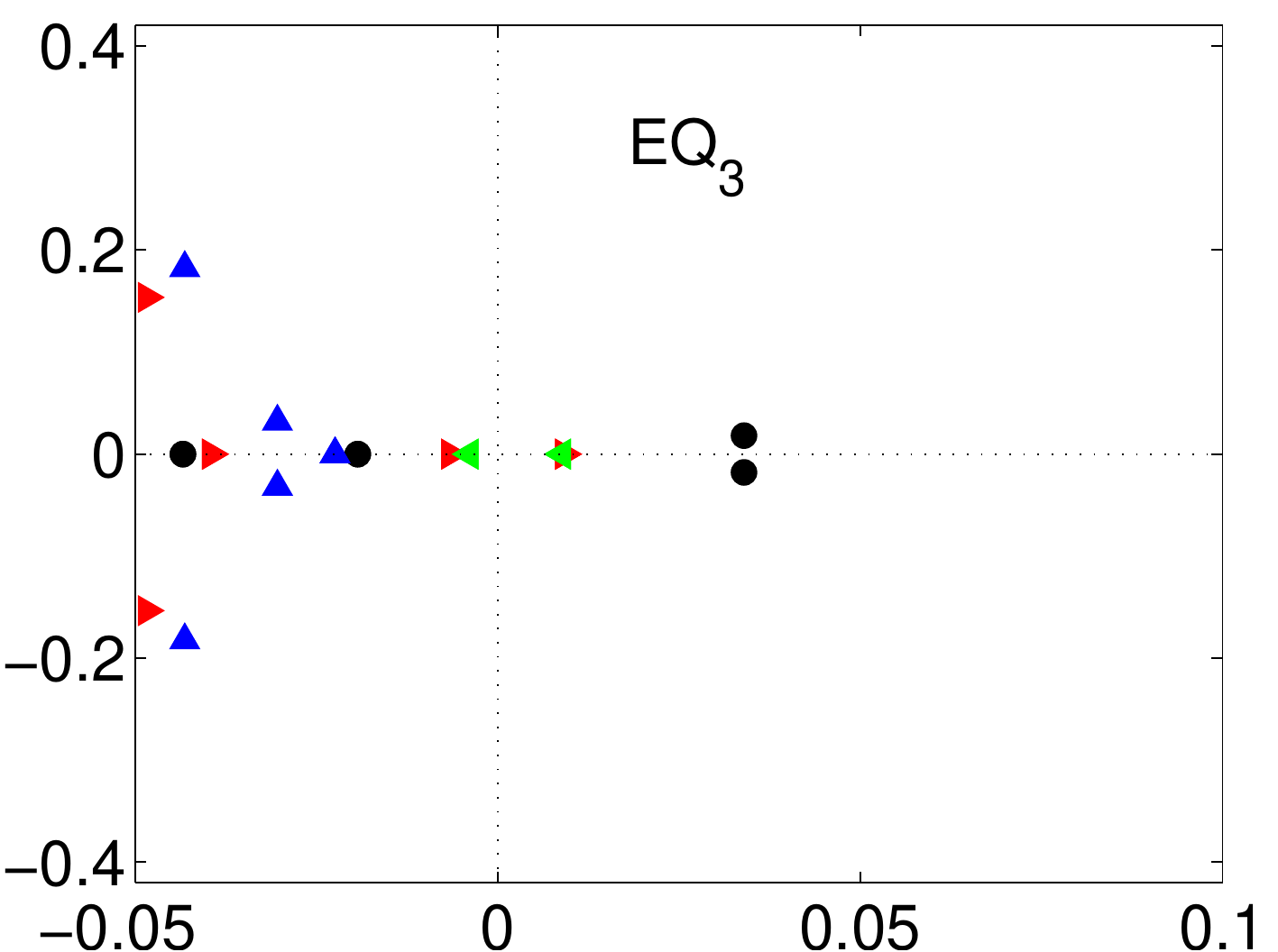}} \hspace{2mm}
{\includegraphics[width=0.44\textwidth]{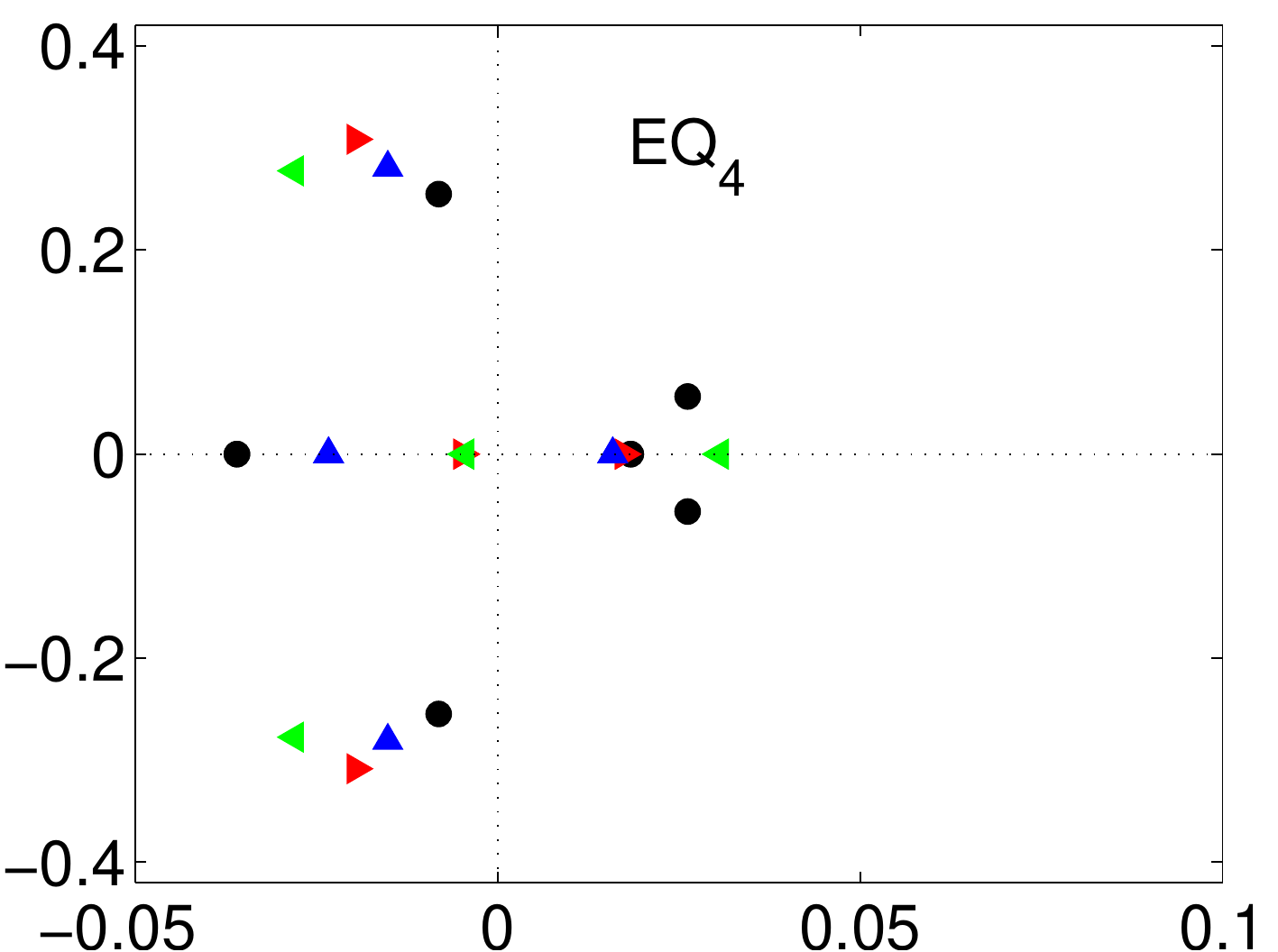}}
\\
\vspace{2mm}
{\includegraphics[width=0.44\textwidth]{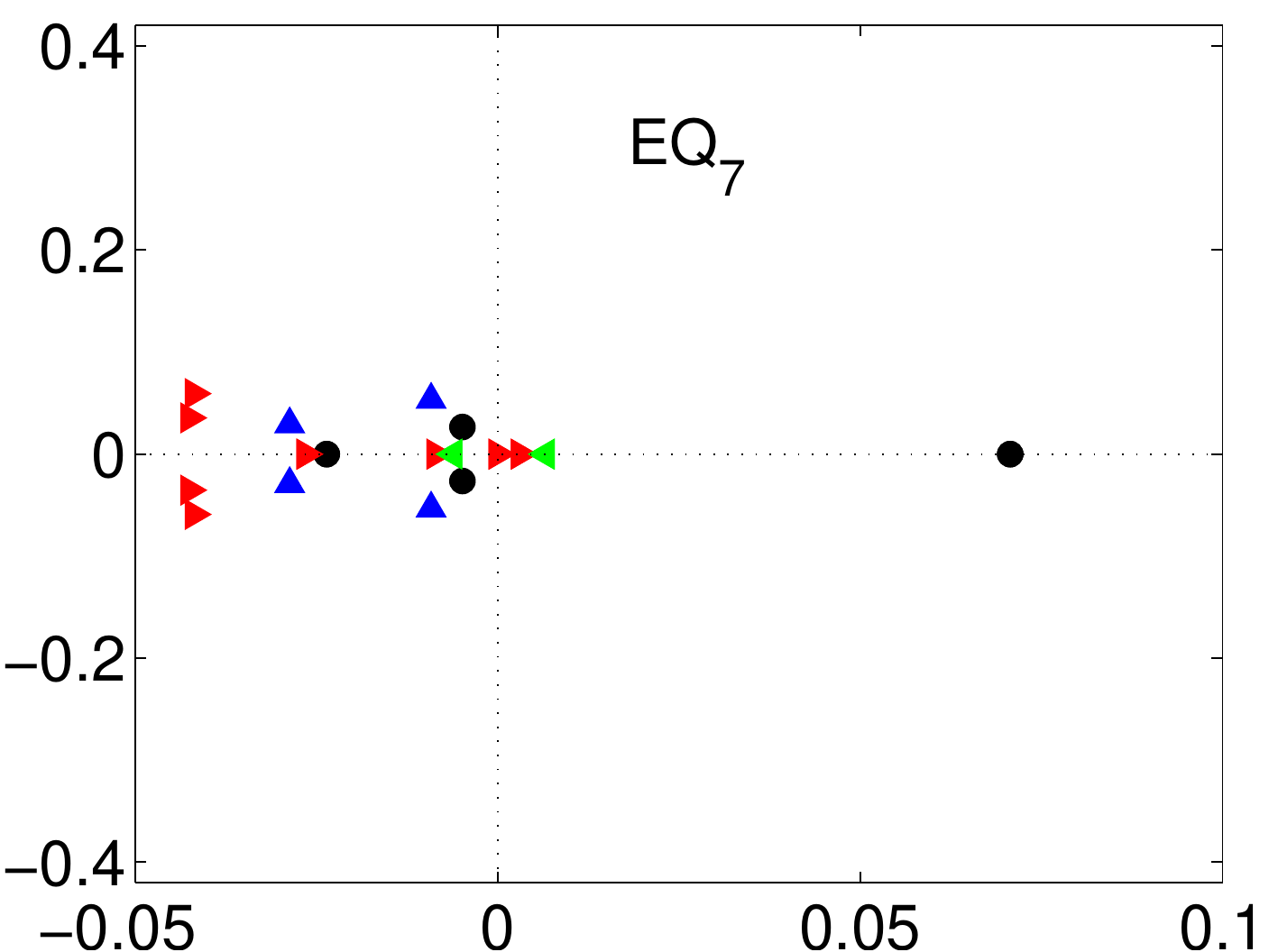}} \hspace{2mm}
{\includegraphics[width=0.44\textwidth]{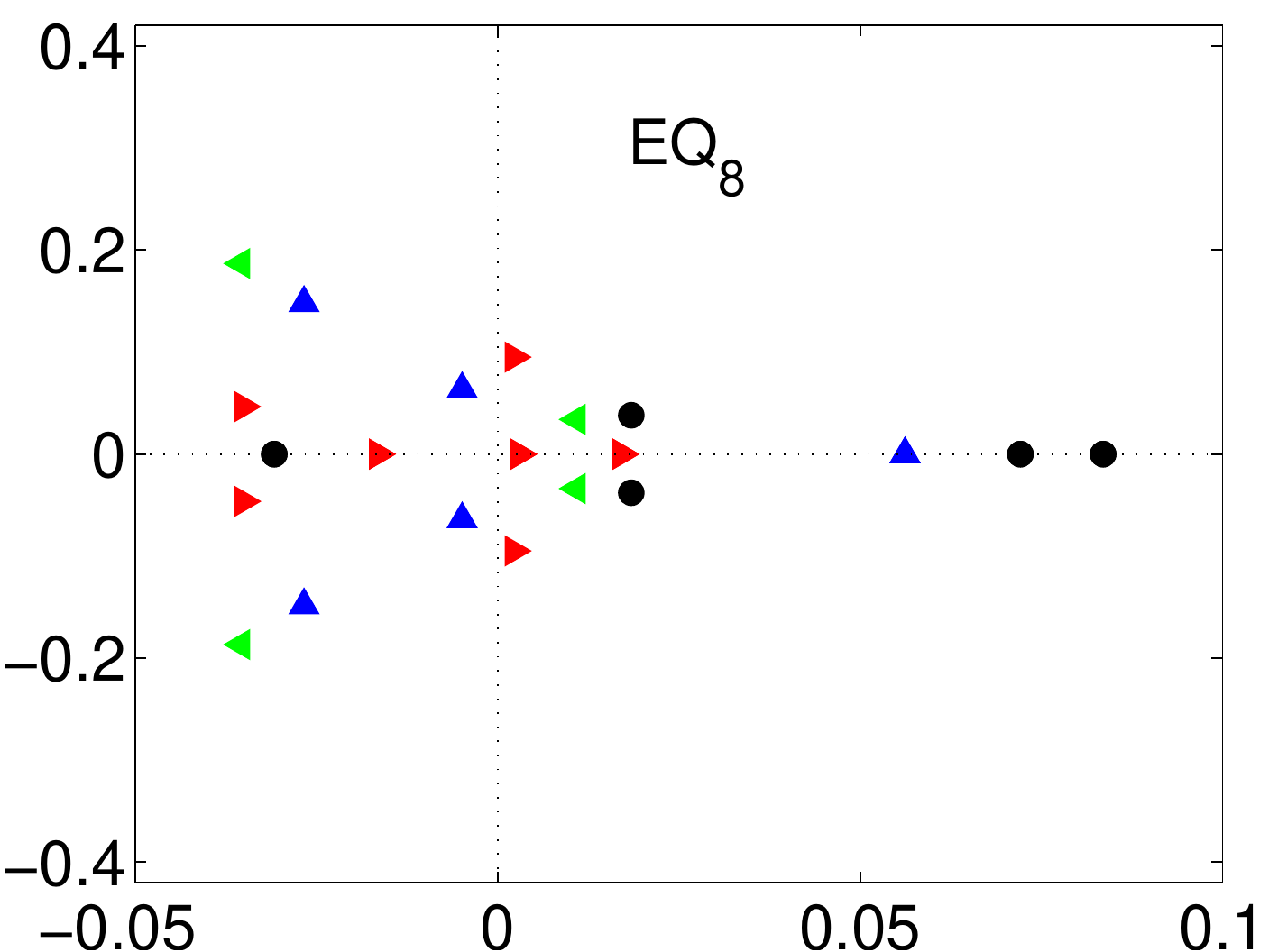}}
\caption[Eigenvalues of \eqva\ 3,4 and 7,8]
{Eigenvalues of \eqva\ \tNNB, \tNB\ and \tEQsev, \tEQeight\ in the $\bNarrow$ cell,
$\Reynolds=400$.
Eigenvalues are plotted according to their symmetries:
{\Large $\bullet$} $+++$, the $S$-invariant subspace,
{\large \colorcomm{\color{red}}{} $\blacktriangleright$} $+--$,
{\large \colorcomm{\color{green}}{} $\blacktriangleleft$} $-+-$, and
{\large \colorcomm{\color{blue}}{} $\blacktriangle$} $--+$, where $\pm$
symbols stand for symmetric/antisymmetric in $s_1,s_2,$ and $s_3$ respectively.
For \tLB, \tUB\ and \tNB\ eigenvalues see
\citetalias{GHCW07} (there referred to as
\ensuremath{\bu_{\text{\tiny LB}}},
\ensuremath{\bu_{\text{\tiny UB}}} and \ensuremath{\bu_{\text{\tiny NB}}},
respectively). For numerical values of all stability eigenvalues see
\cite{HalcrowThesis} and
\HREF{http://channelflow.org}{{\tt channelflow.org}}.
}
\label{f:eqbalambda1}
\end{figure}

Our primary focus is on the $S$-invariant subspace \refeq{shiftRot}
of the $\bNarrow$ cell at $\Reynolds = 400$.
We initiated 28 {\eqv} searches at evenly spaced
intervals $\Delta t = 25$ along a trajectory in the unstable manifold
of \tNB\ that exhibited turbulent dynamics for 800 nondimensionalized
time units after leaving the neighborhood of \tNB\ and before decaying
to laminar flow. \edit{Lower, upper branch pairs are labeled with
consecutive numbers, and the numbers indicate, as closely as possible,
the order of discovery.
We give a name $\text{EQ}_{n}$ to any distinct solution in the
$\bNarrow$ cell at $\Reynolds = 400$, although many of these solutions
can be connected by continuation in \Reynolds\ or
wavenumber.} \tLM\ is the laminar {\eqv}, \tLB\ and \tUB\
are the Nagata lower and upper branch, and \tNB\ is the $\bu_{\text{\tiny NB}}$
solution reported in \citetalias{GHCW07}. The rest are new. Only one of
the 28  searches failed to converge onto an {\eqv}; the successful searches
converged to {\eqva} with frequencies listed in \reftab{t:eqbtable5}.
The higher frequency of occurrence of \tLB\ and \tNB\ suggests that these
are the dynamically most important {\eqva} in the $S$-invariant subspace for
the $\bNarrow$ cell at $\Reynolds = 400$. Stability eigenvalues of known \eqva\
are plotted in \reffig{f:eqbalambda1}. Tables of stability
eigenvalues and other properties of these solutions are given in
\cite{HalcrowThesis}, while the images, movies and full data sets are
available online at \HREF{http://channelflow.org}{{\tt channelflow.org}}.
All {\eqv} solutions have zero spatial-mean pressure gradient,
which was imposed in the flow conditions,
and, due to their symmetry, zero \vCM.

% \begin{figure}
% \centering
% {\includegraphics[width=0.31\textwidth]{EQ0}} \hskip -30ex \tLM \hskip 25ex
% {\includegraphics[width=0.31\textwidth]{EQ1}} \hskip -30ex \tLB \hskip 25ex
% {\includegraphics[width=0.31\textwidth]{EQ2}} \hskip -30ex \tUB \hskip 25ex ~
% \\
% {\includegraphics[width=0.31\textwidth]{EQ3}} \hskip -30ex \tNNB    \hskip 25ex
% {\includegraphics[width=0.31\textwidth]{EQ4}} \hskip -30ex \tNB     \hskip 25ex
% {\includegraphics[width=0.31\textwidth]{EQ5}} \hskip -30ex \tEQfive \hskip 25ex ~
% \\
% %{\includegraphics[width=0.31\textwidth]{uEQ6Re330box}} \hskip -30ex \tEQsix   \hskip 25ex
% {\includegraphics[width=0.31\textwidth]{EQ6}} \hskip -30ex \tEQsix   \hskip 25ex
% {\includegraphics[width=0.31\textwidth]{EQ7}} \hskip -30ex \tEQsev   \hskip 25ex
% {\includegraphics[width=0.31\textwidth]{EQ8}} \hskip -30ex \tEQeight \hskip 25ex ~
% \\
% {\includegraphics[width=0.31\textwidth]{EQ9}}  \hskip -30ex \tEQnine \hskip 25ex
% {\includegraphics[width=0.31\textwidth]{EQ10}} \hskip -30ex \tEQten  \hskip 25ex
% {\includegraphics[width=0.31\textwidth]{EQ11}} \hskip -30ex \tEQelev \hskip 25ex ~
% \caption[$3D$ space plots of known \eqva.]{
% \Eqva\ \tLM\ (laminar solution) through \tEQelev\ in $\bNarrow$ cell.
% Plotting conventions are the same as \reffig{f:bigbox}.
% $\Reynolds=400$  except for \tEQsix, which is shown at $\Reynolds=330$.
% }
% \label{f:eqbaboxes}
% \end{figure}

\begin{figure}
\centering
{\includegraphics[width=0.31\textwidth]{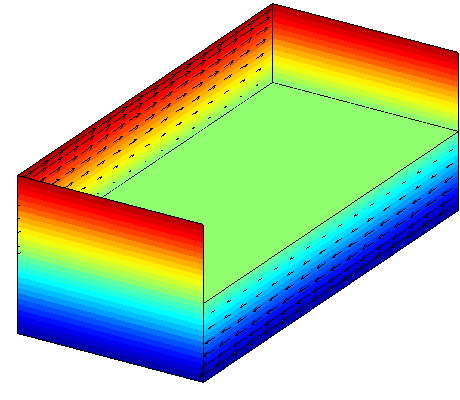}} \hskip -30ex \tLM \hskip 25ex
{\includegraphics[width=0.31\textwidth]{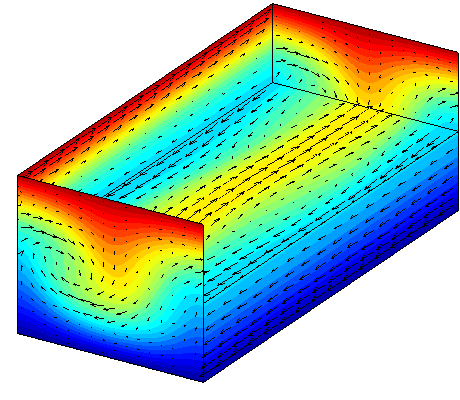}} \hskip -30ex \tLB \hskip 25ex
{\includegraphics[width=0.31\textwidth]{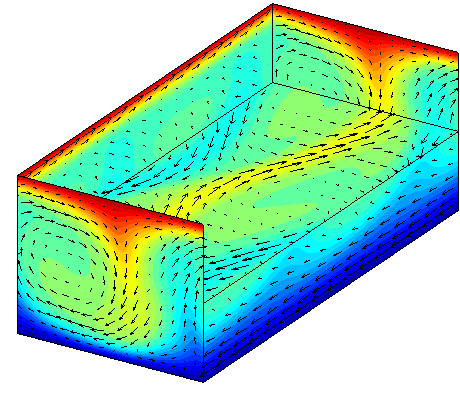}} \hskip -30ex \tUB \hskip 25ex ~
\\
%{\includegraphics[width=0.31\textwidth]{EQ3}} \hskip -30ex \tNNB    \hskip 25ex
{\includegraphics[width=0.31\textwidth]{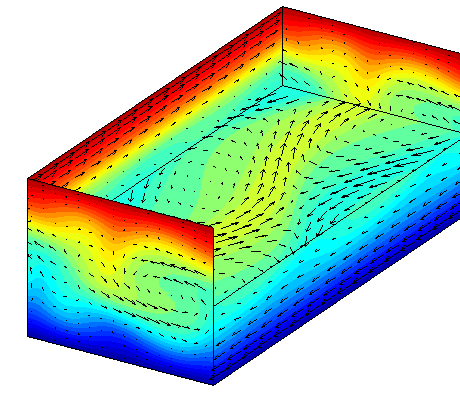}} \hskip -30ex \tNB     \hskip 25ex
{\includegraphics[width=0.31\textwidth]{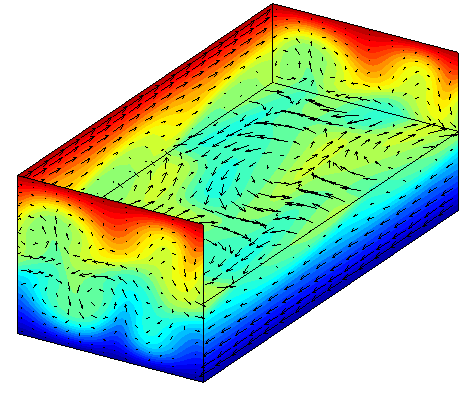}} \hskip -30ex \tEQfive \hskip 25ex ~
{\includegraphics[width=0.31\textwidth]{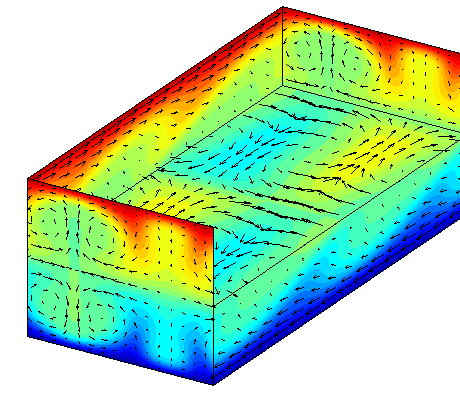}} \hskip -30ex \tEQeight \hskip 25ex ~
\\
%{\includegraphics[width=0.31\textwidth]{uEQ6Re330box}} \hskip -30ex \tEQsix   \hskip 25ex
%{\includegraphics[width=0.31\textwidth]{EQ6}} \hskip -30ex \tEQsix   \hskip 25ex
%{\includegraphics[width=0.31\textwidth]{EQ7}} \hskip -30ex \tEQsev   \hskip 25ex
{\includegraphics[width=0.31\textwidth]{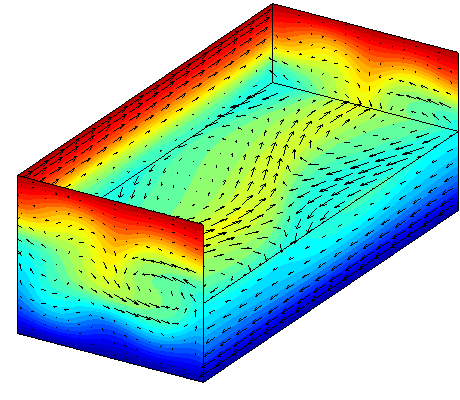}}  \hskip -30ex \tEQnine \hskip 25ex
{\includegraphics[width=0.31\textwidth]{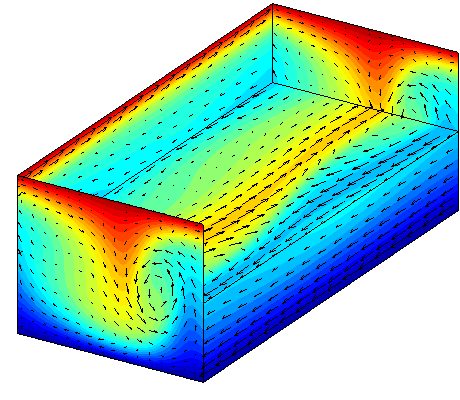}} \hskip -30ex \tEQten  \hskip 25ex
{\includegraphics[width=0.31\textwidth]{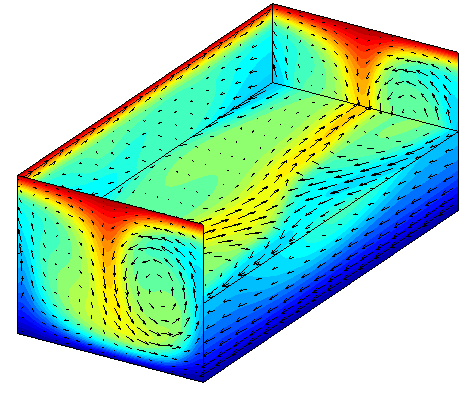}} \hskip -30ex \tEQelev \hskip 25ex ~
\caption[$3D$ space plots of known \eqva.]{
\edit{Equilibrium solutions of plane Couette flow in $\bNarrow = [2\upi/1.14, 2, 2\upi/2.5]
$ at $Re = 400$. Plotting conventions
are the same as \reffig{f:bigbox}. The \colorcomm{color}{gray} scale indicates
\edit{the streamwise velocity $u$, with the front face of the
laminar solution \tEQzero}, $u(y) = y$, serving as a reference. Not all solutions are
shown;
\tEQthree, \tEQsix, \tEQsev, \tEQtwel, and \tEQthirt, are very similar to
\tEQfour,  \tEQfive, \tEQeight, \tEQten, and \tEQelev, respectively.}}
\label{f:eqbaboxes}
\end{figure}

\begin{figure}
\centering
{\footnotesize \tUB}     {\includegraphics[width=0.25\textwidth]{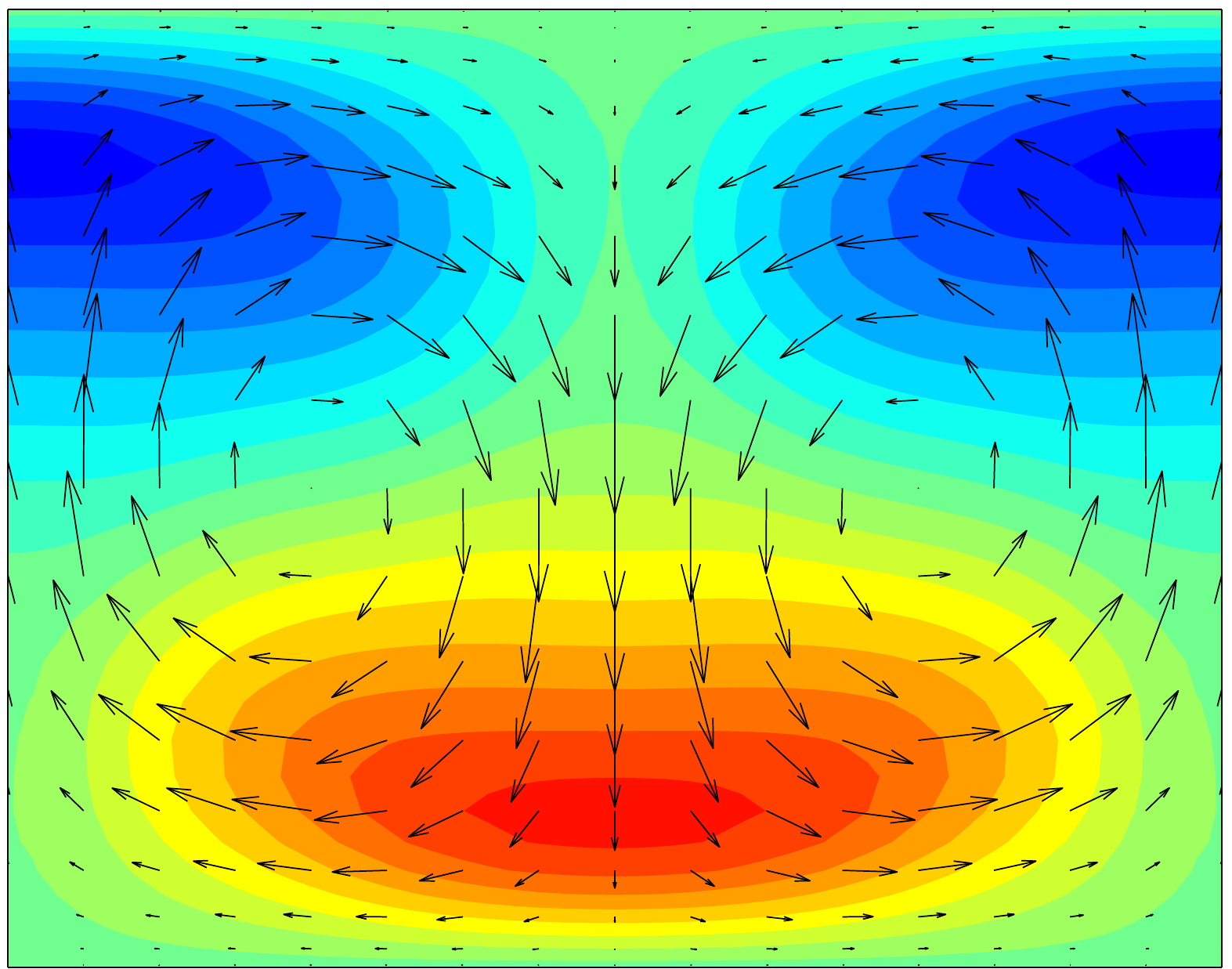}}
{\footnotesize \tNB}     {\includegraphics[width=0.25\textwidth]{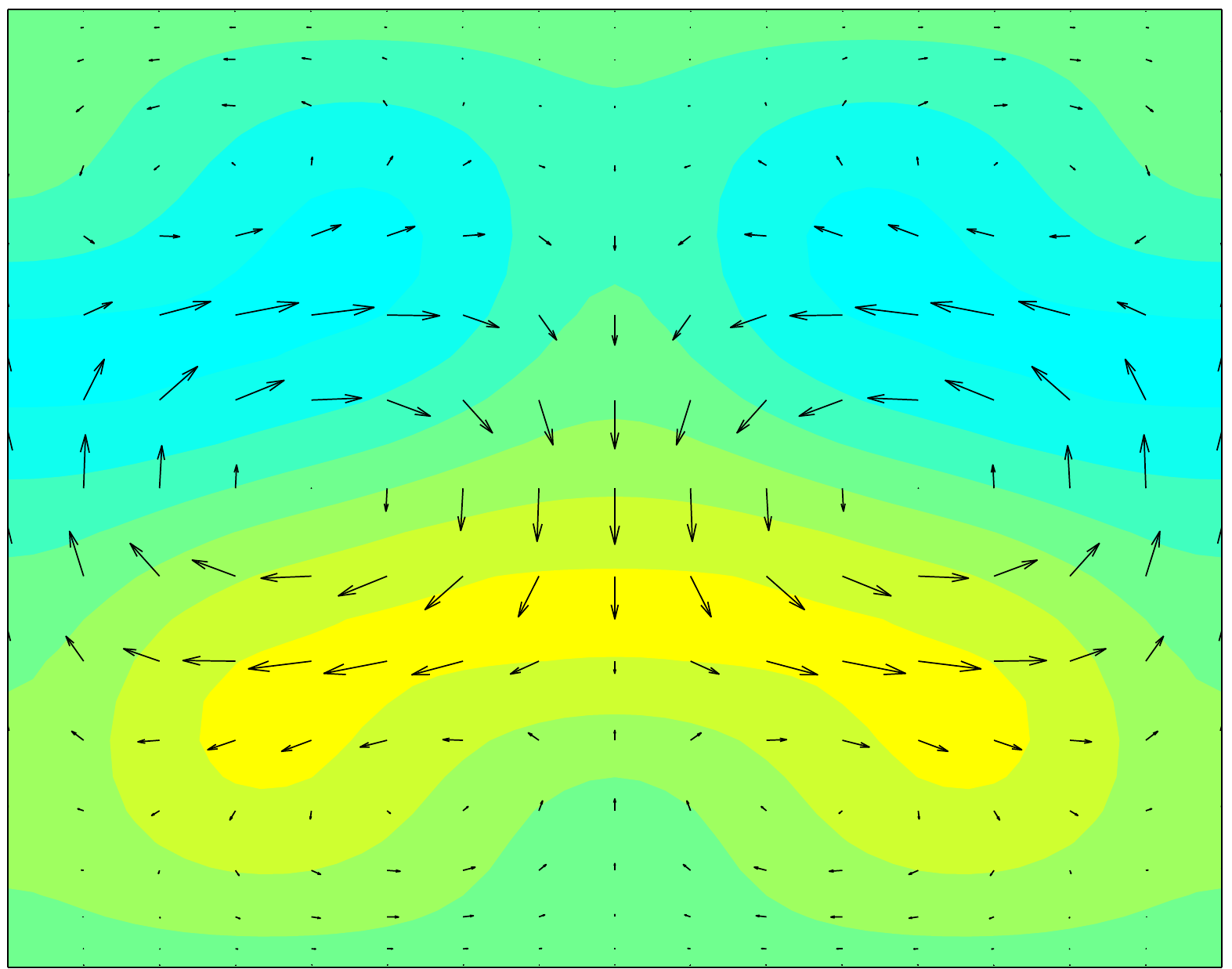}}
{\footnotesize \tEQsix}  {\includegraphics[width=0.25\textwidth]{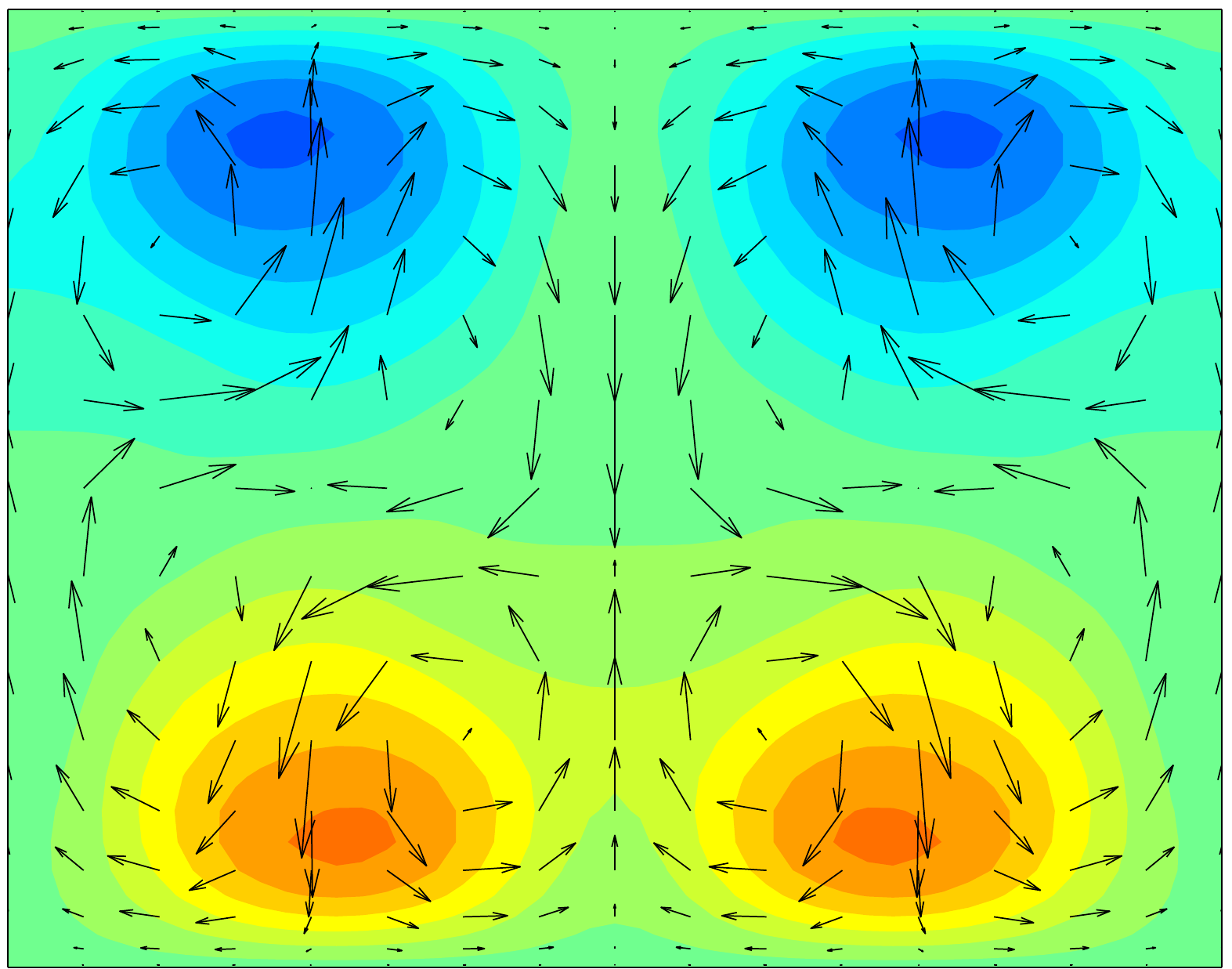}}
\\
{\footnotesize \tLB}     {\includegraphics[width=0.25\textwidth]{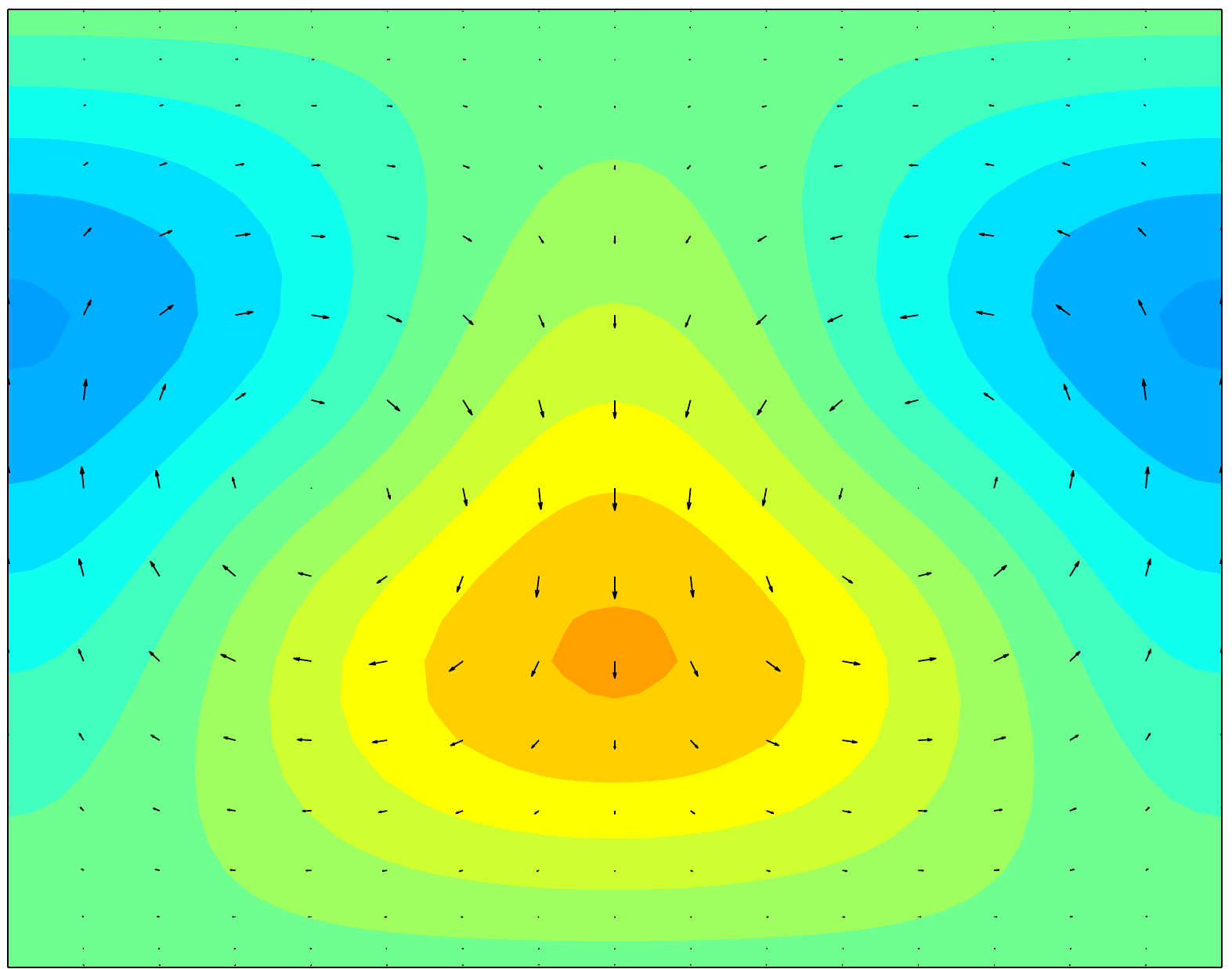}}
{\footnotesize \tNNB}    {\includegraphics[width=0.25\textwidth]{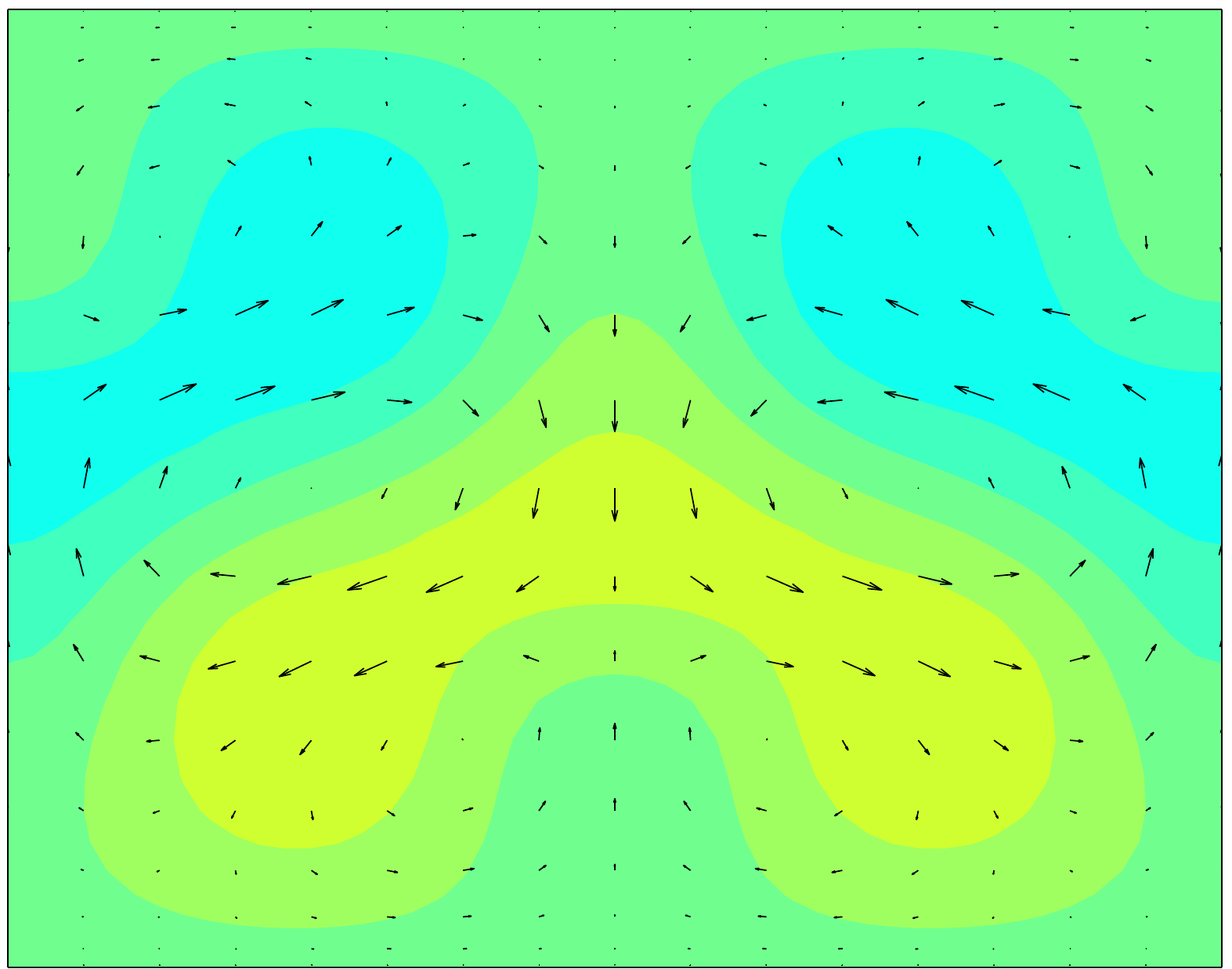}}
{\footnotesize \tEQfive} {\includegraphics[width=0.25   \textwidth]{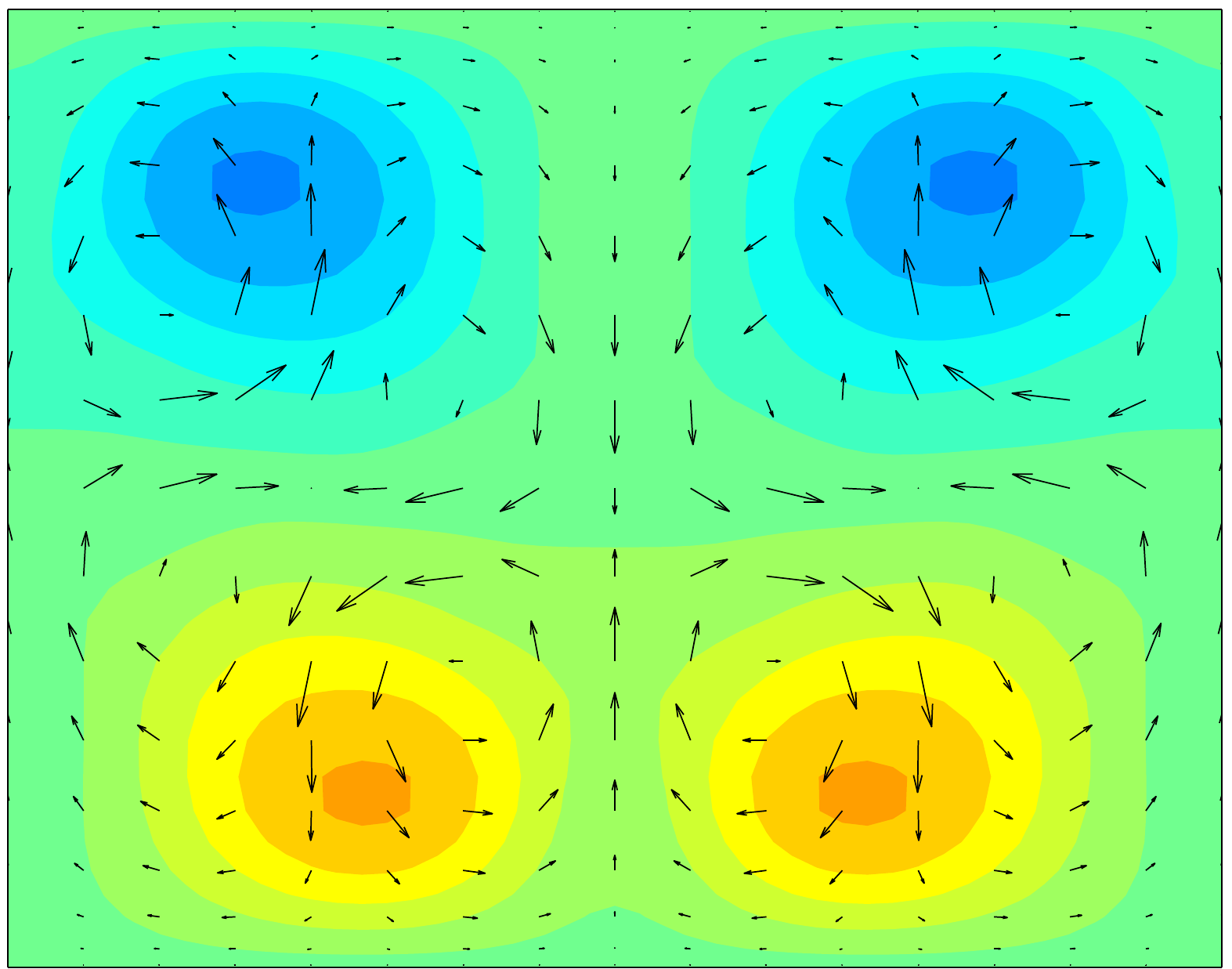}}
\caption{$x$-average of \edit{difference from laminar flow,
$\hbu = \bu - y \hat{\bx}$}, for \eqva\ \tLB-\tEQsix\ in the $\bNarrow$ cell.
The axes are $z$ (horizontal) and $y$ (vertical). Arrows
indicate $[\tilde{v},\tilde{w}]$,
\edit{
  with the same arrow length denoting the same magnitude of in-plane
  velocity in all graphs.
  \colorcomm{The colormap indicates $\tilde{u}$: red/blue is $\tilde{u} =
  \pm 1$, and green is $\tilde{u}=0$.}{Contour lines indicate $\tilde{u}$
  at values $\tilde{u} = \pm (2n+1)/15,$ with negative contours dashed
  and positive solid.}
}
Lower and upper-branch
pairs are grouped together vertically, \eg\ \tLB, \tUB\ are a lower, upper
branch pair.  $\Reynolds=400$  except for $\Reynolds=330$ in \tEQsix.
}
\label{f:eqbaxavg1}
\end{figure}

\begin{figure}
\centering
\hskip 0.8ex
{\footnotesize \tEQeight}{\includegraphics[width=0.25\textwidth]{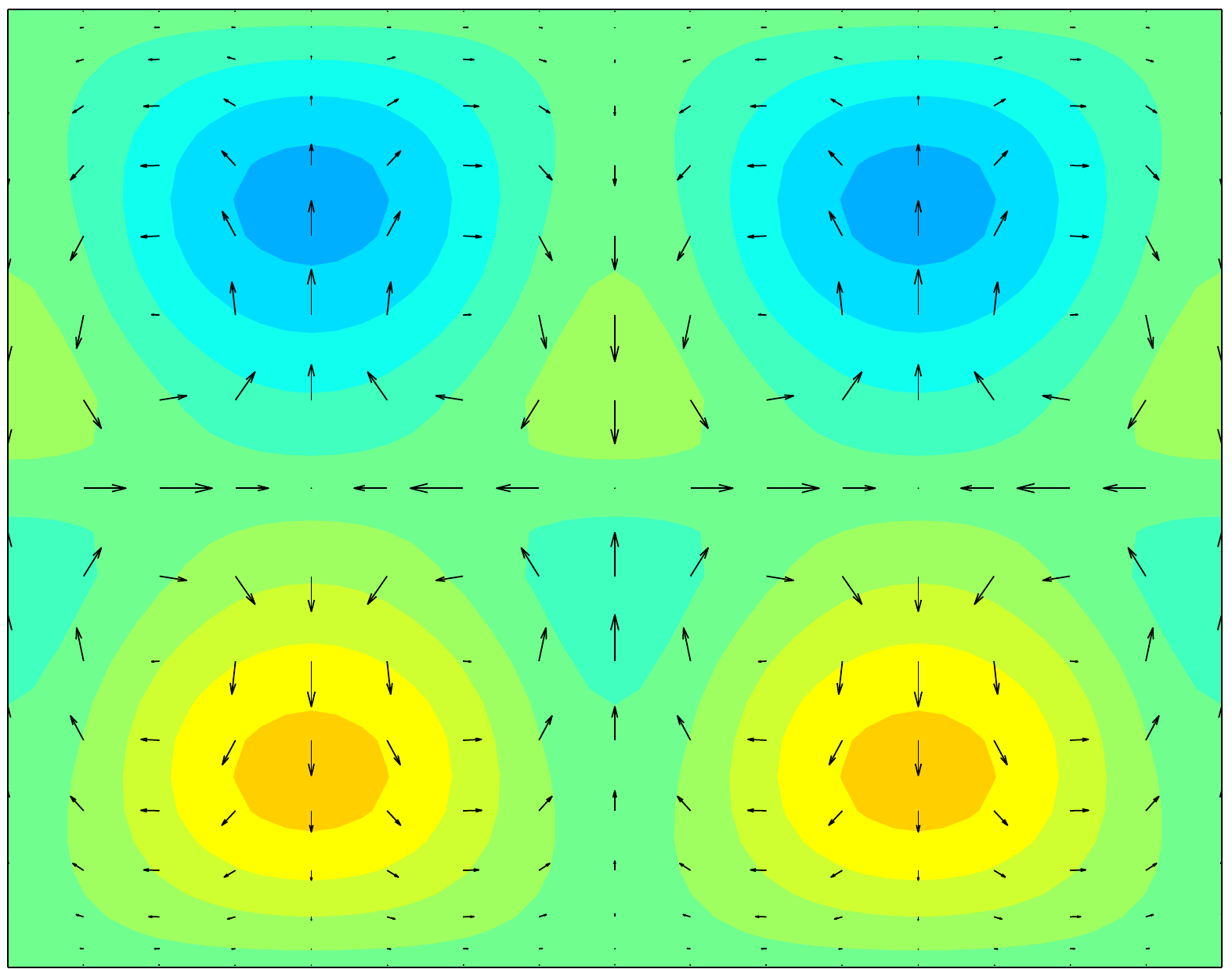}}
\hskip 31ex
{\footnotesize \tEQelev} {\includegraphics[width=0.25\textwidth]{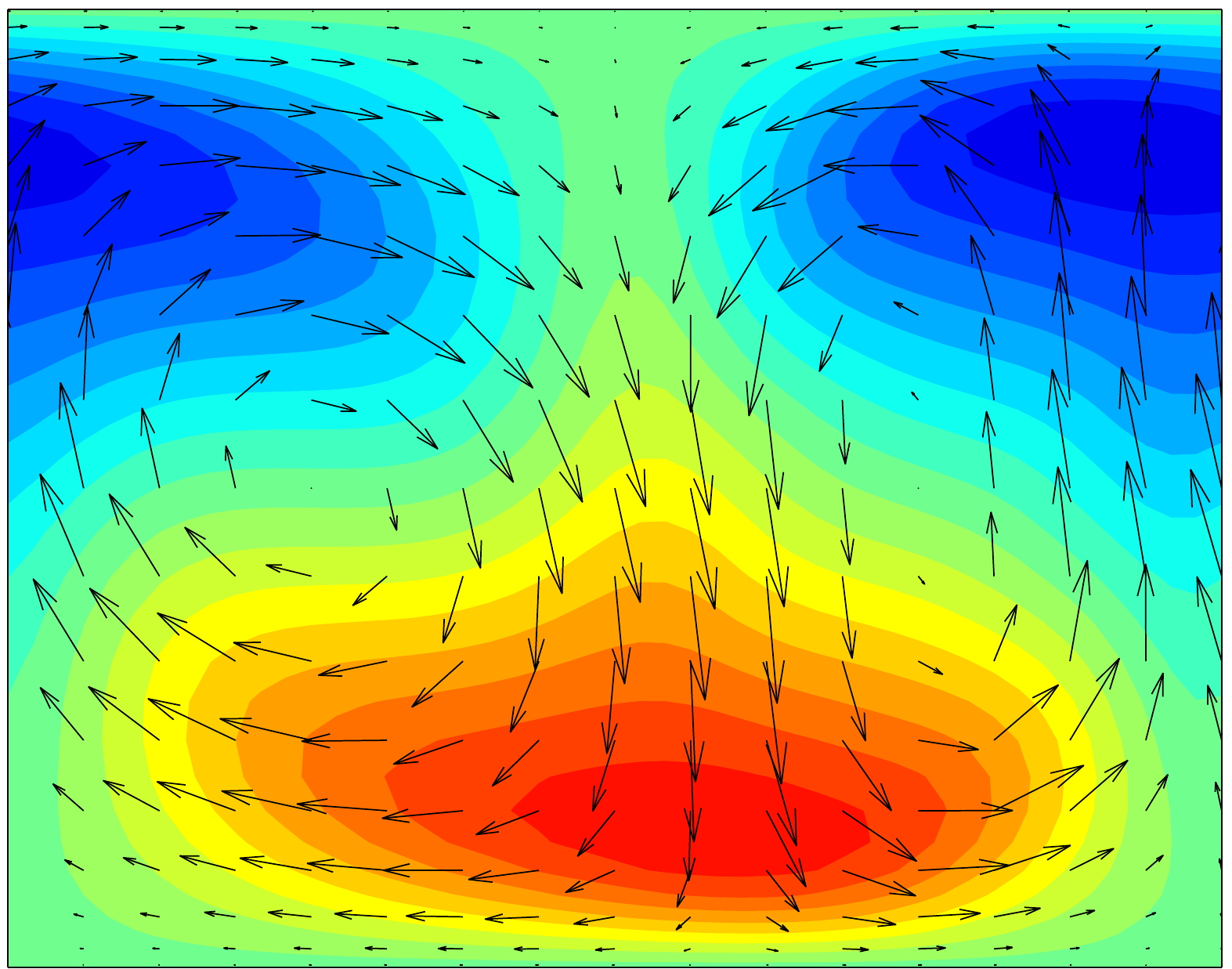}}
\\
{\footnotesize \tEQsev}  {\includegraphics[width=0.25\textwidth]{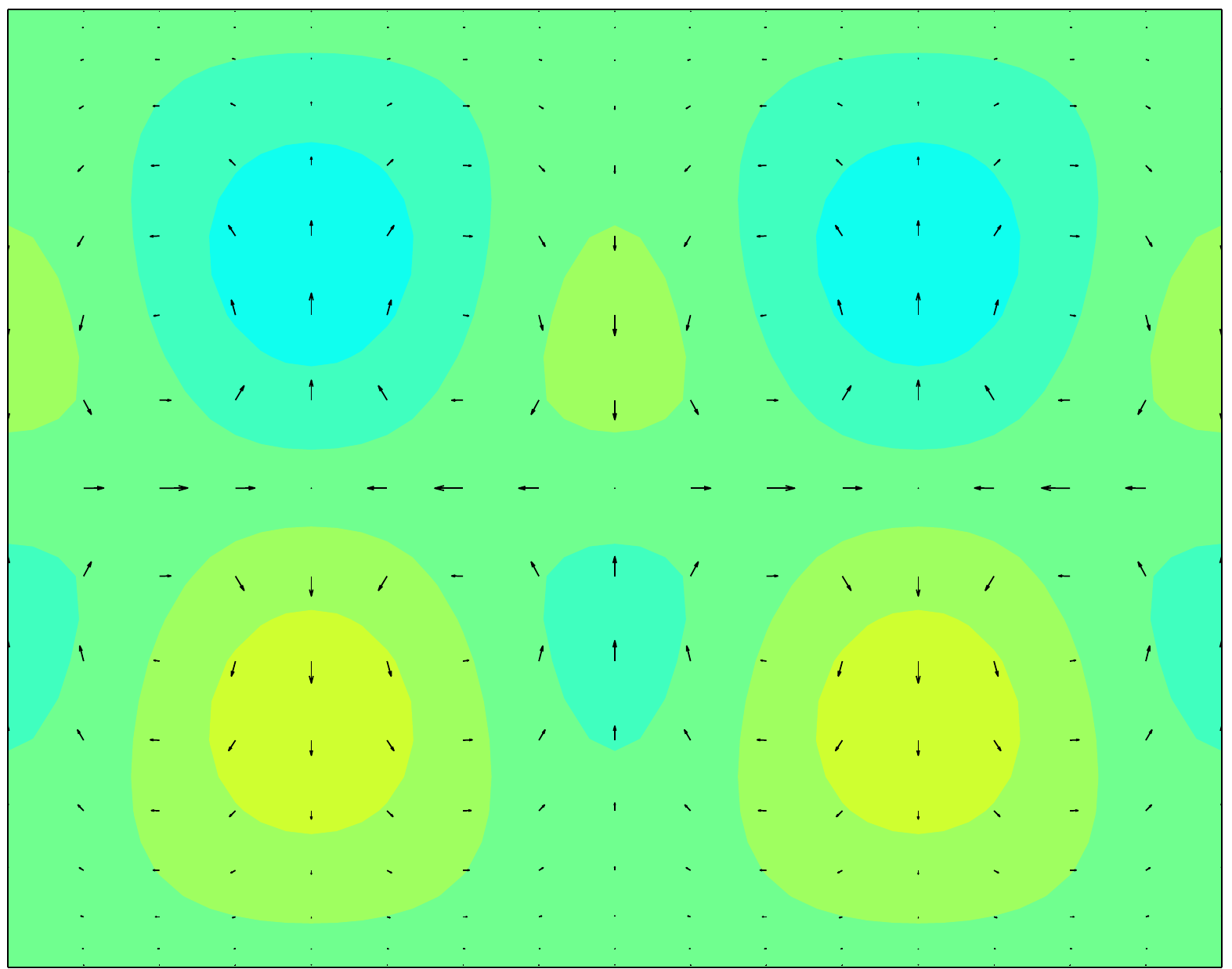}}
{\footnotesize \tEQnine} {\includegraphics[width=0.25\textwidth]{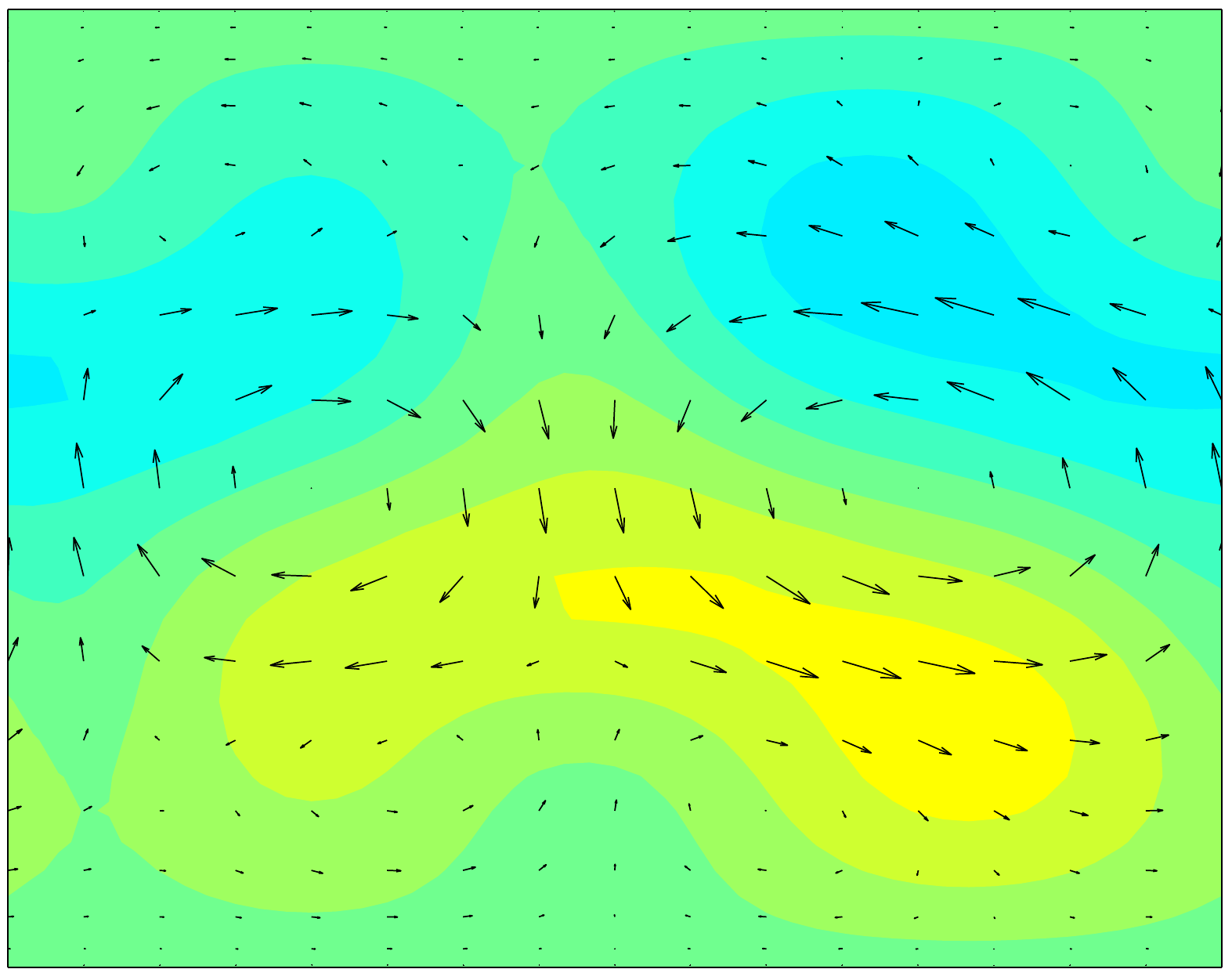}}
{\footnotesize \tEQten}  {\includegraphics[width=0.25\textwidth]{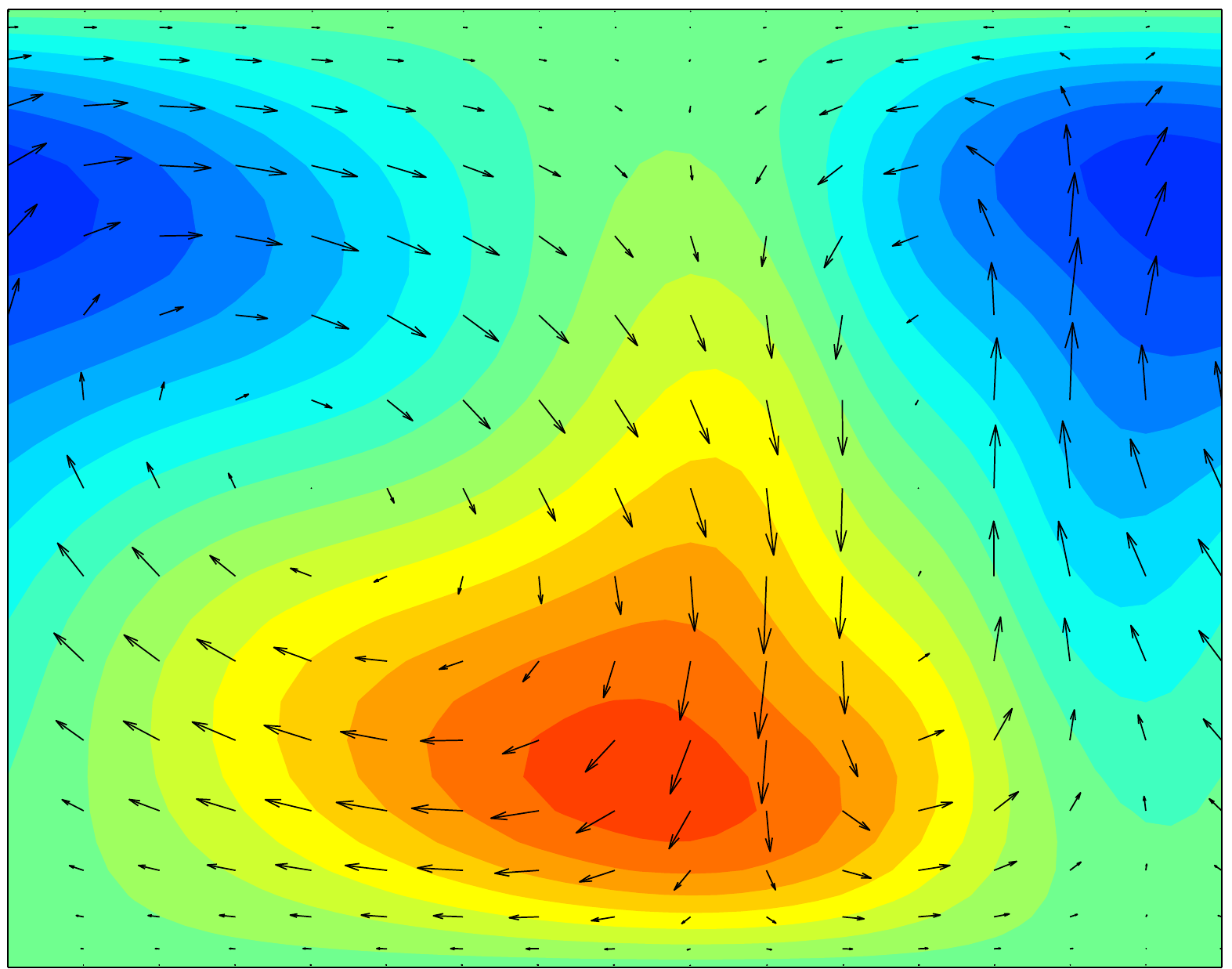}}
\caption[$x$-average cross-sections $\hbu(y,z)$ of \eqva\ 5-11]
{$x$-average of $\hbu$ for \eqva\ \tEQsev-\tEQelev\
\edit{in the $\bNarrow$ cell at $\Reynolds=400$.} Plotting conventions are
the same as in \reffig{f:eqbaxavg1}.}
\label{f:eqbaxavg2}
\end{figure}

\begin{figure}
\centering
 (a)\!\includegraphics[width=0.44\textwidth]{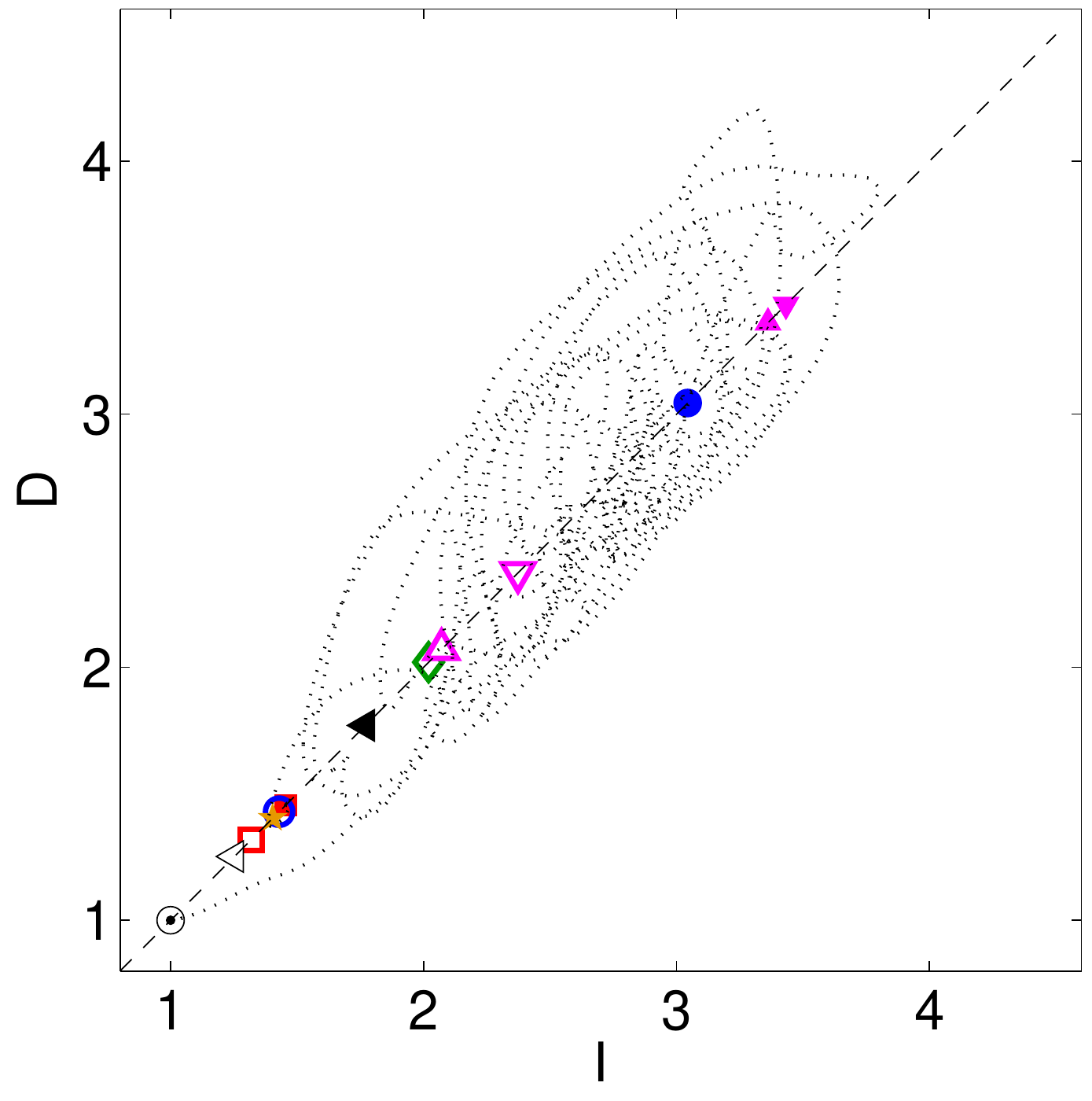}
~(b)\!\includegraphics[width=0.44\textwidth]{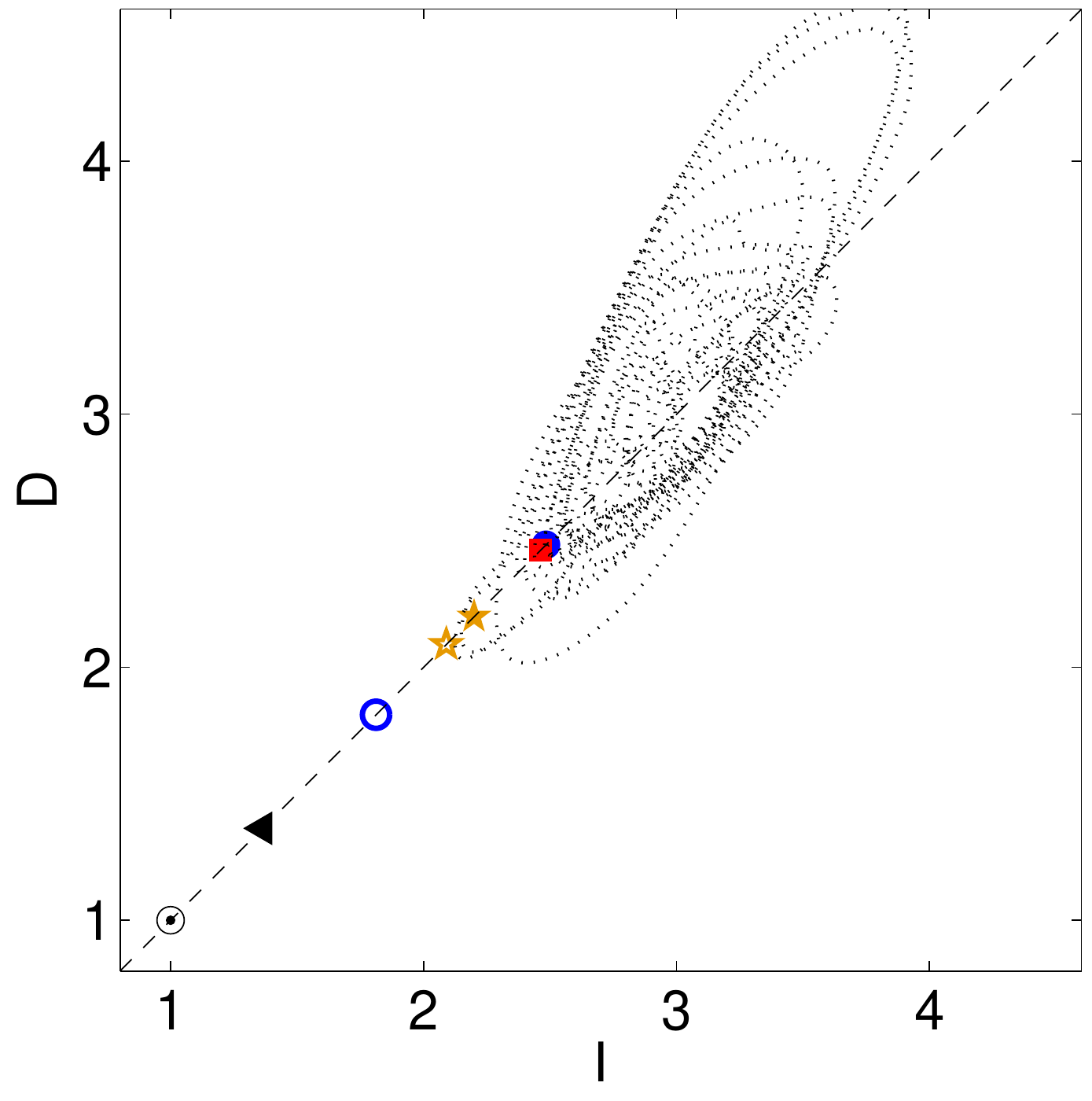}
 \caption[$I-D$ plot.]{
Rate of energy input at the walls $I$ versus dissipation $D$, for
\eqva\ in the (a) \bNarrow\ (b) \bHKW\ cell at $\Reynolds=400$. \edit{
Typical turbulent trajectories marked are with dotted lines.}
\sLM\ \tLM\ is the laminar {\eqv} with $D=I=1$.
In (a),
\sEQone\   \tEQone,
\sEQthree\ \tEQthree,
\sEQfour\  \tEQfour,
\sEQsev\  \tEQsev, and
\sEQnine\  \tEQnine,
are clustered in the range $1.25 < D\!=\!I < 1.55$, and
\sEQtwo\   \tEQtwo,
\sEQfive\  \tEQfive,
\sEQeight\  \tEQeight,
\sEQten\   \tEQten,
\sEQelev\  \tEQelev\,
\edit{
  \sEQtwel\  \tEQtwel\, and
  \sEQthirt\  \tEQthirt\
}
lie in $2 < D\!=\!I < 4$.
\edit{Traveling waves \tTWone, \tTWtwo, and \tTWthree\ (not shown)
are clustered together with \sEQone\  \tEQone, etc.
In (b), the symbols are the same, with both the lower and upper
branches of \tEQnine\ marked with \sEQnine.} \sEQfour\ \tEQfour,
\sEQtwo\ \tEQtwo\ are clustered together near $D = I \approx 2.47$.
}
\label{f:IvD-W03}
\end{figure}

%\medskip\noindent
{\bf \tLB, \tUB\ \eqva.}
This pair of solutions was discovered by \cite{N90}, recomputed by different
methods by
\edit{
\citecomm{Clever and Busse\rf{CB92,CB97}}{\cite{CB92,CB97}}
       }
and \cite{W98,W03}, and found multiple times in searches
\edit{initiated from turbulent simulation data,}
as described above. The lower branch \tLB\ and the upper
branch \tUB\ are born together in a saddle-node bifurcation at
$\Reynolds \approx 218.5$. Just above bifurcation, the two \eqva\ are connected
by a $\tLB \to \tUB$ \hec, see \cite{GHCV08}.
However, at higher values of $\Reynolds$ there appears to be no such simple connection.
The lower branch \tLB\ \eqv\ is discussed in detail in \cite{WGW07}. This
{\eqv} has a 1-dimensional unstable manifold for a  wide range of parameters.
Its stable manifold appears to provide a partial barrier between the basin of
attraction of the laminar state and turbulent states \citep{SGLDE08}.
The upper branch \tUB\ has an 8-dimensional unstable manifold and a dissipation
rate that is higher than the turbulent mean, see \reffig{f:IvD-W03}\,(a).
However, within the $S$-invariant subspace \tUB\ has just one pair of
unstable complex eigenvalues. The two-dimensional $S$-invariant section
of its unstable manifold was explored in some detail in \citetalias{GHCW07}.
It appears to bracket the upper end of turbulence in \statesp, as illustrated
by \reffig{f:WalR400all}.

\edit{
{\bf \tNNB, \tNB.}
The upper branch \tNB\ solution} was found in \citetalias{GHCW07}
and is called $\bu_{\text{\tiny NB}}$ there.
Its lower-branch partner \tNNB\ was found by continuing \tNB\ downwards in
$\Reynolds$ and also by independent searches from samples of turbulent
data. \tNB\ is, with \tLB, the most frequently found {\eqv}, which attests
to its importance in turbulent dynamics.
Like \tLB, \tNB\ serves as a gatekeeper between turbulent flow and the
laminar basin of attraction. As shown in \citetalias{GHCW07}, there is a
heteroclinic connection from \tNB\ to \tLB\ resulting from a complex instability
of \tNB. Trajectories on one side of the heteroclinic connection decay rapidly
to laminar flow; those one the other side take excursion towards turbulence.

\begin{table}
\centering
\begin{tabular}{l|llllccccc}
          & ~$\Reynolds$ & ~~$\Norm{\cdot}$   & ~~$E$
          & ~$D$ & $H$ & $\dim W^u$ & $\dim W^u_{H}$ & acc.
          & freq. \\
 \hline
mean      &     & 0.2828 & 0.087  & 2.926 &  \\
\tLM      &     & 0      & 0.1667 & 1     & $\GPCF$    & ~0 & ~0 &       & 2 \\
\tLB      &     & 0.2091 & 0.1363 & 1.429 & $S$    & ~1 & ~1 & $10^{-6}$ & 7 \\
\tUB      &     & 0.3858 & 0.0780 & 3.044 & $S$    & ~8 & ~2 & $10^{-4}$ & 3 \\
\tNNB     &     & 0.1259 & 0.1382 & 1.318 & $S$    & ~4 & ~2 & $10^{-4}$ & 2 \\
\tNB      &     & 0.1681 & 0.1243 & 1.454 & $S$    & ~6 & ~3 & $10^{-4}$ & 8 \\
\tEQfive  &     & 0.2186 & 0.1073 & 2.020 & $S$    & 11 & ~4 & $10^{-3}$ & 1 \\
\tEQsix   & 330 & 0.2751 & 0.0972 & 2.818 & $S$    & 19 & ~6 & $10^{-3}$ &   \\
\tEQsev   &     & 0.0935 & 0.1469 & 1.252 & $S  \times \{e, \tau_{xz}\}$    & ~3 & ~1 & $10^{-4}$ & 3 \\
\tEQeight &     & 0.1756 & 0.1204 & 3.044 & $S  \times \{e, \tau_{xz}\}$    & 15 & ~2 & $10^{-3} $&   \\
\tEQnine  &     & 0.1565 & 0.1290 & 1.404 & $\{e, \sigma_{xz}\}$ & ~5 & ~3 & $10^{-4}$ & 1 \\
\tEQten   &     & 0.3285 & 0.1080 & 2.373 & $\{e, \sigma_{xz}\}$ & 10 & ~7 & $10^{-4}$ &   \\
\tEQelev  &     & 0.4049 & 0.0803 & 3.432 & $\{e, \sigma_{xz}\}$ & 13 & 10  & $10^{-4}$ & \\
\tEQtwel  &     & 0.3037 & 0.1159 & 2.071 & $\{e, \sigma_{xz}\}$ & ~5 & ~4  & $10^{-5}$ & \\
\tEQthirt &     & 0.4049 & 0.0813 & 3.361 & $\{e, \sigma_{xz}\}$ & 15 & ~9  & $10^{-3}$ & \\
\end{tabular}
\caption[Properties of {\eqv} solutions.]{Properties of \eqv\
solutions for \bNarrow\ cell, $\Reynolds=400$, unless noted otherwise.
The mean values are ensemble and time averages over transient
turbulence. $\Norm{\cdot}$ is the $L^2$-norm of the velocity deviation
from laminar, $E$ is the energy density \refeq{innerproduct}, $D$ is
the dissipation rate, $H$ is the isotropy subgroup, $\dim W^u$ is the
dimension of the {\eqv}'s unstable manifold or the number of its
unstable eigenvalues, and $\dim W^u_H$ is the dimensionality of the
unstable manifold within the $H$-invariant subspace, or the number of
unstable eigenvalues with the same symmetries as the {\eqv}. The
accuracy {\em acc.} of the solution at a given resolution (a $32 \times 33
\times 32$ grid) is estimated by the magnitude of the residual
$\Norm{(f^{T=1}(\bu) - \bu)}/\Norm{\bu}$ when the solution is
interpolated and integrated at higher resolution (a $48 \times 49 \times
48$ grid, $\Delta t = 0.02$). The {\em freq.} column
shows how many times a solution was found among the 28 searches
initiated with samples of the natural measure within the
$S$-invariant subspace.
See also \reffig{f:IvD-W03}\,(a).
    }
\label{t:eqbtable5}
\end{table}

\edit{{\bf \tEQfive, \tEQsix.}}
The lower-branch \tEQfive\ solution was found only once in our 28 searches, and its upper-branch
partner \tEQsix\ only by continuation in Reynolds number. We were only able to
continue \tEQsix\ up to $\Reynolds=335$. At this value it is highly unstable,
with a 19 dimensional unstable manifold, and it is far more dissipative than
a typical turbulent trajectory.

{\bf \tEQsev, \tEQeight} appear together in a saddle
node bifurcation in $\Reynolds$ (see \refsect{s:bifurRe}).
\tEQsev\ / \tEQeight\ might be the same as \cite{Schmi99}'s
`$\sigma$ solutions'. The $x$-average velocity field plots
appear very similar, as do the $D$ versus $\Reynolds$ bifurcation
diagrams. We were not able to obtain Schmiegel's data in order to
make  a direct comparison.
%In this cell, we were not able to continue \tEQeight\
%past $\Re = 270$.
\tEQsev\ is both the closest state to laminar in
terms of disturbance energy and the lowest in terms of drag. It
has one strongly unstable real eigenvalue within the $S$-invariant
subspace and two weakly unstable eigenvalues with $\{s_1,s_3\}$ and
$\{s_2,s_3\}$ antisymmetries, respectively. In this regard, the \tEQ7\
unstable manifold might, like the unstable manifold of \tLB, form part
of the boundary between the laminar basin of attraction and
turbulence. \tEQsev\ and \tEQeight\ are unique among the \eqva\
determined here in that they have the order-8 isotropy subgroup
$S  \times \{e, \tau_{xz}\}$ (see \refsect{s:67-fold}). The action of
the quotient group $G/(S  \times \{e, \tau_{xz}\})$ yields 2 copies of
each, plotted in \reffig{f:reqvaPortrait}. \tEQsev\ and \tEQeight\ are
similar in appearance to \tEQfive\ and \tEQsix, except for the
additional symmetry.

{\bf \tEQnine} is a single lopsided roll-streak pair.
It is produced by a pitchfork bifurcation from \tNB\ at $\Reynolds
\approx 370$ as an $\{s_1,s_2\}$-antisymmetric eigenfunction goes through
marginal stability (the only pitchfork bifurcation we have yet found)
and remains close to \tNB\ at $\Reynolds = 400$.
Thus, it has $\{e, \sigma_{xz}\}$ isotropy. Even though \tEQnine\ is
not $S$-isotropic, we found it from a search initiated on a guess that was
$S$-isotropic to single precision. Such small asymmetries were enough to
draw the Newton-hookstep search algorithm out of the $S$-invariant subspace.

{\bf \tEQten, \tEQelev} are produced in a saddle-node bifurcation
at $\Reynolds \approx 348$ as a lower~/~upper branch pair, and
they lie close to the center of mass of the turbulent repeller,
see \reffig{f:IvD-W03}\,(a). The \edit{velocity fields
have a similar appearance to }
typical turbulent states for this cell size. However, they are
both highly unstable and unlikely to be revisited frequently by
a generic turbulent fluid state.
Their isotropy subgroup $\{e, \sigma_{xz}\}$ is order 2,
so the action of the quotient
group $G/\{e, \tau_{xz}\}$ yields 8 copies of each, which
appear as the 4 overlaid pairs in  projections onto the
\refeq{globalUBframe} basis set, see \reffig{f:WalR400all} and
\reffig{f:reqvaPortrait}. {\bf \tEQtwel, \tEQthirt}
were found by continuation of \tEQten\ in $\gamma$.

\subsection{\Reqva}
\label{s:reqba}

\begin{figure}
\centering
{\includegraphics[width=0.31\textwidth]{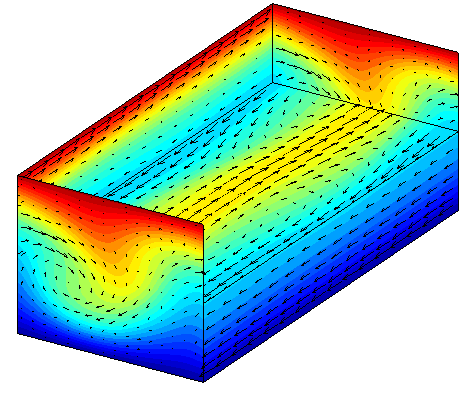}} \hskip -30ex \tTWone   \hskip 25ex
{\includegraphics[width=0.31\textwidth]{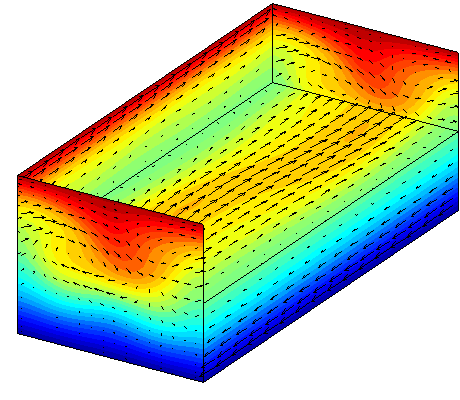}} \hskip -30ex \tTWDone  \hskip 26ex
{\includegraphics[width=0.31\textwidth]{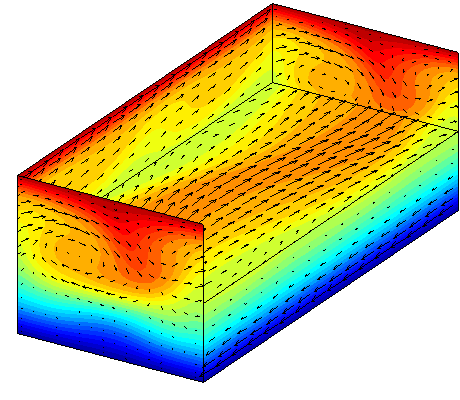}} \hskip -30ex \tTWthree \hskip 25ex ~
  \caption[\Reqva, $\Reynolds=400$.]{
Spanwise \tTWone, streamwise \tTWDone\ and \tTWthree\ \reqva\ in
\bNarrow\ cell, $\Reynolds=400$.}
\label{f:reqvaboxes}
\end{figure}

The first two \reqvD\ solutions reported in the literature were found by
\cite{N97} by continuing \tLB\ \eqv\ to a combined Couette / Poiseuille channel
flow, and then continuing back to \pCf. The result was a pair of streamwise
\reqva\ arising from a saddle-node bifurcation. \cite{Visw07a} found two
\reqva, `D1' and the same solution `D2,' but at a higher $\Reynolds = 1000$,
through an edge-tracking algorithm (\cite{SYE05}, see also
\refsect{s:eqbSols}).
Here we verify Viswanath's solution
and present two new \reqvD\ solutions computed as
symmetry-breaking bifurcations off {\eqv} solutions.
We were not able to compare these
to Nagata's \reqva\ since the data is not available. The
\reqva\ are shown as 3D velocity fields in \reffig{f:reqvaboxes}
and as closed orbits in {\statesp} in \reffig{f:reqvaPortrait}.
Their kinetic
energies and dissipation rates are tabulated in \reftab{t:eqbtable5}. Each
\reqvD\ solution has a zero spatial-mean pressure gradient but
non-zero \vCM\ in the same direction as the wave velocity.
It is likely
each solution could be continued to zero wave velocity but non-zero
spatial-mean pressure gradient.

{\bf \tTWone} \reqv\
 is $s_2$-isotropic and hence spanwise traveling. At $\Reynolds = 400$
its velocity is very small, $\mathbf{c} = 0.00655\;\ez$, and
it has a small but nonzero \vCM, also in the spanwise
direction. This is a curious
property: \tTWone\ induces bulk transport of fluid without a pressure gradient,
and in a direction orthogonal to the motion of the walls.
\tTWone\ was found as
a pitchfork bifurcation from \tLB, and thus lies very close to it in \statesp.
It is weakly unstable, with a $3D$ unstable manifold with two eigenvalues
which are
extremely close to marginal. In this sense \tTWone\ unstable manifold is nearly
one-dimensional, and comparable to \tLB.

{\bf \tTWDone} is a streamwise \reqv\ found by
\cite{Visw07a} and called ${\text{D1}}$ there. It is $s_1$-isotropic, has
a low dissipation rate, and a small but nonzero \vCM\ in the streamwise
direction. Viswanath provided data for this solution; we verified it
with an independent numerical integrator and continued the solution to $\bNarrow$
cell for comparison with the other \reqva. In this cell $\tTWDone$ is fairly stable,
with an eigenspectrum similar to \tTWone's, except with different symmetries.

{\bf \tTWthree}
 is an $s_1$-isotropic streamwise \reqv\ with a
relatively high wave velocity $\mathbf{c} = 0.465\; \ex$ and a nonzero
\vCM\ in the streamwise direction. Its dissipation rate
and energy norm are close to those of \tTWone.

\begin{table}
\centering
\begin{tabular}{l|lllcccccc}
          &  ~~$\Norm{\cdot}$   & ~~$E$
          & ~$D$ & $H$ & $\dim W^u$ & $\dim W^u_{H}$ % & acc.  PC omitted
          & $\mathbf{c}$ & ~mean $\bu$\\
\hline
mean      & 0.2828 & 0.087  & 2.926 &  \\
\tTWone   & 0.2214 & 0.1341 & 1.510 & $\{e, \sigma_x \tau_z\}$ & 3 & 2 &  $0.00655\;\ez$~ &~ $0.00482\;\ez$ \\
\tTWDone  & 0.1776 & 0.1533 & 1.306 & $\{e, \sigma_z \tau_x\}$ & 3 & 2 &  $0.3959~\; \ex$~ &~ $0.0879~\; \ex$ \\
\tTWthree & 0.2515 & 0.1520 & 1.534 & $\{e, \sigma_z \tau_x\}$ & 4 & 2 &  $0.4646~\; \ex$~ &~ $0.1532~\; \ex$ \\
\end{tabular}
\caption[Properties of {\reqvD} solutions.]{Properties of {\reqvD}
solutions for \bNarrow\ cell, $\Reynolds=400$, defined as in \reftab{t:eqbtable5},
with wave velocity $\mathbf{c}$ and \vCM. See also \reffig{f:IvD-W03}\,(a).}
\end{table}

\section{Continuation under Reynolds number}
\label{s:bifurRe}

\begin{figure}
\centering
{\small (a)}\includegraphics[width=0.45\textwidth]{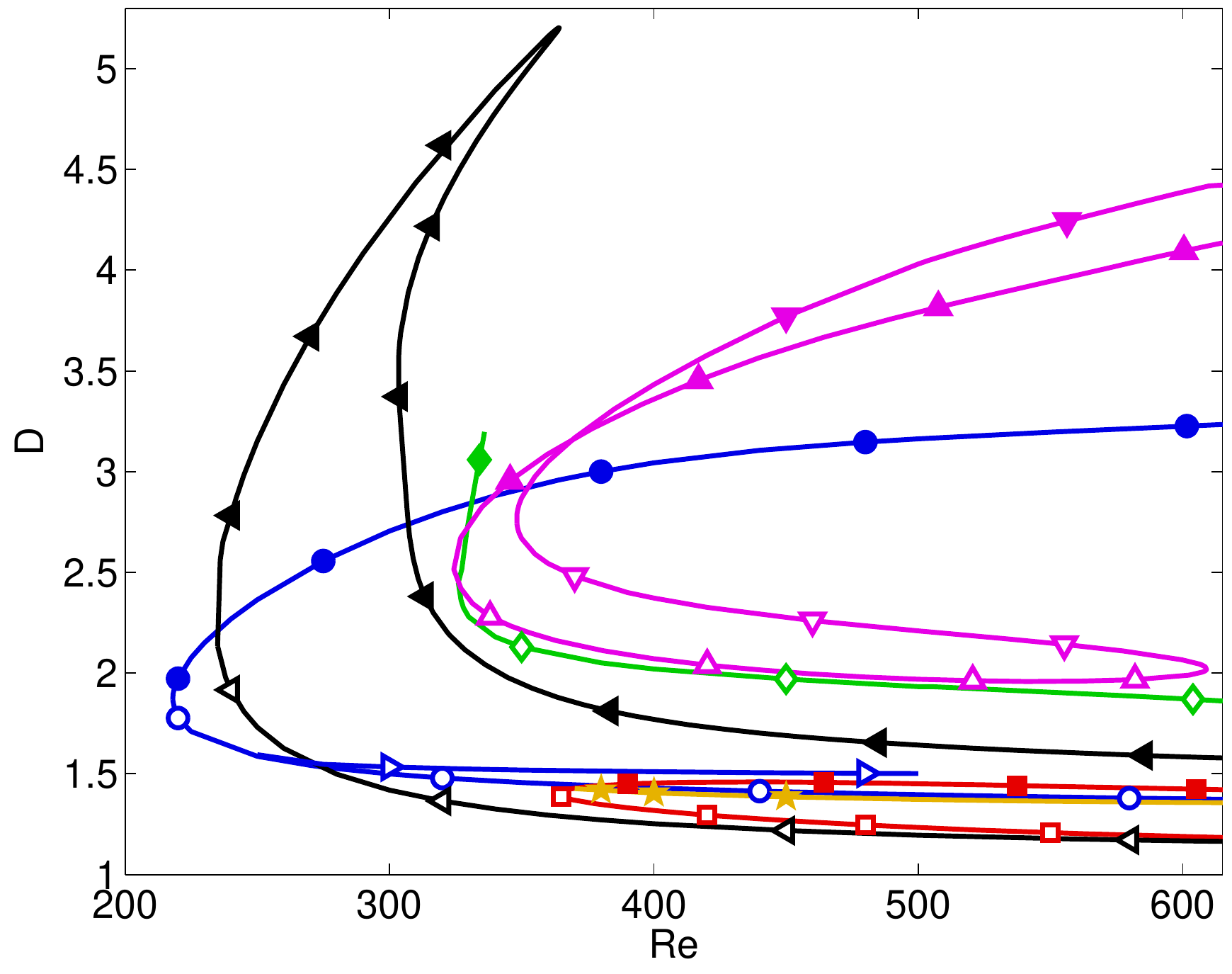}%
~~{\small (b)}\includegraphics[width=0.47\textwidth]{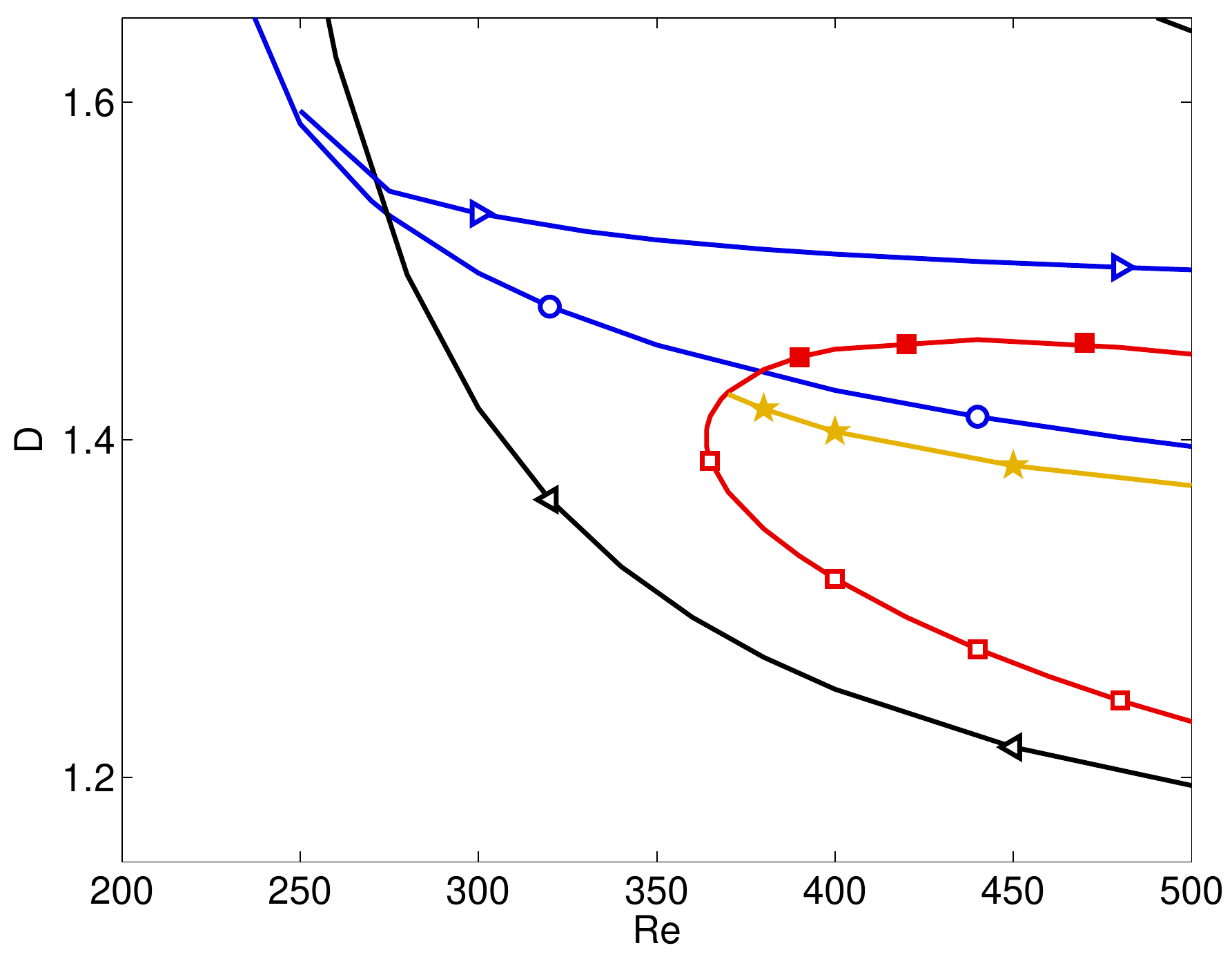}
\caption[Bifurcations of (relative) {\eqva} under variation of \Reynolds.]{
(a) Dissipation $D$ of \eqva\ as a function of Reynolds number for
the $\bNarrow$ cell, \edit{$\alpha, \gamma = 1.14, 2.5$.}
\sLB\ \tLB, \sUB\ \tUB, \sNNB\ \tNNB, \sNB\ \tNB,
\sEQfive\ \tEQfive, \sEQsix\ \tEQsix, \sEQsev\ \tEQsev, \sEQeight\ \tEQeight,
\sEQnine\ \tEQnine, \sEQten\ \tEQten, \sEQelev\ \tEQelev,
\sEQtwel \tEQtwel,
\edit{\sEQthirt \tEQthirt, and \sTWone\ \tTWone.}
(b) Detail of (a). \edit{\tEQone, \tEQtwo, \tEQthree, \tEQfour,
\tEQsev, and \tEQnine\ extend past $\Reynolds= 1000$ (not plotted here).}
}
\label{f:bifurRe}
\end{figure}

The relations between the {\eqv} and {\reqvD} solutions can be clarified
by tracking their properties under changes in \Reynolds\ and spatial
periodicity.
\refFig{f:bifurRe} shows a bifurcation diagram for \eqva\ and \reqva\ in
the $\bNarrow$ cell, with dissipation rate $D$ plotted against $\Re$ as
the bifurcation parameter. A number of independent solution curves are shown
in superposition. This is a 2-dimensional projection from the $\infty$-dimensional
\statesp, thus, unless noted otherwise, the apparent intersections of the solution
curves do not represent bifurcations; rather, each curve is a family of
solutions with an upper and lower branch, beginning with a saddle-node
bifurcation at a critical Reynolds number.

The first saddle-node bifurcation gives birth to the Nagata lower branch
\tLB\ and upper branch \tUB\ \eqva, at $\Re \approx 218.5$.
\tLB\ has a single $S$-isotropic unstable eigenvalue (and additional
subharmonic instabilities that break the $S$-isotropy of \tLB).
Shortly after bifurcation, \tUB\ has an unstable complex eigenvalue pair within the
$S$-invariant subspace and two unstable real eigenvalues leading out of that space.
As indicated by
the gentle slopes of their bifurcation curves, the \cite{N90} solutions are
robust with respect to Reynolds number. The lower branch solution has been
continued past $\Reynolds =10,000$ and has a single unstable eigenvalue
throughout this range \citep{WGW07}.

\edit{\tEQsev\ and \tEQeight\ are formed in a saddle-node bifurcation
at $\Reynolds \approx 235$. The \tEQeight\ bifurcation curve is unusual in that
it increases rapidly from the bifurcation point to a maximum dissipation of
$D=5.2$ at $\Reynolds=364$, and then turns rapidly but smoothly back down to
much lower dissipation at higher Reynolds numbers. This behavior persists
when examined at higher spatial and temporal resolutions.
}

\tNNB\ and \tNB\ were discovered in independent Newton searches and subsequently
found by continuation to be lower and upper branches of a saddle-node
bifurcation occurring at $\Reynolds \approx 364$. \tNNB\ has a leading unstable
complex eigenvalue pair within the $S$-invariant subspace. Its remaining two unstable
eigen-directions are nearly marginal and lead out of this space.

\tEQsix\ was found by continuing \tEQfive\ backwards in $\Reynolds$ around the
bifurcation point at $\Reynolds \approx 326$. We were not able to continue \tEQsix\
past $\Reynolds=335$. At this point it has a nearly marginal stable pair of eigenvectors
whose isotropy group is $\GPCF$, which rules out a
bifurcation to {\reqva} along these modes. Just beyond  $\Reynolds = 335$
the dynamics in this region appears to be roughly periodic, suggesting that \tEQsix\
undergoes a supercritical Hopf bifurcation here. At $\Reynolds \approx 348$,
\tEQten, \tEQelev\ are born in a saddle node bifurcation, similar in character
to the \tLB\ / \tUB\ bifurcation, \edit{as are \tEQtwel, \tEQthirt.}

\refFig{f:bifurRe}{\textit(b)} shows several symmetry-breaking bifurcations.
At $\Reynolds \approx 250$, \tTWone\ bifurcates from \tLB\ in a
subcritical pitchfork as an $s_2$-symmetric, $s_1,s_3$-antisymmetric
eigenfunction of \tLB\ becomes unstable, resulting in a
spanwise-moving \reqv. At $\Reynolds \approx 370$, the $\tEQnine$
\eqv\ bifurcates off \tNB\ along an \edit{$s_1,s_2$-antisymmetric,
$s_3$-symmetric} eigenfunction of \tNB.
Since $s_3$ symmetry fixes
phase in both $x$ and $z$,  this solution bifurcates off \tNB\ as an
{\eqv} rather than a {\reqv}.

\edit{
\refFig{f:spanwise-eqva}\textit{(a)} shows the instability of each
{\eqv} as a function of Reynolds number.
As a measure of instability we use
the sum of the real parts of the {\eqv}'s unstable eigenvalues,
\ie\ the local exponential rate of stretching of the
{\eqv}'s unstable manifold. Several observations can be made.
Each lower-branch solution (open symbol) is less unstable than its upper-branch
counterpart (closed symbol). Lower branch solutions tend towards
{\em lesser instability} as Reynolds number increases, upper branch solutions
become {\em more} unstable. Since lower/upper branches of a given solution are
defined by lower/higher dissipation rates, this implies that lower instability
and lower dissipation go hand in hand. However this relation does not
generally hold between different solution branches: \tEQsev\ has lower
dissipation than \tEQone (\reffig{f:bifurRe}\textit{(a)}) but is more unstable
(\refFig{f:spanwise-eqva}\textit{(a)}). Several of the lower-branch
solutions have very slowly decreasing instability over the range of Reynolds
numbers shown; for these, the {\em number} of unstable eigenvalues is constant
or slowly decreasing as well.}
\tEQone\ has three unstable eigenvalues shortly after bifurcation and just one
for $270 \leq \Reynolds \leq 10,000$,
\edit{
\tEQsev\ has six after bifurcation and three for $340 \leq \Reynolds \leq 800$,
\tEQthree\ has four from bifurcation onwards, $363.9 \leq \Reynolds \leq 800$, and
\tEQtwel\ has eight from bifurcation onwards, $324.4 \leq \Reynolds \leq 600$.
The upper limits of these ranges are merely the endpoints of our calculations.
\cite{CB97} show that at certain wavenumbers, \tEQtwo\ is stable for a small range
of Reynolds numbers just after bifurcation. We did not look for or find regions of
stability for any of  the new solutions, though we expect such regions exist for
carefully tuned parameters. For example, a stable periodic orbit has been found
in the related Kuramoto-Sivashinsky system \citep{lanCvit07}.
For the upper branch
solutions generally both the numbers of unstable eigenvalues as well
as their sums increase with $\Reynolds$. Lastly, it should be remembered that
it is not at all clear how much the instability of an \eqv\ has to do with
the Lyapunov exponents of the turbulent flow - that depends on how close and how
frequently a typical trajectory visits the neighborhood of a given \eqv.
}

\section{Continuation under spanwise wavenumber}
\label{s:aspect}

\begin{figure}
\centering
{\small (a)} \includegraphics[width=0.45\textwidth]{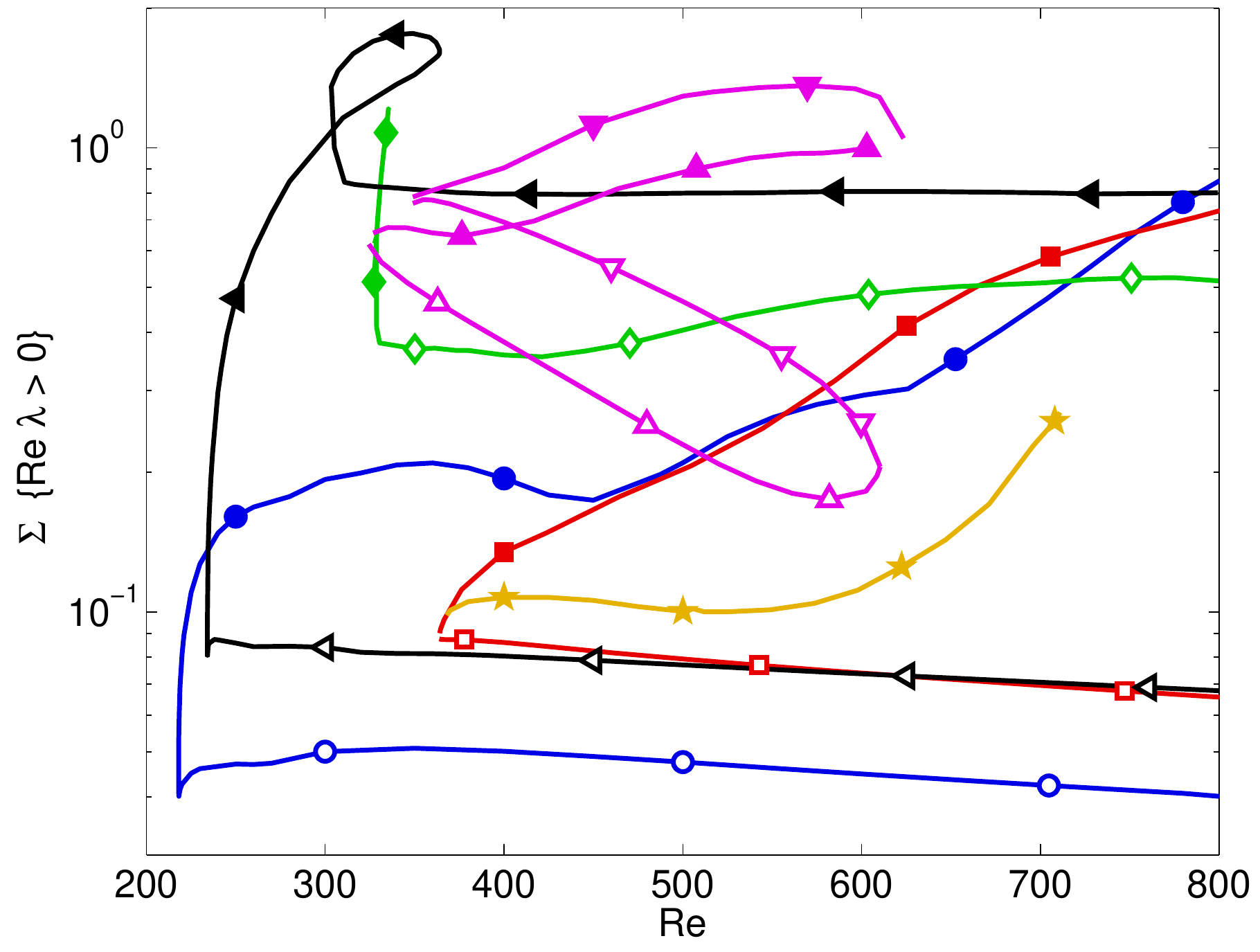}%
{\small (b)} \includegraphics[width=0.45\textwidth]{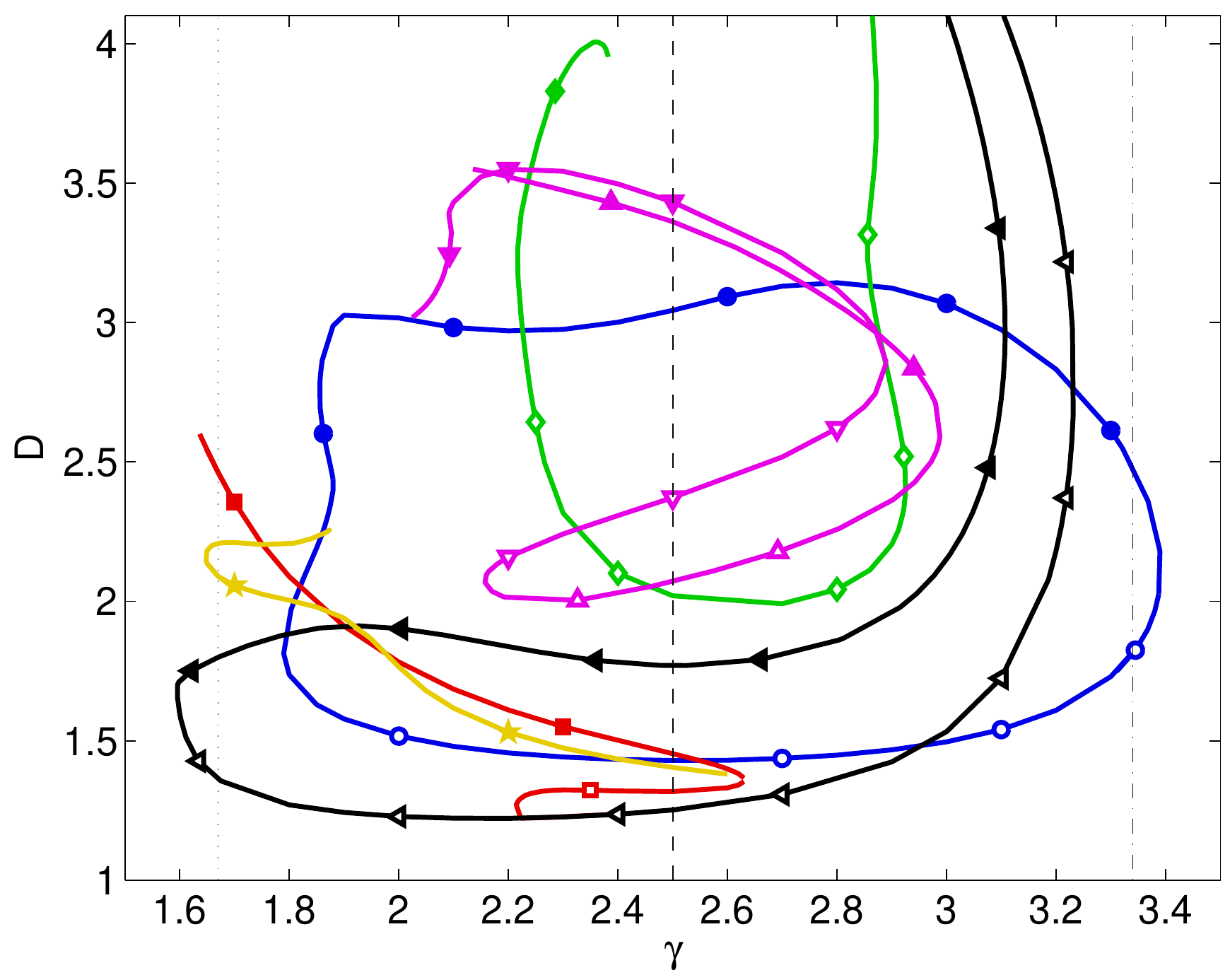}%
\caption{
\edit{
(a) Sum of the real parts of unstable eigenvalues of equilibria as a
function of $\Reynolds$, at $\alpha,\gamma = 1.14,2.5$ (the
\bNarrow\ cell).
(b) Dissipation $D$ of {\eqva} as a function of spanwise wavenumber
$\gamma$, with fixed streamwise wavenumber $\alpha = 1.14$ and Reynolds
number $\Reynolds = 400$. Symbols are \sLB\ \tLB,  \sUB\ \tUB,
\sNNB\ \tNNB,   \sNB\ \tNB,   \sEQfive\ \tEQfive, \sEQsix\ \tEQsix,
\sEQsev\ \tEQsev\,   \sEQeight\ \tEQeight, \sEQnine\ \tEQnine,
\sEQten\ \tEQten,   \sEQelev\ \tEQelev.
\sEQtwel\ \tEQtwel,   \sEQthirt\ \tEQthirt.
The vertical dotted line marks the fundamental wavenumber
$\gamma = 1.67 = 2 \upi/L_z$ of the \bHKW\ cell ($L_z=3.76$), the
dashed  line the fundamental $\gamma = 2.5$ of \bNarrow\ ($L_z=2.51$),
and the dot-dashed line the first harmonic $\gamma = 3.34$
of \bHKW. The intersection of the \tLB, \tUB\ curves with
$\gamma = 3.34$ indicates that these solutions exist as ``doubled''
states in the \bHKW\ cell, as shown in \reffig{f:HKW-eqva}.
The \tEQsev\ and \tEQeight\ curves join smoothly near $(\gamma, D) = (2.6,
5.4)$. We were not able to continue \sEQsix\ \tEQsix\ past $\gamma = 2.38$,
nor \sEQfour\ \tEQfour\ below $\gamma = 1.64$.
}}
\label{f:spanwise-eqva}
\end{figure}

In this section we examine changes in solutions under variation in spanwise
periodicity.
\refFig{f:spanwise-eqva}\textit{(b)} shows the dissipation $D$ of the solutions
as a function of spanwise wavenumber $\gamma$. Only a few of the
intersections in this plot indicate bifurcations; the rest are artifacts
of the projection onto the
\edit{
$(\gamma, D)$ plane. The true bifurcations are
\tEQelev\ branching off from \tEQtwo\ near $(\gamma, D) = (2,3)$;
\tEQthree\ from \tEQsev\ near $(\gamma, D) = (2.2, 1.2)$; and
\tEQnine\ from \tEQfour\ near $(\gamma, D) = (2.6,1.4)$.
Continuation in $\gamma$ also shows that the \tEQten, \tEQelev\ and
\tEQtwel, \tEQthirt\ solution curves, which appear to be independent in
\reffig{f:bifurRe}, are connected by a saddle-node bifurcation
between \tEQten\ and \tEQtwel\ near $(\gamma, D) = (2.15, 2.05)$.}

\begin{figure}
\centering
{\includegraphics[width=0.30\textwidth]{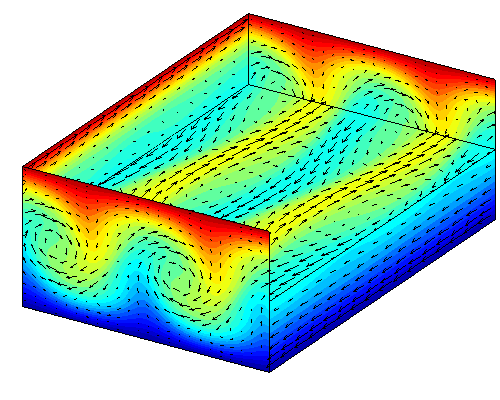}} \hskip -28ex $2 \times$\tEQone   \hskip 22ex
{\includegraphics[width=0.30\textwidth]{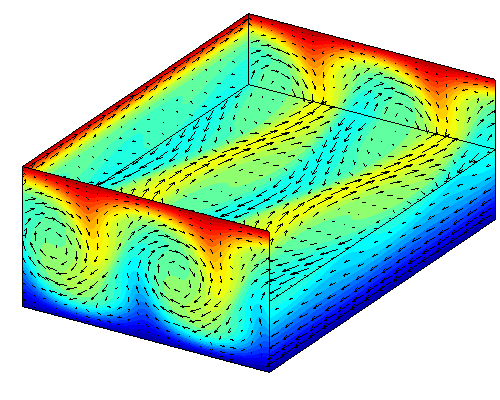}} \hskip -28ex $2 \times$\tEQtwo   \hskip 22ex
{\includegraphics[width=0.30\textwidth]{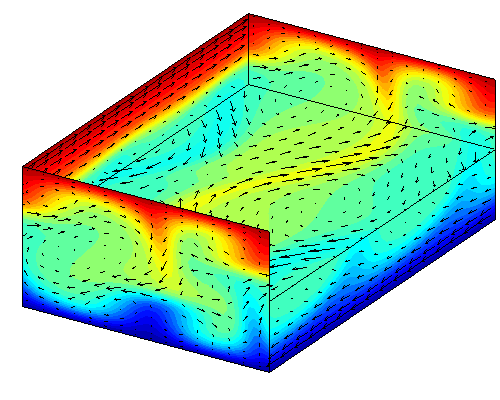}}  \hskip -28ex \tNB      \hskip 25ex ~
 \\
{\includegraphics[width=0.30\textwidth]{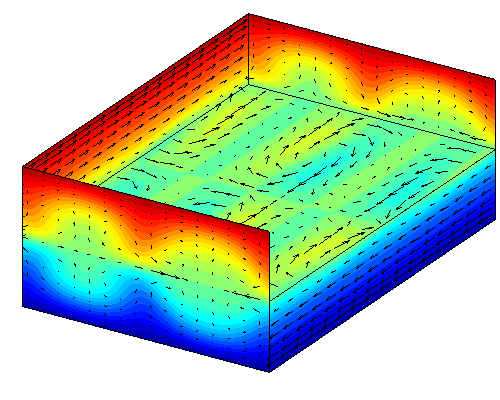}}  \hskip -28ex \tEQsev   \hskip 25ex
{\includegraphics[width=0.30\textwidth]{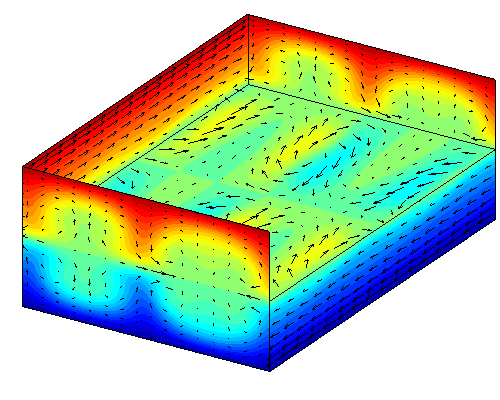}}  \hskip -28ex \tEQeight \hskip 25ex
{\includegraphics[width=0.30\textwidth]{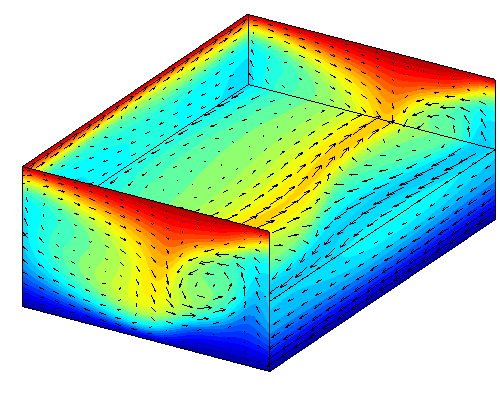}}  \hskip -28ex \tEQnine  \hskip 25ex ~
%{\includegraphics[width=0.30\textwidth]{EQ9bhkw}}  \hskip -28ex \tEQnine b  \hskip 25ex ~
\caption[Physical space plots of HKW cell \eqva.]{
  Equilibria in the \bHKW\ cell of \cite{HaKiWa95}, $\Reynolds = 400$.
 \tNB, \tEQsev, \tEQnine, and the spanwise-doubled \eqv\ solutions
 $2 \times \tLB$ and $2 \times \tUB$.}
\label{f:HKW-eqva}
\end{figure}

\edit{In addition to these bifurcations,}
we are interested in connecting the solutions for
\bNarrow\ discussed in \refsect{s:eqba}, to the wider \bHKW\ cell of
\cite{HaKiWa95}, which empirically exhibits turbulence for long time scales
at $\Reynolds = 400$. Of the \eqva\ discussed above, only \tNB, \tEQsev,
\edit{\tEQeight}, and \tEQnine\ could be continued at $\Reynolds = 400$
from \edit{$\gamma = 2.5$ of} \bNarrow\ down to the fundamental wavenumber
\edit{$\gamma = 1.67$ of} \bHKW. \tEQnine\ appears in a saddle-node
bifurcation just below $\gamma = 1.67$.
The other \eqva\ terminate in saddle-node bifurcations above $\gamma = 1.67$
%  or they bifurcate into pairs of \reqva\ in pitchfork bifurcations.
\edit{or at bifurcations from other solution curves.
However, the \tLB\ and \tUB\ solutions can be continued upwards in
$\gamma$ to the first harmonic $\gamma = 3.34 = 2 \times \pi/L_z$ of \bHKW\
($L_z = 1.67$). These solutions then appear in \bHKW\ as as spanwise
``doubled'' states $2 \times \tLB$ and  $2 \times \tUB$.}
\refFig{f:HKW-eqva} shows 3D velocity fields for {\eqva} in
\bHKW: \tNB, \tEQsev, \edit{\tEQeight}, and \tEQnine\ and the
spanwise doubled $2 \times \tLB$ and $2 \times \tUB$. The upper branch
of \tEQnine\, not shown, is very similar to \tEQnine.
The properties of these solutions are listed in \reftab{t:eqbtableHKW}.

The low dissipation values of the \bHKW\ {\eqva} in \reffig{f:IvD-W03}\,(b)
suggest that they are not involved in turbulent dynamics, except perhaps
as gatekeepers to the laminar {\eqv}. We suspect that the {\eqva}
as yet undiscovered, or the already known \po\ solutions \citep{GHC08a}
do play a key role in organizing turbulent dynamics.
However, unlike the \bNarrow\ cell, we were not able to find any {\eqva}
for \bHKW\ cell from initial guesses sampled from %long-time
turbulent trajectories within the $S$-invariant subspace. This is curious,
contrasted to our success in finding {\eqva} from such
guesses in \bNarrow, and it suggests that the aspect ratios of the
\bHKW\ cell are the most incommensurate (fit the intrinsic widths of rolls
least well) compared to the roll and streak scales of
spanwise-infinite domains, which are apparent (approximately) in the
simulation of \reffig{f:bigbox}.
\edit{
The stability calculations by
\citecomm{Clever and Busse\rf{CB97}}{\cite{CB97}} indicate that the
Nagata solutions prefer a 2:1 streamwise to spanwise aspect ratio.
Hence a study of changes in solutions under variation in both
streamwise and spanwise periodicities
might shed further light on the physical nature of
these solutions.
       }

\begin{table}
\centering
\begin{tabular}{l|lllcccc}
          & ~~$\Norm{\cdot}$   & ~~$E$
          & ~$D$ & $H$ & $\dim W^u$ & $\dim W^u_{H}$  & acc.\\
\hline
mean          & 0.40   & 0.15   & 3.0      &           &    &   &  \\
\tLM          & 0      & 0.1667 & 1        & $\GPCF$   & ~0 & ~0 &  \\
$2\times$\tLB & 0.2458 & 0.1112 & 1.8122   & $H$ & ~5 & ~3 & $10^{-5}$  \\
$2\times$\tUB & 0.3202 & 0.0905 & 2.4842   & $H$ & ~6 & ~2 & $10^{-5}$   \\
\tNB          & 0.2853 & 0.0992 & 2.4625   & $S$       & 40 & 13   & $10^{-3}$ \\
\tEQsev       & 0.1261 & 0.1433 & 1.3630   & $S \times \{e, \tau_{xz}\}$ & ~6 & ~2 &  $10^{-3}$ \\
\tEQeight     & 0.1969 & 0.1186 & 1.7967   & $S \times \{e, \tau_{xz}\}$ & 19 & ~1 &  $10^{-3}$ \\
\tEQnine      & 0.3159 & 0.1175 & 2.0900   & $\{e, \sigma_{xz}\}$  & 11 & ~0  & $10^{-4}$ \\
\tEQnine (upper)     & 0.3276 & 0.1119 & 2.2000   & $\{e, \sigma_{xz}\}$  & 16 & ~5  & $10^{-4}$\\
\end{tabular}
\label{t:eqbtableHKW}
\caption[Properties of invariant solutions.]{
Properties of \eqv\ and \reqv\ solutions for \bHKW\ cell, $\Reynolds=400$,
defined as in \reftab{t:eqbtable5}.
\edit{For $2\times$\tLB and $2\times$\tUB, $H$ is the 8th-order group
generated by $\tau_z, \sigma_z \tau_x,$ and $\sigma_{xz} \tau_z^{1/4}$.}
See also \reffig{f:IvD-W03}\,(b).
}
\end{table}

\section{Conclusion and perspectives}
\label{s:conclusions}

As a turbulent flow evolves, every so often we catch a glimpse of a
familiar pattern. For any finite spatial resolution, the flow
approximately follows for a finite time a pattern belonging to a
finite alphabet of admissible fluid states, represented here by a set
of \eqv\ and \reqv\ solutions of \NS.
These are not the `modes' of the fluid; \edit{they do not
provide a decomposition of the flow into a sum of components at
different wavelengths, or a projection basis for low-dimensional
modeling.} Each solution spans the whole range of
physical scales of the turbulent fluid,
from the outer wall-to-wall scale, down to the viscous
dissipation scale. Numerical computations require sufficient
resolution to cover all of these scales, so no \edit{global}
dimension reduction is likely.
The role of invariant solutions of \NS\ is, instead,
to partition the $\infty$-dimensional \statesp\
into a finite set of neighborhoods visited by
a typical long-time turbulent fluid state.

Motivated by the recent observations of \recurrStr
s in experiments and numerical studies, we undertook here an
exploration of the hierarchy of all known \eqva\ and \reqva\ of
fully-resolved \pCf\ in order to describe the spatio-temporally
chaotic dynamics of transitionally turbulent fluid flows. Turbulent
\pC\ dynamics visualized in \statesp\ appears pieced together from
close visitations to \cohStr s connected by transient
interludes, as can be seen in \cite{GibsonMovies} animations of
\reffig{f:bigbox}. The $3D$ fluid states explored by the small
aspect-ratio \eqva\ and their unstable manifolds studied in this paper
are strikingly similar to states observed in larger
aspect-ratio simulations, such as \reffig{f:bigbox}.

For \pCf\ \eqva, \reqva\ and periodic solutions embody
a vision of turbulence as a repertoire of recurrent spatio-temporal
patterns explored by turbulent dynamics. The new \eqva\ and \reqva\
that we present here form the backbone of this repertoire. Currently,
a taxonomy of these myriad states eludes us, but emboldened by successes
in applying periodic orbit theory to the simpler, warm-up
Kuramoto-Sivashinsky problem \citep{Christiansen:97,lanCvit07,SCD07},
we are optimistic. Given a set of \eqva, the next step is to
understand how the dynamics interconnects the neighborhoods of the
invariant solutions discovered so far; a task that we address in
\citecomm{\rf{GHCV08}}{ \cite{GHCV08}} which discusses their
heteroclinic connections, and \citecomm{\rf{GHC08a}}{\cite{GHC08a}}
which discusses their \po\ solutions.

The reader might rightfully wonder what the small-aspect
periodic cells studied here have to do with physical \pCf\
and wall-bounded shear flows in general, with large
aspect ratios and physical spanwise-streamwise boundary conditions.
Indeed, the outstanding issue
that must be addressed in future work is the
small-aspect cell periodicities imposed for computational efficiency.
So far, most computations of invariant solutions have focused on
spanwise-streamwise (axial-streamwise in case of the pipe flow)
periodic cells barely large enough to allow for sustained turbulence.
Such small cells introduce dynamical artifacts
such as lack of structural stability and cell-size dependence of the
sustained turbulence states.
However, every solution that we find is also a solution of the
infinite aspect-ratio problem, \ie, a solution whose finite
$[L_x,2,L_z]$ cell tiles the infinite $3D$ \pCf. As we saw in
\refsect{s:aspect}, under a continuous variation of spanwise
length $L_z$ such solutions come in continuous families whose
fundamental wavelengths reflect the roll and streak instability
scales observed in large-aspect systems such as \reffig{f:bigbox}.
Here we can draw the inspiration from pattern-formation theory,
where the most unstable wavelengths from a continuum of unstable
solutions set the scales observed in simulations.

\begin{acknowledgments}
We would like to acknowledge F.\ Waleffe for providing his \eqv\
solution data and for his very generous guidance through the course of
this research. We greatly appreciate discussions with D.\ Viswanath,
his guidance in numerical algorithms, and for providing
his \reqv\ data. We are indebted to G.~Kawahara, L.S.~Tuckerman,
B.~Eckhardt, D.~Barkley, and
J.~Elton for inspiring discussions. P.C., J.F.G.\ and J.H.\ thank
G.~Robinson,~Jr.\ for support.
J.F.G.\ was partly supported by NSF grant DMS-0807574.
J.H.\ thanks R.~Mainieri and T.~Brown,
Institute for Physical Sciences, for partial support. Special thanks
to the Georgia Tech Student Union which generously funded our access
to the Georgia Tech Public Access Cluster Environment (GT-PACE),
essential to the computationally demanding Navier-Stokes calculations.
            \edit{
[Note added in proof: since submission of this article
it has come to our attention that
\citecomm{Itano and Generalis\rf{ItGe09}}{ \cite{ItGe09}} have
independently determined the
{\tEQsev}, {\tEQeight} \eqva.]
            }
\end{acknowledgments}

\bibliographystyle{jfm}

\bibliography{halcrow}

\begin{thebibliography}{50}
\expandafter\ifx\csname natexlab\endcsname\relax\def\natexlab#1{#1}\fi

\bibitem[Canuto {\em et~al.\/}(1988)Canuto, Hussaini, Quarteroni \&
  Zang]{Canuto88}
{\sc Canuto, C., Hussaini, M.~Y., Quarteroni, A. \& Zang, T.~A.} 1988 {\em
  Spectral Methods in Fluid Dynamics\/}. Springer-Verlag.

\bibitem[{Cherhabili} \& {Ehrenstein}(1997)]{Cherh97}
{\sc {Cherhabili}, A. \& {Ehrenstein}, U.} 1997 {Finite-amplitude equilibrium
  states in plane {C}ouette flow}. {\em J. Fluid Mech.\/} {\bf 342}, 159--177.

\bibitem[Christiansen {\em et~al.\/}(1997)Christiansen, Cvitanovi\'{c} \&
  Putkaradze]{Christiansen:97}
{\sc Christiansen, F., Cvitanovi\'{c}, P. \& Putkaradze, V.} 1997
  Spatio-temporal chaos in terms of unstable recurrent patterns. {\em
  Nonlinearity\/} {\bf 10}, 55--70.

\bibitem[Clever \& Busse(1992)]{CB92}
{\sc Clever, R.~M. \& Busse, F.~H.} 1992 Three-dimensional convection in a
  horizontal layer subjected to constant shear. {\em J.\ Fluid Mech.\/} {\bf
  234}, 511--527.

\bibitem[Clever \& Busse(1997)]{CB97}
{\sc Clever, R.~M. \& Busse, F.~H.} 1997 {Tertiary and quaternary solutions for
  plane {C}ouette flow}. {\em J.\ Fluid Mech.\/} {\bf 344}, 137--153.

\bibitem[Cvitanovi{\'c} {\em et~al.\/}(2009)Cvitanovi{\'c}, Davidchack \&
  Siminos]{SCD07}
{\sc Cvitanovi{\'c}, P., Davidchack, R.~L. \& Siminos, E.} 2009 On state space
  geometry of the kuramoto-sivashinsky flow in a periodic domain. {\tt
  arXiv:0709.2944}, SIAM J. Appl. Dynam. Systems, to appear.

\bibitem[Duguet {\em et~al.\/}(2008)Duguet, Pringle \& Kerswell]{duguet08}
{\sc Duguet, Y., Pringle, C. C.~T. \& Kerswell, R.~R.} 2008 Relative periodic
  orbits in transitional pipe flow. {\tt arXiv:0807.2580}.

\bibitem[Ehrenstein {\em et~al.\/}(2008)Ehrenstein, Nagata \& Rincon]{Ehren08}
{\sc Ehrenstein, U., Nagata, M. \& Rincon, F.} 2008 Two-dimensional nonlinear
  plane {P}oiseuille-{C}ouette flow homotopy revisited. {\em Phys. Fluids\/}
  {\bf 20}, 064103--1--4.

\bibitem[Faisst \& Eckhardt(2003)]{FE03}
{\sc Faisst, H. \& Eckhardt, B.} 2003 Traveling waves in pipe flow. {\em Phys.
  Rev. Lett.\/} {\bf 91}, 224502.

\bibitem[Frisch(1996)]{frisch}
{\sc Frisch, U.} 1996 {\em Turbulence\/}. Cambridge, UK: Cambridge University
  Press.

\bibitem[Gibson(2008{\natexlab{{\em a\/}}})]{channelflow}
{\sc Gibson, J.~F.} 2008{\natexlab{{\em a\/}}} {C}hannelflow: a spectral
  {N}avier-{S}tokes simulator in {C}++. {\em Tech. Rep.\/}. Georgia Inst. of
  Technology, {\tt {Channelflow.org}}.

\bibitem[Gibson(2008{\natexlab{{\em b\/}}})]{GibsonMovies}
{\sc Gibson, J.~F.} 2008{\natexlab{{\em b\/}}} Movies of plane {C}ouette. {\em
  Tech. Rep.\/}. Georgia Institute of Technology, {\tt
  ChaosBook.org/tutorials}.

\bibitem[Gibson \& Cvitanovi{\'c}(2009)]{GHC08a}
{\sc Gibson, J.~F. \& Cvitanovi{\'c}, P.} 2009 Periodic orbits of plane
  {C}ouette flow. In preparation.

\bibitem[Gibson {\em et~al.\/}(2008)Gibson, Halcrow \& Cvitanovi{\'c}]{GHCW07}
{\sc Gibson, J.~F., Halcrow, J. \& Cvitanovi{\'c}, P.} 2008 Visualizing the
  geometry of state-space in plane {C}ouette flow. {\em J.\ Fluid Mech.\/} {\bf
  611}, 107--130, {\tt arXiv:0705.3957}.

\bibitem[Gilmore \& Letellier(2007)]{GL-Gil07b}
{\sc Gilmore, R. \& Letellier, C.} 2007 {\em The Symmetry of Chaos\/}. Oxford:
  Oxford Univ. Press.

\bibitem[Golubitsky \& Stewart(2002)]{golubitsky2002sp}
{\sc Golubitsky, M. \& Stewart, I.} 2002 {\em {The symmetry perspective}\/}.
  Boston: Birkh{\"a}user.

\bibitem[Halcrow(2008)]{HalcrowThesis}
{\sc Halcrow, J.} 2008 Geometry of turbulence: An exploration of the
  state-space of plane {C}ouette flow. PhD thesis, School of Physics, Georgia
  Inst. of Technology, Atlanta, {\tt ChaosBook.org/projects/theses.html}.

\bibitem[Halcrow {\em et~al.\/}(2009)Halcrow, Gibson, Cvitanovi{\'c} \&
  Viswanath]{GHCV08}
{\sc Halcrow, J., Gibson, J.~F., Cvitanovi{\'c}, P. \& Viswanath, D.} 2009
  Heteroclinic connections in plane {C}ouette flow. {\em J.\ Fluid Mech.\/}
  {\bf 621}, 365--376, {\tt arXiv:0808.1865}.

\bibitem[Hamilton {\em et~al.\/}(1995)Hamilton, Kim \& Waleffe]{HaKiWa95}
{\sc Hamilton, J.~M., Kim, J. \& Waleffe, F.} 1995 Regeneration mechanisms of
  near-wall turbulence structures. {\em J. Fluid Mech.\/} {\bf 287}, 317--348.

\bibitem[Harter(1993)]{Harter93}
{\sc Harter, W.~G.} 1993 {\em Principles of Symmetry, Dynamics, and
  Spectroscopy\/}. New York: Wiley.

\bibitem[Hof {\em et~al.\/}(2004)Hof, van Doorne, Westerweel, Nieuwstadt,
  Faisst, Eckhardt, Wedin, Kerswell \& Waleffe]{science04}
{\sc Hof, B., van Doorne, C. W.~H., Westerweel, J., Nieuwstadt, F. T.~M.,
  Faisst, H., Eckhardt, B., Wedin, H., Kerswell, R.~R. \& Waleffe, F.} 2004
  Experimental observation of nonlinear traveling waves in turbulent pipe flow.
  {\em Science\/} {\bf 305}~(5690), 1594--1598, {\tt
  www.sciencemag.org/cgi/reprint/305/5690/1594.pdf}.

\bibitem[Hoyle(2006)]{hoyll06}
{\sc Hoyle, R.} 2006 {\em Pattern Formation: An Introduction to Methods\/}.
  Cambridge: Cambridge Univ. Press.

\bibitem[Itano \& Generalis(2009)]{ItGe09}
{\sc Itano, T. \& Generalis, S.~C.} 2009 Hairpin vortex solution in planar
  {Couette} flow: A tapestry of knotted vortices. {\em Phys. Rev. Lett.\/} {\bf
  102}, 114501.

\bibitem[Itano \& Toh(2001)]{IT01}
{\sc Itano, T. \& Toh, S.} 2001 The dynamics of bursting process in wall
  turbulence. {\em J.\ Phys.\ Soc.\ Japan\/} {\bf 70}, 701--714.

\bibitem[{J.E. Dennis, Jr.,} \& Schnabel(1996)]{DS}
{\sc {J.E. Dennis, Jr.,} \& Schnabel, R.~B.} 1996 {\em Numerical Methods for
  Unconstrained Optimization and Nonlinear Equations\/}. Philadelphia: SIAM.

\bibitem[Jim{\'e}nez {\em et~al.\/}(2005)Jim{\'e}nez, Kawahara, Simens, Nagata
  \& Shiba]{JKSNS05}
{\sc Jim{\'e}nez, J., Kawahara, G., Simens, M.~P., Nagata, M. \& Shiba, M.}
  2005 Characterization of near-wall turbulence in terms of equilibrium and
  bursting solutions. {\em Phys. Fluids\/} {\bf 17}, 015105.

\bibitem[Kim {\em et~al.\/}(1971)Kim, Kline \& Reynolds]{KKR71}
{\sc Kim, H., Kline, S. \& Reynolds, W.} 1971 {The production of turbulence
  near a smooth wall in a turbulent boundary layer}. {\em J. Fluid Mech.\/}
  {\bf 50}, 133--160.

\bibitem[Kleiser \& Schumann(1980)]{Kleiser80}
{\sc Kleiser, L. \& Schumann, U.} 1980 Treatment of incompressibility and
  boundary conditions in 3-{D} numerical spectral simulations of plane channel
  flows. In {\em Proc. 3rd GAMM Conf. Numerical Methods in Fluid Mechanics\/}
  (ed. E.~Hirschel), pp. 165--173. GAMM, Viewweg, Braunschweig.

\bibitem[Lan \& Cvitanovi{\'c}(2008)]{lanCvit07}
{\sc Lan, Y. \& Cvitanovi{\'c}, P.} 2008 Unstable recurrent patterns in
  {K}uramoto-{S}ivashinsky dynamics. {\em Phys. Rev. E\/} {\bf 78}, 026208,
  {\tt arXiv.org:0804.2474}.

\bibitem[Marsden \& Ratiu(1999)]{MarRat99}
{\sc Marsden, J.~E. \& Ratiu, T.~S.} 1999 {\em Introduction to Mechanics and
  Symmetry\/}. New York, NY: Springer-Verlag.

\bibitem[Nagata(1990)]{N90}
{\sc Nagata, M.} 1990 Three-dimensional finite-amplitude solutions in plane
  {C}ouette flow: bifurcation from infinity. {\em J. Fluid Mech.\/} {\bf 217},
  519--527.

\bibitem[Nagata(1997)]{N97}
{\sc Nagata, M.} 1997 {Three-dimensional traveling-wave solutions in plane
  {C}ouette flow}. {\em Phys. Rev. E\/} {\bf 55}, 2023--2025.

\bibitem[Peyret(2002)]{Peyret02}
{\sc Peyret, R.} 2002 {\em Spectral Methods for Incompressible Flows\/}.
  Springer-Verlag.

\bibitem[Pringle \& Kerswell(2007)]{Pringle07}
{\sc Pringle, C.~T. \& Kerswell, R.~R.} 2007 Asymmetric, helical, and
  mirror-symmetric traveling waves in pipe flow. {\em Phys.\ Rev.\ Lett.\/}
  {\bf 99}, 074502.

\bibitem[{Rincon}(2007)]{Rincon07}
{\sc {Rincon}, F.} 2007 {On the existence of two-dimensional nonlinear steady
  states in plane {C}ouette flow}. {\em Phys. Fluids\/} {\bf 19}, 4105--+, {\tt
  arXiv:0706.1165}.

\bibitem[Schmiegel(1999)]{Schmi99}
{\sc Schmiegel, A.} 1999 Transition to turbulence in linearly stable shear
  flows. PhD thesis, Philipps-Universit{\"a}t Marburg, available on {\tt
  archiv.ub.uni-marburg.de/diss/z2000/0062}.

\bibitem[Schneider {\em et~al.\/}(2008)Schneider, Gibson, Lagha, Lillo \&
  Eckhardt]{SGLDE08}
{\sc Schneider, T., Gibson, J., Lagha, M., Lillo, F.~D. \& Eckhardt, B.} 2008
  Laminar-turbulent boundary in plane {Couette} flow. {\em Phys. Rev. E.\/}
  {\bf 78}, 037301, {\tt arXiv:0805.1015}.

\bibitem[Skufca {\em et~al.\/}(2006)Skufca, Yorke \& Eckhardt]{SYE05}
{\sc Skufca, J.~D., Yorke, J.~A. \& Eckhardt, B.} 2006 Edge of chaos in a
  parallel shear flow. {\em Phys. Rev. Lett.\/} {\bf 96}~(17), 174101.

\bibitem[Tuckerman \& Barkley(2002)]{TuckBar03}
{\sc Tuckerman, L.~S. \& Barkley, D.} 2002 Symmetry breaking and turbulence in
  perturbed plane {C}ouette flow. {\em Theoretical and Computational Fluid
  Dynamics\/} {\bf 16}, 91--97, {\tt arXiv:physics/0312051}.

\bibitem[Viswanath(2007)]{Visw07b}
{\sc Viswanath, D.} 2007 Recurrent motions within plane {C}ouette turbulence.
  {\em J. Fluid Mech.\/} {\bf 580}, 339--358, {\tt arXiv:physics/0604062}.

\bibitem[Viswanath(2008)]{Visw07a}
{\sc Viswanath, D.} 2008 The dynamics of transition to turbulence in plane
  {Couette} flow. In {\em Mathematics and Computation, a Contemporary View. The
  Abel Symposium 2006\/}, {\em Abel Symposia\/}, vol.~3. Berlin:
  Springer-Verlag, {\tt arXiv:physics/0701337}.

\bibitem[Waleffe(1990)]{Waleffe90}
{\sc Waleffe, F.} 1990 Proposal for a self-sustaining mechanism in shear flows.
  Center for Turbulence Research, Stanford University/NASA Ames, unpublished
  preprint (1990).

\bibitem[Waleffe(1995)]{W95a}
{\sc Waleffe, F.} 1995 Hydrodynamic stability and turbulence: beyond transients
  to a self-sustaining process. {\em Stud. Applied Math.\/} {\bf 95}, 319--343.

\bibitem[Waleffe(1997)]{W97}
{\sc Waleffe, F.} 1997 On a self-sustaining process in shear flows. {\em Phys.
  Fluids\/} {\bf 9}, 883--900.

\bibitem[Waleffe(1998)]{W98}
{\sc Waleffe, F.} 1998 Three-dimensional coherent states in plane shear flows.
  {\em Phys. Rev. Lett.\/} {\bf 81}, 4140--4143.

\bibitem[Waleffe(2001)]{W01}
{\sc Waleffe, F.} 2001 Exact coherent structures in channel flow. {\em J. Fluid
  Mech.\/} {\bf 435}, 93--102.

\bibitem[Waleffe(2002)]{W02}
{\sc Waleffe, F.} 2002 Exact coherent structures and their instabilities:
  Toward a dynamical-system theory of shear turbulence. In {\em Proceedings of
  the International Symposium on ``Dynamics and Statistics of Coherent
  Structures in Turbulence: Roles of Elementary Vortices''\/} (ed. S.~Kida),
  pp. 115--128. National Center of Sciences, Tokyo, Japan.

\bibitem[Waleffe(2003)]{W03}
{\sc Waleffe, F.} 2003 Homotopy of exact coherent structures in plane shear
  flows. {\em Phys. Fluids\/} {\bf 15}, 1517--1543.

\bibitem[Wang {\em et~al.\/}(2007)Wang, Gibson \& Waleffe]{WGW07}
{\sc Wang, J., Gibson, J.~F. \& Waleffe, F.} 2007 Lower branch coherent states
  in shear flows: transition and control. {\em Phys. Rev. Lett.\/} {\bf
  98}~(20).

\bibitem[Wedin \& Kerswell(2004)]{WK04}
{\sc Wedin, H. \& Kerswell, R.~R.} 2004 Exact coherent structures in pipe flow:
  travelling wave solutions. {\em J. Fluid Mech.\/} {\bf 508}, 333--371.

\end{thebibliography}

\end{document}